\documentclass[12pt,letterpaper]{article}

\usepackage{hyperref}

\usepackage{graphicx}
\usepackage{color}
\usepackage[usenames,dvipsnames]{xcolor}
\newcommand{\HW}[1]{\textcolor{Black}{#1}}  
\newcommand{\HWn}[1]{\textcolor{Black}{#1}} 
\newcommand{\SFS}[1]{\textcolor{Black}{#1}}  

\usepackage{natbib}
\usepackage{amsmath}
\usepackage{amssymb}
\usepackage{algorithm}
\usepackage{url}
\usepackage{fancyhdr}
\usepackage{booktabs}
\usepackage{enumerate}

\usepackage{threeparttable} 
\usepackage{siunitx} 
\usepackage{pdflscape} 
\usepackage{adjustbox} 
\usepackage{bm}
\usepackage{multibib}
\newcites{supp}{Supplementary References}

\usepackage{subcaption}

\newtheorem{thm}{Theorem}

\newtheorem{prop}[thm]{Proposition}

\newcommand{\omegav}{\boldsymbol{\omega}}

\newcommand{\blind}{1}

\newcommand{\Amult}{\boldsymbol{A}} 

\newcommand{\betac}{\beta} 

\newcommand{\betad}{d} 
\newcommand{\betav}{\boldsymbol{\betac}}

\newcommand{\betat}[1]{\betav_{#1}} 
\newcommand{\bfz}{{\mathbf{0}}}
\newcommand{\Betadis}[1]{\mathcal{B}\left(#1\right)}
\newcommand{\Bino}[1]{\mbox{\rm BiNom}\left(#1\right)}
\newcommand{\bs}{b} 
\newcommand{\Bs}{B} 
\newcommand{\br}{{\mathbf{\bs}}} 
\newcommand{\Br}{{\mathbf{\Bs}}} 
\newcommand{\Bt}[1]{\mathbf{B}_{#1}}

\newcommand{\Count}[1]{\#\{#1\}}

\newcommand{\deltatilde}{\tilde \delta}

\newcommand{\Diag}[1]{\mbox{\rm Diag}\left(#1\right)}
\newcommand{\Dir}[1]{ \mathcal{D}\left(#1\right)}

\newcommand{\error}{\epsilon} 

\newcommand{\errordiff}{\varepsilon}

\newcommand{\EVfs}{\mathrm{EV}}
\newcommand{\Exp}[1]{\mathcal{E}\left(#1\right)}

\newcommand{\Ferror}{F_{\errordiff}}
\newcommand{\ferror}{f_{\errordiff}}
\newcommand{\filtg}{\mathbf{g}}

\newcommand{\gammatilde}{\tilde \gamma}

\newcommand{\Gamfun}[1]{\Gamma (#1)}
\newcommand{\Gammainv}[1]{\mathcal{G}^{-1} \left(#1\right)}
\newcommand{\GenLogistic}[2]{\mathcal{GL}_{\mbox{\tiny #1}}\left(#2\right)}

\newcommand{\identm}{{\mathbf I}}

\newcommand{\iid}{\mbox{\rm i.i.d.}}
\newcommand{\im}[1]{^{(#1)}}
\newcommand{\indic}[1]{I\{#1\}}

\newcommand{\kfK}{{\mathbf{K}}}  
\newcommand{\kfm}{\hat{\betav}}  
\newcommand{\kfP}{{\mathbf{P}}}  
\newcommand{\kfQc}{Q}  
\newcommand{\kfQ}{{\mathbf{\kfQc}}}  
\newcommand{\kfwc}{ w}
\newcommand{\kfw}{{\mathbf{\kfwc}}}  
\newcommand{\Kt}[1]{\kfK_{#1}}

\newcommand{\labset}{L}
\newcommand{\lambdav}{\boldsymbol{\lambda}}
\newcommand{\Logistic}{\mathcal{LO}}
\newcommand{\logit}{\mbox{\rm logit}}

\newcommand{\mb}{\boldsymbol{m}}

\newcommand{\mr}{{\mathbf{m}}} 

\newcommand{\mbg}{m_b}
\newcommand{\mbgk}{m_{kb}}
\newcommand{\mg}{m_\gamma}
\newcommand{\mgk}{m_{k \gamma}}
\newcommand{\Mulnom}[1]{\mbox{\rm MulNom}\left(#1\right)}

\newcommand{\new}{^\mathrm{new}}
\newcommand{\newS}{^{\footnotesize LS}}
\newcommand{\Normal}[1]{ \mathcal{N}\left(#1\right)}
\newcommand{\Normalpdf}{\phi}
\newcommand{\Normalcdf}{\Phi}
\newcommand{\Normult}[2]{ \mathcal{N} _{#1}\left(#2\right)}

\newcommand{\omegaH}{\omega}

\newcommand{\PG}[1]{\mathcal{PG}\left(#1\right)}

\newcommand{\pl}{\pi}  
\newcommand{\Probsym}{\mbox{\rm Pr}}
\newcommand{\Prob}[1]{\Probsym (#1)}
\newcommand{\Pt}[2]{\kfP_{#1|#2}} 

\newcommand{\Qrcm}{{\mathbf{\kfQ}}}

\newcommand{\rvX}{X}

\newcommand{\scale}{\omega}
\newcommand{\scalev}{\boldsymbol{\omega}}

\newcommand{\shift}{^{\footnotesize L}}
\newcommand{\Star}[1]{\tilde #1}
\newcommand{\Stc}[2]{S_{#1|#2}} 

\newcommand{\trans}[1]{#1 ^{\top}} 

\newcommand{\Uniform}[1]{\mathcal{U}\left[#1\right]}

\newcommand{\wt}[1]{\kfw_{#1}}

\newcommand{\Xbeta}{{\mathbf \Xz}}
\newcommand{\xt}[1]{{\betav}_{#1}}

\newcommand{\xthat}[2]{\kfm_{#1|#2}} 
\newcommand{\Xz}{x}

\newcommand{\yc}{y}
\newcommand{\ym}{{\mathbf \yc}}
\newcommand{\ypro}{{u}}
\newcommand{\yprodiff}{z}
\newcommand{\yprodiffv}{\mathbf{\yprodiff}}
\newcommand{\yprohat}[2]{\hat{\yprodiff}_{#1|#2}} 

\newcommand{\yprow}{{w}}
\newcommand{\yprowtilde}{\tilde{w}}
\newcommand{\yprov}{{v}}
\newcommand{\yproztilde}{\tilde{z}}

\newcommand{\yprovall}{{\mathbf{\ypro}}}
\newcommand{\yprovtilde}{\tilde{v}}

\newcommand{\zs}{z}
\newcommand{\zv}{\boldsymbol{\zs}} 

\begin{document}

\def\spacingset#1{\renewcommand{\baselinestretch}
{#1}\small\normalsize} \spacingset{1}


\if1\blind
{
  \title{\bf Ultimate {P}\'{o}lya  {G}amma Samplers -- Efficient {MCMC} for possibly imbalanced binary and categorical
  data}
  \author{
  \large Gregor Zens (WU Wien) \\[6pt]
  \large Sylvia Fr\"uhwirth-Schnatter~(WU Wien)\\[6pt]
     \large Helga Wagner (JKU Linz)
 }
  \date{{\small \today}}
  \maketitle
} \fi
\if0\blind
{
  \bigskip
  \bigskip
  \bigskip
  \begin{center}
    \title{\bf  Ultimate {P}\'{o}lya  {G}amma Samplers -- Efficient {MCMC} for possibly imbalanced binary and categorical
  data}
  \end{center}
  \medskip
} \fi
\begin{abstract}

Modeling binary and categorical data is one of the most commonly encountered tasks of applied statisticians and econometricians. While Bayesian methods in this context have been available for decades now, they often require a high level of familiarity with Bayesian statistics or suffer from issues such as low sampling efficiency. To contribute to the accessibility of Bayesian models for binary and categorical data, we introduce novel latent variable representations based on P\'{o}lya-Gamma random variables for a range of commonly encountered logistic regression models. From these latent variable representations, new Gibbs sampling algorithms for binary, binomial, and multinomial logit models are derived. All models allow for a conditionally Gaussian likelihood representation, rendering extensions to more complex modeling frameworks such as state space models straightforward. However, sampling efficiency may still be an issue in these data augmentation based estimation frameworks. To counteract this, novel marginal data augmentation strategies are developed and discussed in detail. The merits of our approach are illustrated through extensive simulations and real data applications.

\end{abstract}
\noindent%

\newpage
\spacingset{1.5}

\setcounter{page}{2}

\section{Introduction}

Applied statisticians and econometricians commonly have to deal with modeling binary or categorical outcome variables.
 Widely used tools for analyzing such data  include probit as well as binary, multinomial, and binomial logit regression models. Bayesian approaches toward inference are very useful  in this context,
 \HW{as} they allow to easily 
 extend the standard regression framework to more complex settings such as random effects or state space models. However, as opposed to regression models with Gaussian outcomes, their implementation \HW{can be} demanding from a computational viewpoint \HW{\citep{cho-rid:lea}}.

 One strategy  to implement sampling-based inference relies on importance sampling \citep{zel-ros:bay}
 or various types of Metropolis-Hastings (MH) algorithms \citep{ros-etal:bay}, exploiting directly the non-Gaussian likelihood. However, these algorithms often require
 careful tuning and substantial experience with Bayesian computation, especially in more complex frameworks like state space models. 

  Routine Bayesian computation for these type of data more often relies on Markov Chain Monte Carlo (MCMC) algorithms based on
  data augmentation (DA, \citealp{tan-won:cal}). As shown by the seminal paper of \citet{alb-chi:bay_ana}, the binary probit model admits a latent variable representation where the latent variable equation is linear in the unknown parameters,
   with an error term following a standard normal distribution. As simulating the latent variables is easy when the parameters are known, the latent variable representation admits a straightforward Gibbs sampler using one level of DA, where the unknown parameters are sampled from a conditionally Gaussian model. This strategy works also for more complex models, such as probit state space or random effects models.\footnote{There is also an active literature on posterior simulation tools for probit and logit regression models that does not rely on DA. For instance, \citet{durante2019conjugate} introduces a framework for conjugate analysis of the probit model that has been generalized subsequently, see \citet{anceschi2022bayesian} for a review. \citet{sen2020efficient} use a sampling framework for logistic regression based on \SFS{piecewise} deterministic Monte Carlo processes. We provide a discussion of these and other alternative methods in Appendix A.1.}

However, MCMC estimation based on DA is less straightforward for a logit model which still admits a latent
variable representation that is linear in the unknown parameters, but exhibits an error term that follows a logistic distribution.
Related latent variable representations with non-Gaussian errors exist for multinomial logit (MNL) models (\citealp{fru-fru:dat}) and logistic regression models for binomial outcomes \citep{fus-etal:eff}. While the latent variables usually can be easily sampled,
sampling the unknown parameters is more involved due to the non-Gaussian 
error terms.

A common solution relies on a scale-mixture representation of the non-Gaussian error distribution and introduces the corresponding scale parameters as a second level of DA. Conveniently, the unknown model parameters can then be sampled from a conditionally Gaussian regression model. Examples include a representation of the logistic distribution involving the Kolmogoroff-Smirnov distribution \citep{hol-hel:bay} and highly accurate finite scale-mixture approximations \citep{fru-fru:aux,fru-fru:dat,fru-etal:imp}. A seminal paper in this context is \citet{pol-etal:bay_inf} which avoids any explicit latent variable representation. They derive the \textit{P\'{o}lya-Gamma sampler} that exploits a 
mixture representation of the non-Gaussian likelihood of the marginal model based on the P\'{o}lya-Gamma distribution and works with a single level of DA.

In the present paper, we propose a new sampling scheme involving the P\'{o}lya-Gamma distribution. 
Instead of working with the marginal model, we introduce a new  
mixture representation of the logistic distribution based on the
P\'{o}lya-Gamma distribution in the latent variable representation of the logit model. Similar to \citet{hol-hel:bay} and \citet{fru-fru:dat}, we use DA and introduce the P\'{o}lya-Gamma mixing variables 
 as a second set of latent variables.
 Our new P\'{o}lya-Gamma mixture representation 
 has the advantage that the joint 
 posterior distribution of all 
 augmented variables is easy to sample from, as the P\'{o}lya-Gamma mixing variable follows a tilted P\'{o}lya-Gamma distribution conditional on the latent utilities. 
 This allows to sample the unknown model parameters from a conditionally Gaussian model, facilitating posterior simulation in complex frameworks such as state space or random effects models.

A commonly encountered challenge when working with MCMC methods based on DA is poor mixing. For binary and categorical regressions, this issue is especially pronounced for imbalanced data, where the success probability is either close to zero or one for the majority of the observations, see the excellent work of \citet{joh-etal:mcm}. Neither the original  P\'{o}lya-Gamma  sampler of \citet{pol-etal:bay_inf} with a single level of DA, nor our new  P\'{o}lya-Gamma  sampler  with two levels of DA, are an exception to this rule.

To resolve this issue, we introduce imbalanced marginal data augmentation (iMDA) as a \textit{boosting} strategy
to make our new sampler as well as the original probit sampler of \citet{alb-chi:bay_ana} robust to possibly imbalanced data. This strategy is inspired by earlier work
on marginal data augmentation (MDA) for binary and
categorical data \citep{liu-wu:par,mcc-etal:bay, van-men:art,ima-van:bay}. Starting from
a latent variable  representation of the binary model,
we expand the latent variable representation with the help of two unidentified `working parameters'. One parameter is a global scale parameter for the latent variable, which has been shown to improve mixing considerably by \citet{liu-wu:par}, among others. However, this strategy alone does not resolve slow mixing when dealing with highly imbalanced data. To address this, we introduce an additional, unknown location parameter, which improves mixing considerably in the case of imbalanced data. As iMDA only works in the context of a latent variable representation, this strategy cannot be applied to the original P\'{o}lya-Gamma  sampler of \citet{pol-etal:bay_inf} due to the lack of such a representation. In comparison, our new P\'{o}lya-Gamma representation of the logit model is very generic and is easily combined with iMDA, not only for binary regression models, but also for more flexible models such as binary state space models. 
We refer to a sampling strategy combining a P\'{o}lya-Gamma 
mixture representation with iMDA as an \textit{ultimate} P\'{o}lya-Gamma (UPG) sampler due to its efficiency.

A further contribution of the present paper is to show that such an UPG \HW{sampler} 
can 
be derived for other 
non-Gaussian regression problems, including models for \SFS{categorical} and binomial data. \HW{For the MNL model, commonly 
a logit model  based on a (partial) differenced random utility model (dRUM) representation
is applied to sample the category specific parameters, see e.g. \citet{hol-hel:bay,fru-fru:dat} or \citet{pol-etal:bay_inf}.
Utilizing this
partial dRUM representation, we derive a new sampler for the MNL model in the present paper.  
Since the latent variable equation is linear in the unknown parameters and involves a logistic error distribution, we use 
once more the P\'{o}lya-Gamma mixture representation of the logistic distribution and introduce the  mixing variables as additional latent variables.} For binomial models, a latent variable representation  which did not involve a choice equation was introduced  by
 \citet{fus-etal:eff}. Since an explicit choice equation is needed to apply iMDA, we derive a new
 latent variable representation for binomial data which 
 involves error terms that follow generalized logistic distributions. 
 We introduce P\'{o}lya-Gamma 
 mixture representations of these distributions and utilize the resulting 
 auxiliary variables as an additional latent layer.
 Both for MNL models and for binomial models, this DA scheme leads to a conditionally Gaussian posterior and allows to sample all unknowns through efficient block moves. Again, we apply iMDA to derive UPG samplers which mix well, also in the context of imbalanced data. 
 
 \HW{Overall, we find that the various algorithms show} highly competitive performance when compared to alternative DA frameworks, which we demonstrate via extensive simulation studies. In addition, we present real world data examples that further illustrate the merits of \HW{our approach}. 
 The underlying algorithms for probit regression and logistic regression models for binary, \SFS{categorical} and binomial outcomes have been made available in the R package \texttt{UPG}, which is available on \textit{CRAN} \citep{zens2021efficient}.  

The remainder of the paper is structured as follows. Section~\ref{section:binary}
introduces the UPG sampler. 
This sampling strategy is extended
 to \SFS{categorical} data in Section~\ref{section:MNL} and to binomial data in  Section~\ref{section:binomial}.
 In Section~\ref{sec:sim_da}, the UPG 
 sampler 
 is compared to alternative DA algorithms. 
   Section~\ref{section:appext} applies the framework to    binary state space models and discusses the utility of the approach in the context of mixture-of-experts models.  Section~\ref{section:conclude} concludes.

\section{Ultimate P\'{o}lya-{G}amma samplers for binary data} \label{section:binary}

\subsection{Latent variable representations for binary data}  \label{section:latbinary}

Models for a vector of $N$ binary observations $\ym=(y_1,\dots,y_N)$ are defined by
\begin{equation}\label{eq:binm}
  \Prob{y_i=1|\lambda_i}= \Ferror (\log \lambda_i),
  \end{equation}
  where $\lambda_i$ depends on exogenous variables and unknown parameters $\betav$, e.g.,
  $ \log \lambda_{i}= \Xbeta_{i} \betav$ in a standard binary regression model.
  Choosing  the cdf $\Ferror(\errordiff)=\Normalcdf (\errordiff)$ of the standard normal distribution
 leads to the probit model $\Prob{y_i=1|\lambda_i} = \Normalcdf( \log \lambda_i)$, whereas
  the cdf $\Ferror(\errordiff)=e^\errordiff/(1+e^\errordiff)$ of the logistic distribution 
leads to the  logit model $$\Prob{y_i=1|\lambda_i}= \lambda_i/(1+ \lambda_i) .$$

\noindent
A latent variable representation of model (\ref{eq:binm}) involving a latent utility  $\yprodiff_{i}$ is given by:
 \begin{eqnarray}
y_i=\indic{ \yprodiff_{i}>0}, && \displaystyle   \yprodiff_{i} = \log \lambda_i + \errordiff_{i},
 \quad  \errordiff_{i} \sim \ferror(\errordiff_i),\label{eq:binlat}
\end{eqnarray}
where  $\ferror(\errordiff)= \Ferror' (\errordiff) = \Normalpdf (\errordiff)$ is equal to the standard normal  pdf for a probit model and equal to $\ferror(\errordiff)= e^{\errordiff}/(1+e^{\errordiff})^2$ for a logit model.

In Bayesian inference,  the set of observed data
$\ym=(y_1, \dots,y_N)$ can be augmented with the latent variables $\zv =(\yprodiff_{1}, \dots, \yprodiff_{N})$ in (\ref{eq:binlat}) to obtain the set of complete data $(\zv,\ym)$, facilitating the implementation of MCMC algorithms. As shown by \citet{alb-chi:bay_ana},
this single level of DA involving  $\zv$ leads to a straightforward Gibbs sampler for the probit model. \HW{With $ \log \lambda_{i}= \Xbeta_{i} \betav$, the
following two-step sampling  {\it Scheme~1} can be set up under a Gaussian prior $p(\betav)$:}

\begin{itemize} \itemsep -2mm
\item[(Z)] Given $\betav$, sample the latent variables $\yprodiff_{i}$ for each $i=1, \dots,N$
independently  from $p(\yprodiff_{i}|\betav, \ym)$
    (see Appendix A.4.1);
       \item[(P)] sample the unknown parameters $\betav$ conditional on  
   $\zv$
   from the Gaussian posterior $p(\betav|\zv,\ym)$ derived from  regression model (\ref{eq:binlat}).
   \end{itemize}
Two main challenges are associated with such MCMC schemes, namely slow convergence and a lack of closed form posteriors for the unknown parameters, such as $p(\betav|\zv,\ym)$, outside of probit models. We address both issues in the present paper.

First, to boost MCMC convergence, we rely on MDA in the spirit of \citet{liu-wu:par}. In that paper, the scale-based transformation $\tilde{\yprodiff}_{i}=\sqrt{\delta}\yprodiff_{i}$, depending on a `working parameter' $\delta$, is used to define the expanded probit regression model
\begin{eqnarray}
y_i=\indic{ \tilde\yprodiff_{i}>0 }, && \displaystyle   \tilde\yprodiff_{i} = \sqrt{\delta} \Xbeta_{i} \betav + \tilde\errordiff_{i},
 \quad  \tilde \errordiff_{i} \sim \Normal{0,\delta}.\label{til:binlat}
\end{eqnarray}
 In model (\ref{til:binlat}), the  likelihood $p(\tilde{\zv}|\delta)$ of $\tilde{\zv}=(\tilde{\yprodiff}_{1}, \dots,
  \tilde{\yprodiff}_{N})$, marginalized w.r.t.~$\betav$,  is available in closed form and yields
an inverse Gamma posterior $p(\delta|\tilde{\zv})$ under a
conjugate prior $p(\delta)$. Assuming prior independence of $\delta$ and $\betav$, this allows to rescale the latent variables $\zv$ without involving $\betav$.
 Specifically, a draw $\Star{\delta}$ from
 the working prior $p(\delta)$ is used to `propose' a scale-move $\tilde{\yprodiff}_{i}=\sqrt{\Star{\delta}}\yprodiff_{i}$ in system (\ref{til:binlat}), based solely on prior information. Then, an updated value $\delta \new$ is sampled
 from the posterior $p(\delta|\tilde{\zv})$ and the proposed scale-move is immediately `corrected' (using a posteriori information) via the inverse transformation  $\yprodiff_{i} \new = \tilde{\yprodiff}_{i}/\sqrt{\delta \new}$, before
 $\betav$ is updated conditional on $\zv \new$.
  This  extends {\it Scheme~1}  to {\it Scheme~2}:
\begin{itemize} \itemsep -2mm
\item[(Z)] Sample from $p(\zv|\betav,\ym)$ as in \textit{Scheme 1};
\item[(B-S)] move from $\zv$ to $\zv \new$ using a scale-based expansion move under prior $p(\delta)$;
\item[(P)] sample from $p(\betav| \zv \new ,\ym)$  as in \textit{Scheme 1}.
   \end{itemize}
The boosted \textit{Scheme~2} always provides better convergence results than \textit{Scheme~1}, see  \citet{van-men:art} and \cite{hob-mar:the} for further theoretical results. Indeed, as an example in \citet{liu-wu:par} illustrates, Step~(B-S) improves efficiency considerably in cases where the coefficient of determination in the latent regression model is large, as long as the data are balanced. However, DA schemes are in general known to be slowly mixing for imbalanced data sets where only a few cases with $y_i=1$ or $y_i=0$ among the $N$ data points are observed \citep{joh-etal:mcm}. Indeed, sampling under \textit{Scheme~2} is still highly inefficient in such cases, as will be illustrated in Section~\ref{intuition}.

A first major contribution of this paper is to protect DA algorithms for binary and categorical data against imbalanced data by using, in addition to a scale-based transformation, a location-based expansion $\tilde{\yprodiff}_{i}=\yprodiff_{i} + \gamma$, depending on a `working parameter' $\gamma$, to define the expanded version
\begin{eqnarray}
y_i=\indic{ \tilde\yprodiff_{i}>\gamma  }, && \displaystyle   \tilde\yprodiff_{i} = \gamma  + \log \lambda_i + \errordiff_{i}, \label{loc:binlat}
\end{eqnarray}
 of the binary regression model (\ref{eq:binlat}).  
 
 As opposed to (\ref{til:binlat}), the choice equation in (\ref{loc:binlat}) depends on $\gamma$ and defines a likelihood $p(\ym|\gamma, \tilde{\zv})$. In a probit regression model, the  likelihood $p(\tilde{\zv}|\gamma)$ \SFS{of the latent data}, marginalized w.r.t.~$\betav$,  is available in closed form. In combination with the likelihood $p(\ym|\gamma, \tilde{\zv})$
 and a Gaussian working prior $p(\gamma)$, a Gaussian posterior $p(\gamma|\tilde{\zv},\ym)$, truncated  the interval  $[L,U)$ defined by, respectively, the maximum utility $L$ of the outcomes where $y_i=0$ and the minimum utility $U$ of the outcomes where $y_i=1$, is obtained.  
 Assuming prior independence of $\gamma$ and $\betav$ then allows to shift the latent variables $\zv$ without involving $\betav$. Similar to the scale-based expansion, a location-move $\tilde{z}_i = z_i + \tilde{\gamma}$ is proposed using 
 \SFS{a draw $\tilde{\gamma}$ from the working prior $p(\gamma)$}, 
 before being immediately `corrected' via the inverse transformation $z_i \new = \tilde{z}_i - \gamma \new = z_i + \tilde{\gamma} - \gamma \new$ using 
 \SFS{a draw $\gamma \new$ from the posterior distribution $p(\gamma|\tilde{\zv},\ym)$,}
 see Section~\ref{sec:mcmcdetails}
 for further details. Subsequently, the regression coefficients $\betav$ are sampled conditional on $\zv \new$. We find that performing such a location-based expansion step before a scale-based transformation yields dramatic improvement compared to \textit{Scheme~1} and \textit{Scheme~2}, also in cases where the data are imbalanced, see Section~\ref{intuition} and Section~\ref{sec:sim_da} for further illustration.

A second main contribution of the paper is to take
 location-based and scale-based parameter expansion
  beyond the probit regression model by introducing new
  latent variable representations for binary,   binomial and multinomial logit  models.
For binary logit models, a second level of DA is introduced to deal with the logistic error term. For this, we apply a new  
mixture representation
 of the logistic distribution, 
  \begin{eqnarray}  \label{PGHW}
   \ferror(\errordiff_i)= e^{\errordiff_i}/(1+e^{\errordiff_i})^2 = \frac{1}{4}\int  e ^{- \omegaH _i \, \errordiff_i^2/2} p(\omegaH_i) d\,\omegaH_i ,
    \end{eqnarray}
   where   $\omegaH_i \sim \PG{2,0}$ follows a  P\'{o}lya-Gamma distribution \citep{pol-etal:bay_inf},
   see Appendix A.2.1 and A.2.2 for details. 
    This  representation is very convenient, as the conditional posterior $ \omegaH_i~|~\errordiff_{i} \sim \PG{2, |\errordiff_{i}|}$    of $\omegaH_i$
  given $\errordiff_{i}$ is a tilted  P\'{o}lya-Gamma distribution
 which is easy to sample from, see \citet{pol-etal:bay_inf}.
  For a binary logit model with $ \log \lambda_{i}= \Xbeta_{i} \betav$,
   this new representation allows constructing a  P\'{o}lya-Gamma sampler that extends \textit{Scheme~1} in the following way:
   \begin{itemize} \itemsep -2mm
   \item[(Z)] sample the latent variable $\yprodiff_{i}$ from $p(\yprodiff_{i}|\betav,y_i)$ independently for each $i$
   in the latent variable model (\ref{eq:binlat})
   (see Algorithm~\ref{alg:bin} and Appendix A.4.1) 
        and sample the scale
    parameter $\omegaH_i$  conditional on  $\yprodiff_{i}$ and $\betav$ from
   $\omegaH _i| \yprodiff_{i}, \betav \sim \PG{2, |\yprodiff_{i} - \Xbeta_{i} \betav |}$;

       \item[(P)] sample the unknown parameters $\betav$ conditional on  the latent variables
   $\yprodiffv=(\yprodiff_{1}, \dots, \yprodiff_{N})$ and $\scalev=( \scale_1,\dots, \scale_N)$ from the
   conditionally Gaussian posterior  $p(\betav|\scalev, \yprodiffv, \ym)$.
            \end{itemize}
  While this scheme is easy to implement, it can be slowly mixing, like any such sampler. To deal with this issue, we additionally include the two parameter expansion steps introduced above, performing first a location-based and then a scale-based transformation. We refer to the resulting sampling scheme as \textit{Scheme~3} and provide full theoretical and computational details  in Section~\ref{sec:mcmcdetails}. In later sections, we extend this strategy to logistic regression models for \SFS{categorical} and binomial outcomes.

While our boosting  strategy is inspired by \citet{liu-wu:par} and related to earlier work
on MDA for binary and
categorical data \citep{mcc-etal:bay, van-men:art,ima-van:bay}, it generalizes this literature in several aspects. Importantly, it works for any binary data model with a latent variable representation. In addition, freeing the location of the threshold $\gamma$ in model (\ref{loc:binlat}) 
 leads to an MCMC scheme that is well mixing, even in cases of extremely imbalanced data, see much of the remainder of this article for further illustration.
 %
%
A related strategy to improve mixing behaviour in the context of data augmentation algorithms is outlined in \citet{dua-etal:sca}. In their contribution, the authors use location and scale parameters to reparametrize augmented likelihood functions in binary and count data regression models. These calibration parameters have to be set manually, and the authors propose an involved optimization procedure based on large sample arguments and approximations to determine suitable values. The resulting algorithms show efficiency gains that are comparable to the marginal data augmentation proposed in this article when analyzing data sets with many observations and rare outcomes. A potential downside of the approach of \citet{dua-etal:sca} is that the optimization procedure relies on the inverse of the observed Fisher information and the sampler uses Metropolis-Hastings updates. Both may result in scaling issues when many covariates are present. In such settings, a pure data augmentation approach as proposed in this article may prove more effective. Importantly, our approach is also fully automatic and does not rely on any approximations in the sense that tuning-free and exact Gibbs updates for the location and scale parameters are derived. We give details on the resulting posterior simulation scheme in Subsection~\ref{sec:mcmcdetails}. Before presenting these details, we illustrate the specific roles of the location and scale parameters $\gamma$ and $\delta$ using heuristic arguments in the next subsection.

\subsection{Illustration and intuition} \label{intuition}

\begin{figure}[t!]
  \centering
\begin{subfigure}{0.57\textwidth}
  \centering
 \includegraphics[width=\linewidth]{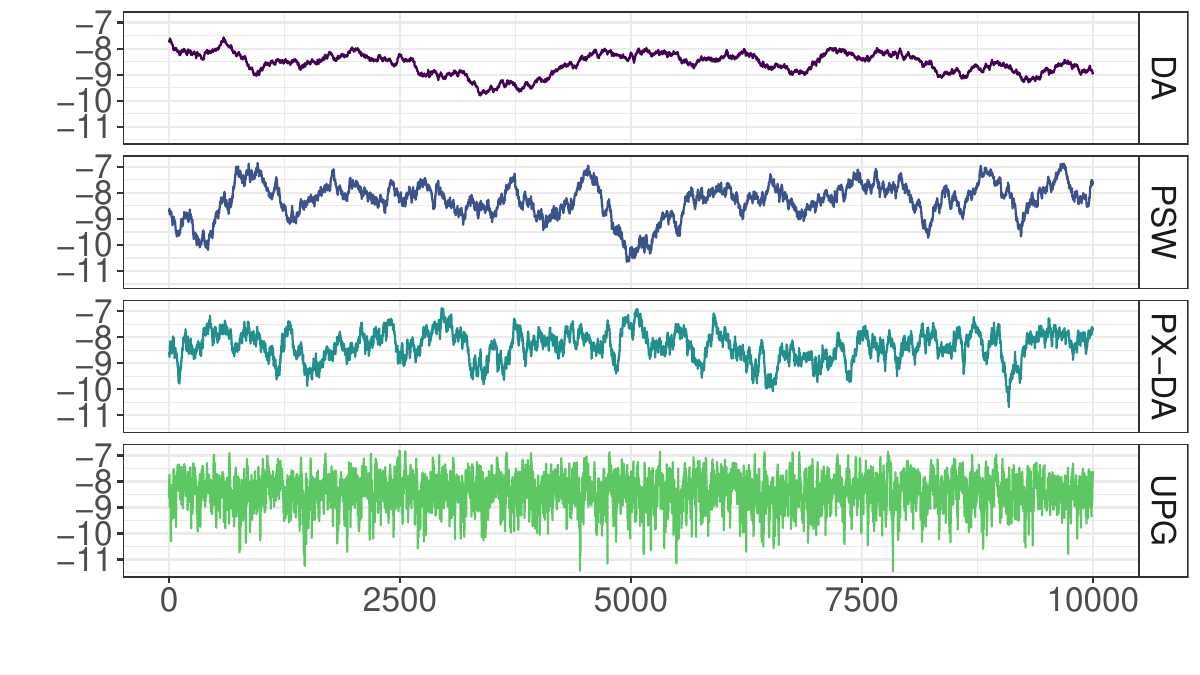}
\end{subfigure}\hfill
\begin{subfigure}{0.42\textwidth}
  \centering
  \includegraphics[width=\linewidth]{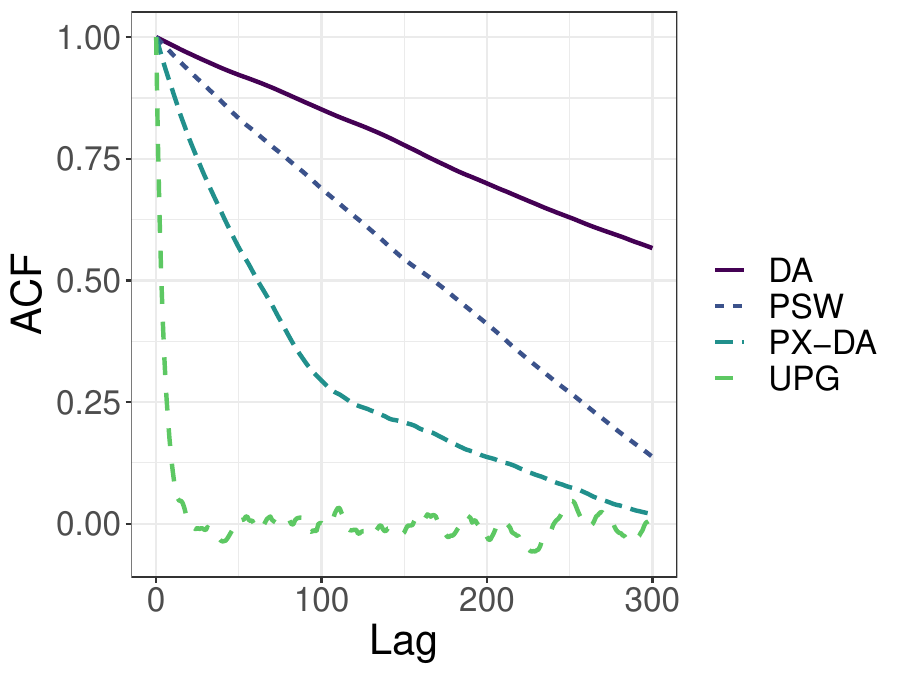}
\end{subfigure}\\
\caption{MCMC draws and corresponding autocorrelation functions of an intercept only logistic regression model fitted  using plain data augmentation (DA, \textit{Scheme 1}), the original P\'{o}lya-{G}amma sampler (PSW), a MDA sampler with scale-based expansion (PX-DA, \textit{Scheme 2}) and a DA sampler with scale- and location-based expansion (UPG, \textit{Scheme 3}). Two out of $N = 10,000$ binary observations are non-zero.}
\label{fig:traceplots_acf}
\end{figure}

As a first illustration of the potential merits of the proposed iMDA scheme in imbalanced logistic regression settings, we compare estimation efficiency of the popular P\'{o}lya-{G}amma sampler from \citet{pol-etal:bay_inf} with a plain DA sampler as in \textit{Scheme 1}, a scale-based parameter expansion scheme (as in \textit{Scheme 2}) and the proposed approach based on location-based and scale-based expansion (as in \textit{Scheme 3}) in \autoref{fig:traceplots_acf}. A more systematic comparison 
will be given in Section \ref{sec:sim_da}. It is clearly visible that the UPG sampler 
outperforms all other samplers in terms of efficiency. Notably, these efficiency gains are realized despite 
introducing two layers of latent auxiliary variables, which usually increases autocorrelation in the posterior draws significantly. This is counteracted by our novel iMDA strategy based on the working parameters $\gamma$ and $\delta$. 

We start with the role of $\delta$, the working parameter used for scale-based expansion of the latent utility equation. Broadly speaking, this scale-based expansion will be highly effective in scenarios where the coefficient of determination in the latent utility model is high. In such settings, the current parameter draw almost perfectly determines the location of the latent utilities and vice versa. As a result, the MCMC chain is only able to move very slowly. To resolve this issue, $\delta$ artificially decreases the coefficient of determination via increasing the 
error variance in the latent utility equation. In turn, this decreases the dependency of the latent utilities and the regression coefficients, directly enabling larger steps of the Markov chain. In other words, $\delta$ is used to 
make 
 the posterior of the latent utilities in the expanded model 
 more diffuse than the posterior of the utilities in the original model. Similar as well as more formal arguments and further illustration of such scale-based expansion steps have been discussed for instance in \citet{liu-wu:par} or \citet{ima-van:bay}. 

However, a scale-based expansion alone is usually not enough to fully resolve the issue that step sizes become small relative to the range of the high posterior density region in imbalanced data settings \citep{joh-etal:mcm}. This can be seen from the unsatisfactory performance of 
the PX-DA sampler 
in \autoref{fig:traceplots_acf} and has also been discussed in \citet{dua-etal:sca}. In our approach, this issue is effectively offset through the location-based expansion of the latent utility model. In this subsection, we aim to illustrate the mechanism behind this strategy 
through a small numerical exercise and defer full  
details to Section~\ref{sec:mcmcdetails}.  

\begin{figure}[t]
  \centering
\begin{subfigure}{0.3\textwidth}
  \centering
  \includegraphics[width=\linewidth]{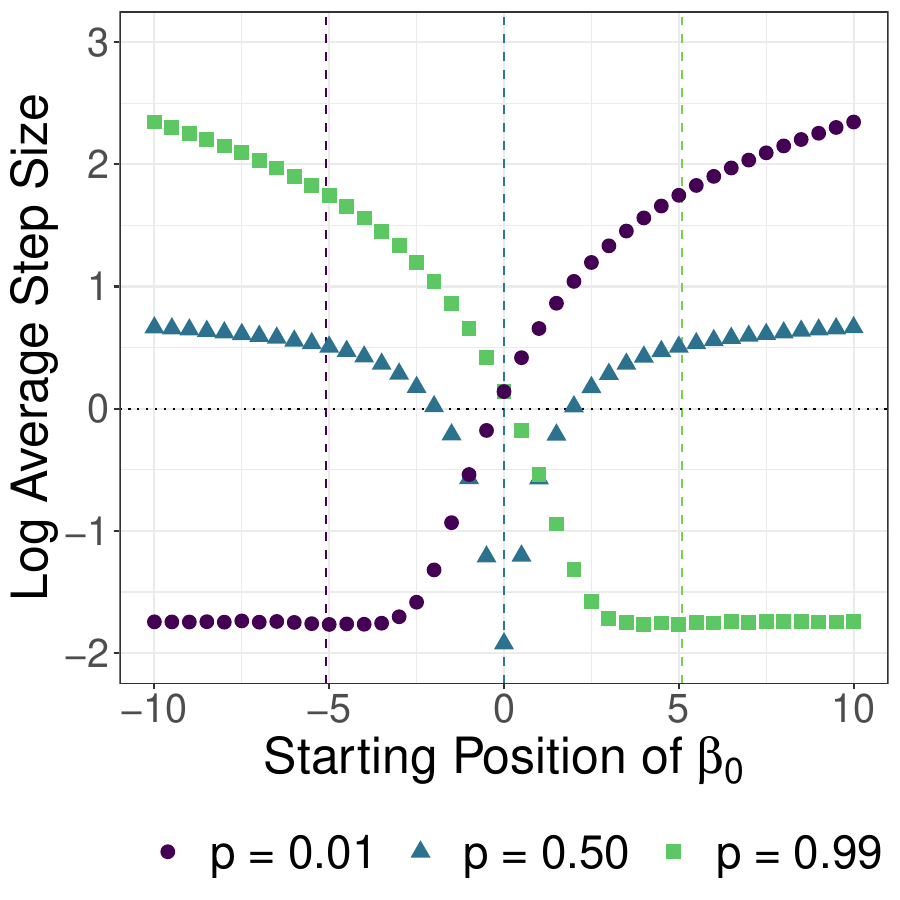}
\end{subfigure}
\begin{subfigure}{0.3\textwidth}
  \centering
  \includegraphics[width=\linewidth]{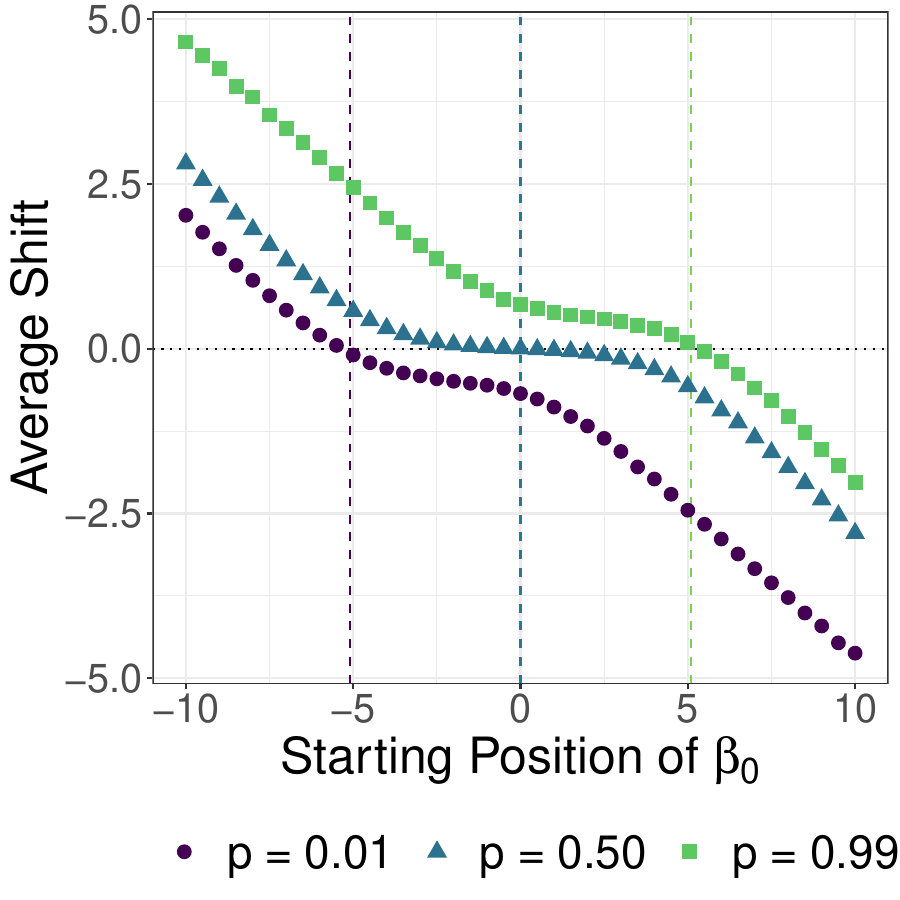}
\end{subfigure}
\begin{subfigure}{0.3\textwidth}
  \centering
  \includegraphics[width=\linewidth]{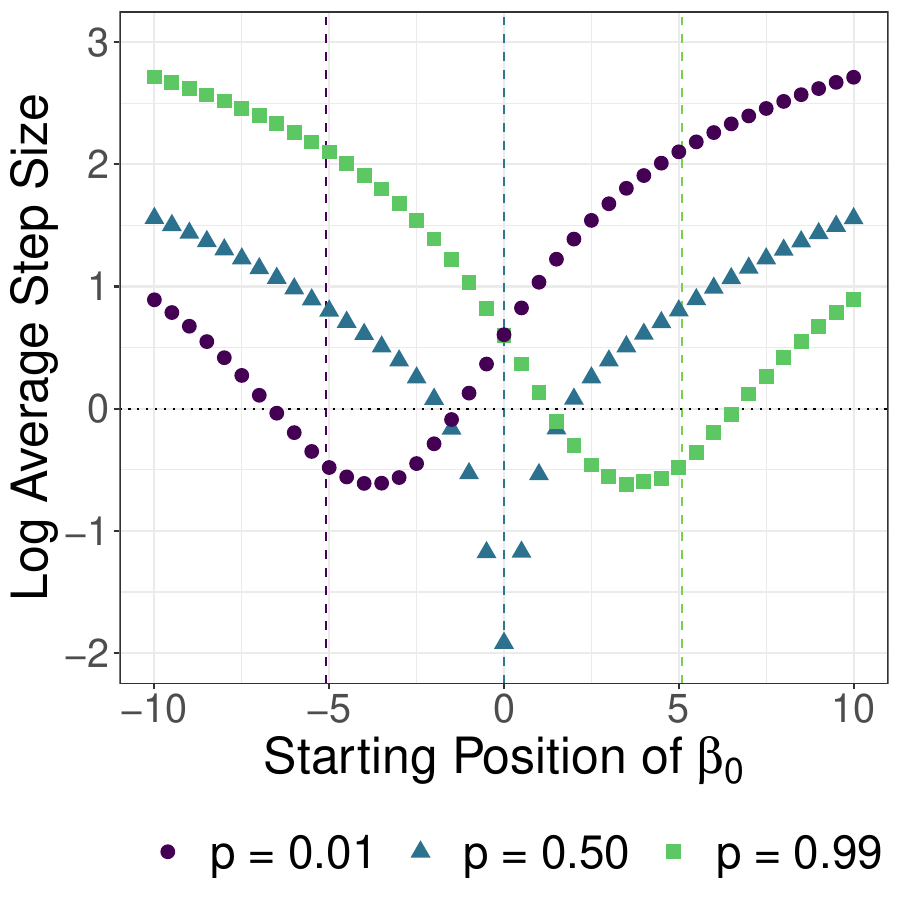}
\end{subfigure}

\caption{Illustration of the mechanism behind the location-based expansion. (left) Log average step size of a plain DA sampler. (middle) Realized shift of utilities. (right) Log average step size of a sampler with location-based expansion. Dotted lines are the means of the posterior distribution of $\beta_0$ under a $N(0,100)$ prior.}
\label{fig:imda_intuition}
\end{figure}

To investigate how the location-based expansion influences step sizes of the Markov chain, we consider three data sets with $N = 100$ observations each. One data set is balanced, while the others are imbalanced, with success probabilities $99\%$ and $1\%$, respectively. We simulate $25,000$ replications of a single MCMC iteration for a grid of starting positions of the intercept $\beta_0$, using $\mathcal{N}(0, 100)$ prior distributions for both $\beta_0$ and $\gamma$. For each starting position and for each replication, we save the absolute step size of a plain DA sampler ({\em Scheme 1}) and the step size of a sampler with an additional location-based expansion step, as well as the realized shift $\tilde{\gamma} - \gamma \new $ in the sampler including the location-based expansion step.

The results are summarized in Figure~\ref{fig:imda_intuition}. The left panel shows the log average step size of the plain DA scheme. It is evident that step sizes decrease significantly when exploring posterior regions that reach far into the positive (negative) part of the real line in imbalanced scenarios with high (low) success probabilities. The purpose of the location-based expansion is to counteract this issue via shifting the utilities by $\tilde{\gamma} - \gamma \new$, directly leading to larger step sizes of the Markov chain. The average shift for each data set and value of $\beta_0$ is depicted in the middle panel of Figure~\ref{fig:imda_intuition}. The magnitude of the shift, $|\tilde{\gamma} - \gamma \new|$, is equivalent to the increase in step size in the location-expanded sampler. While step sizes increase everywhere, the improvement is particularly large in the tails of the posterior density in imbalanced data sets, where standard DA algorithms are usually highly inefficient. In addition, the shift-move evidently acts as a `push into the right direction' that systematically leads the Markov chain back towards the highest posterior density region, effectively avoiding staying in the tails of the posterior distribution for too long. The log average step sizes of 
the PX-DA 
sampler are shown in the right panel of Figure~\ref{fig:imda_intuition}. As expected from the preceding discussion, the most significant step size improvements are observed in the tail regions of the posterior distribution in the imbalanced cases.


\subsection{MCMC details for binary logit regression models} \label{sec:mcmcdetails}

The latent utility representation of the  binary logit model is
 \begin{eqnarray}
y_i=\indic{ \yprodiff_{i}>0}, && \displaystyle   \yprodiff_{i} = \Xbeta_{i} \betav + \errordiff_{i},
 \quad  \errordiff_{i} \sim  \Logistic,\label{eq:binreg1}
\end{eqnarray}
where \SFS{$\Logistic$ is the
logistic distribution. We}
assume $ \betav \sim \Normult{\betad}{\bfz,\Amult_0 }$ follows a multivariate Gaussian distribution a priori, \SFS{where} $\Amult_0$ is either fixed  or equipped with a hierarchical structure, e.g., to define
a shrinkage prior (see e.g., \citealp{pii-veh:spa}). The first block of the MCMC scheme consists of two steps that simulate the two sets of latent variables, $\yprodiffv$ and $\scalev$. Given $\betav$ and the outcome $y_i$,
we sample  $\yprodiff_{i}$  for each $i$ from  $p(\yprodiff_{i}|  \lambda _i, y_i)$
in the logistic model (\ref{eq:binreg1}) where  $\log \lambda _i=\Xbeta_{i} \betav$. Then, the P\'{o}lya-Gamma scale parameters are simulated from $\omegaH _i| \yprodiff_{i}, \betav \sim \PG{2, |\yprodiff_{i} - \Xbeta_{i} \betav |}$. 

For given latent variables, a location-based parameter expansion step, based on a working prior  $p(\gamma) = \Normal{0,G_0}$, is then applied. For this, a prior draw $\Star{\gamma}  \sim \Normal{0,G_0} $ is used to `propose', for each $i=1,\dots,N$, a location move $\tilde{\yprodiff}_{i} = \yprodiff_{i}+ \Star{\gamma}$ in the expanded model
\begin{eqnarray}
y_i=\indic{ \tilde{\yprodiff}_{i}>\gamma}, \qquad \tilde{\yprodiff}_{i} = \gamma + \Xbeta_{i} \betav +  \errordiff_{i}, 
 \label{expand3}
\end{eqnarray}
while $\omegaH_i$ is unaffected. Conditional on the latent variables
 $\tilde{\zv}=(\tilde{\yprodiff}_{1}, \dots, \tilde{\yprodiff}_{N})$
 and $\scalev=( \scale_1,\dots, \scale_N)$, but marginally w.r.t.~$\betav$,
the conditional distribution  $ \gamma|\scalev, \tilde{\zv} \sim \Normal{g_N,G_N}$  is  Gaussian   where:
\begin{eqnarray}
&& \displaystyle G_N= (G_0^{-1} + \sum_{i=1}^N \scale_i  -  \trans{\mbg}  \Br_N \mbg )^{-1}, \quad
  g_N = G_N ( \mg - \trans{\mbg}\Br_N  \mr_N (\tilde{\zv}) ),
 \label{GVVV}\\
 &&   \displaystyle\Br_N = (\Amult_0^{-1} +  \sum_{i=1}^N \scale_i \trans{\Xbeta_{i}} \Xbeta_{i}  )^{-1},  \quad  \mr_N (\tilde{\zv})   = \sum_{i=1}^N \scale_i \trans{\Xbeta_{i}} \tilde{\yprodiff}_i , \quad \mbg =  \sum_{i=1}^N \scale_i \trans{\Xbeta_{i}}, \quad \mg= \sum_{i=1}^N \scale_i \tilde{\yprodiff}_i,
  \nonumber
\end{eqnarray}
as is easily shown, see Appendix A.4.2. 
  Since the choice equation in (\ref{expand3}) depends on $\gamma$,
  $p(\gamma|\scalev, \tilde{\zv})$ has to be combined with the likelihood $p(\ym|\gamma,\tilde{\zv})$
  of the observed outcomes $\ym=(y_1,\dots,y_N)$ to define the posterior $p(\gamma|\scalev, \tilde{\zv},\ym )$.
  The derivation of the likelihood $p(\ym|\gamma,\tilde{\zv})$  is a generic step in our sampler which
  does not involve the specification of $\lambda_i$:
   \begin{eqnarray} \label{ipost34}
p(\ym|\gamma, \tilde{\zv}) \propto
\prod_{i:y_i=0}  \indic{\gamma \geq\tilde{\yprodiff}_{i}}
\prod_{i:y_i=1} \indic{\gamma < \tilde{\yprodiff}_{i}} \propto   \indic{ L(\Star{\gamma})  \leq \gamma <  U(\Star{\gamma})} ,
\end{eqnarray}
 where $\indic{\cdot}$ is the indicator function and $L(\Star{\gamma})=\max_{i:y_i=0} \tilde{\yprodiff}_{i} = \max_{i:y_i=0} \yprodiff_{i} + \Star{\gamma} $ and
 $U(\Star{\gamma})=\min_{i:y_i=1} \tilde{\yprodiff}_{i}= \min_{i:y_i=1} \yprodiff_{i} + \Star{\gamma}$. 
  If no  outcome $y_i=0$ is observed, then    $L (\Star{\gamma})=-\infty$;
 if no outcome $y_i=1$ is observed, then  $U (\Star{\gamma})= + \infty$.
Hence,
  $p(\gamma|\scalev, \tilde{\zv},\ym ) \propto
  p(\ym|\gamma,\tilde{\zv}) p(\gamma|\scalev, \tilde{\zv})
  $ is equal to a truncated version of the Gaussian posterior (\ref{GVVV}):
  \begin{eqnarray}  \label{GVVVtr}
\gamma|\scalev, \tilde{\zv},\ym  \sim \Normal{g_N,G_N} \indic{  L(\Star{\gamma})  \leq \gamma <  U(\Star{\gamma}) }.
\end{eqnarray}
An updated working parameter $\gamma \new$ is sampled from (\ref{GVVVtr}) and the proposed location-based move is `corrected' based on a posteriori information by defining the shifted utilities $ \yprodiff_{i} \shift  = \tilde{\yprodiff}_{i} - \gamma \new = \yprodiff_{i} + \Star{\gamma}-\gamma \new$. 

\begin{algorithm}[t]
  \caption{The ultimate P\'{o}lya-Gamma sampler for binary data.}  \label{alg:bin}
  \footnotesize
  Choose starting values for  $\lambdav=(\lambda_1, \dots, \lambda_N)$  and repeat the
  following steps: \vspace*{-2mm}
 \begin{itemize} \itemsep -1mm
  \item[(Z)]   For each $i=1, \dots, N$,
  sample $\yprodiff_{i}= \log \lambda_i +\Ferror^{-1} ( y_i +    U_i (1-y_i -\pi_i))$
   in model (\ref{eq:binlat}), where $U_i \sim \Uniform{0,1}$, $\pi_i 
   =\Ferror(\log \lambda_i)$, and $\Ferror^{-1}(p) = \Normalcdf^{-1}(p)$ for
the probit and  $\Ferror^{-1}(p)= \log p - \log (1- p)$ for the logit model.
  For a logit model, sample   $ \omegaH_i | \yprodiff_{i}, \log \lambda_i  \sim \PG{2, |\yprodiff_{i}-\log \lambda_i |}$.

   \item[(B-L)] Location-based parameter expansion:
   sample $\Star{\gamma}  \sim \Normal{0,G_0} $ and
 propose utilities $\tilde{\yprodiff}_{i} = \yprodiff_{i}+ \Star{\gamma}$ for $i=1,\dots,N$.  Sample  $\gamma \new $ from $ \gamma| \scalev, \tilde{\zv},\ym$ and
 define shifted utilities $ \yprodiff_{i} \shift = \tilde{\yprodiff}_{i} - \gamma \new $.
 For a binary regression model,   $p(\gamma|\scalev, \tilde{\zv},\ym )$ is given by the truncated Gaussian-posterior in (\ref{GVVVtr}).

 \item[(B-S)] Scale-based parameter expansion:
sample  $\Star{\delta} \sim \Gammainv{d_0,D_0}$ and
sample  $\delta \new $ from $ \delta | \Star{\delta}, \zv \shift, \scalev$.
   Define rescaled utilities  $ \yprodiff_{i} \newS = \sqrt{\Star{\delta}/\delta \new} \yprodiff_{i} \shift $.
 For a binary regression model,  $ \delta | \Star{\delta}, \zv \shift, \scalev \sim \Gammainv{d_N,D_N(\Star{\delta}) }$ is an inverse Gamma distribution,
     with  $d_N$ and $D_N(\Star{\delta})$ given by (\ref{oszbet}).

  \item [(P)] Sample the unknown parameter in  $\log \lambda_{i}$ conditional on
   $ \zv \newS$.   For a binary regression model, $\betav|\delta \new, \Star{\delta}, \zv \shift,\scalev
 \sim \Normal{\sqrt{\Star{\delta}/\delta \new}  \Br_N \mr_N (\zv \shift)  ,  \Br_N }$ where
 $\mr_N (\zv \shift)$ and $\Br_N$
  are given by (\ref{GVVV}).
\end{itemize}
\end{algorithm}

This location-based move is followed by a scale-based expansion, using
an inverse Gamma $\Gammainv{d_0,D_0}$ working prior $p(\delta)$.
 Similar to before, $\Star{\delta} $ 
 is sampled from  $p(\delta)$ and used to propose, for each $i=1,\dots,N$, a scale-based move $\tilde{\yprodiff}_{i} = \sqrt{\Star{\delta}} \yprodiff_{i} \shift $ in the expanded model
\begin{eqnarray}
&& \displaystyle
y_i=\indic{ \tilde{\yprodiff}_{i}>0}, \qquad \tilde{\yprodiff}_{i} = \sqrt\delta \Xbeta_{i} \betav + \sqrt\delta \errordiff_{i}.
 \label{expand4}
\end{eqnarray}
Conditional on the  P\'{o}lya-Gamma scale parameters $\omegaH_i$, it follows that
 \begin{eqnarray*}  
 p(\tilde{\yprodiff}_{i}| \omegaH_i, \delta, \betav ) \propto \frac{1}{\sqrt\delta}
 \exp\left\{  - \frac{\omegaH_i}{2} \left( \frac{\tilde{\yprodiff}_{i}}{\sqrt\delta} - \Xbeta_{i} \betav \right) ^2 \right\}=  \frac{1}{\sqrt\delta}\exp\left\{  - \frac{\omegaH_i}{2} \left( \sqrt \frac{\Star{\delta}}{\delta} \yprodiff_{i} \shift - \Xbeta_{i} \betav \right) ^2 \right\}.
     \end{eqnarray*}
 Hence, conditional on $\delta$, $\Star{\delta}$ and the shifted utilities 
 $\zv \shift=( \yprodiff_{1} \shift, \dots, \yprodiff_{N} \shift)$,
the posterior  $\betav|\delta, \Star{\delta}, \zv \shift,\scalev  \sim
 \Normal{\sqrt{ \Star{\delta} /\delta} \br_N  ,  \Br_N } $ is
Gaussian with $ \br_N = \Br_N \mr_N (\zv \shift)$ and $\mr_N (\zv \shift)$ and $\Br_N$, as in (\ref{GVVV}).
Furthermore, conditional on
 $\zv \shift$, but marginally w.r.t.~$\betav$, 
the posterior $\delta| \Star{\delta}, \zv \shift,\scalev \sim \Gammainv{d_N,D_N(\Star{\delta}) }$ is inverse Gamma
with following moments:
 \begin{eqnarray} \label{oszbet}
&& d_N=d_0+  \frac{N}{2},  \qquad D_N (\Star{\delta})= D_0 +  \frac{\Star{\delta}}{2}
\left( \sum_{i=1}^N \scale_i (\yprodiff_{i} \shift -\Xbeta_{i} \br_N )^2 +
 \trans{\br_N } \Amult_0^{-1} \br_N \right).
\end{eqnarray}
 An updated working parameter $\delta \new$ is sampled from
 $\Gammainv{d_N,D_N(\Star{\delta}) }$ and the proposed scale-based move is corrected by defining the rescaled utilities
 $ \yprodiff_{i} \newS = \sqrt{\Star{\delta}/\delta \new}\yprodiff_{i} \shift$.
 This concludes the scale-based expansion and $\betav|\zv \newS,\scalev$ is sampled conditional on
 $\zv \newS$ or, equivalently, from
 the Gaussian posterior $\betav|\delta \new, \Star{\delta}, \zv \shift,\scalev
 \sim \Normal{\sqrt{\Star{\delta}/\delta \new} \br_N  ,  \Br_N }$.
 As Algorithm~\ref{alg:bin} illustrates,
 many steps in this ultimate P\'{o}lya-Gamma (UPG) sampler are generic and easily extended to more complex models
for binary data, as will be illustrated in Section~\ref{section:appext}.

\section{\mbox{Ultimate P\'{o}lya-{G}amma samplers for \SFS{categorical} data}} \label{section:MNL}

Let  $\left\{y_{i}\right\}$, $i=1,\dots, N$, be a sequence of categorical data,
 where  $y_{i}$ is equal to one of at least three unordered
categories. The categories are  labeled by $\labset=\{0,\dots,m\}$,
and for any $k$ the set of all categories but $k$ is denoted by
$\labset_{-k}= L  \setminus \{k\}$.
 We assume that the observations are mutually independent and that for each $k \in L$ the
probability of $y_{i}$ taking the value $k$ depends on covariates $\Xbeta_{i}$   in the following way:
 \begin{eqnarray} \label{mraglam}
\Prob{y_{i}=k |\betav_0,\dots,\betav_m}
=\pl_{ki} (\betav_0,\dots,\betav_m)  
= \displaystyle  \frac{ \exp(\Xbeta_{i} \betav_{k})}
 {\displaystyle\sum_{l=0}^m \exp(\Xbeta_{i}\betav_{l})},
 \end{eqnarray}
 where $\betav_0, \dots, \betav_m$ are category specific  unknown
 parameters of dimension $\betad$.  To make the model identifiable, the parameter $\betav_{k_0}$ of a baseline
category $k_0$ is set equal to $\bfz$: $\betav_{k_0}=\bfz$. Thus, the
parameter $\betav_k$ is relative to the baseline category $k_0$ in terms of the change in
log-odds. In the following, we assume without
loss of generality that $k_0=0$.
 A more general version of the multinomial logit (MNL)  model (again with baseline   $k_0=0$) reads:
 \begin{eqnarray}  \label{mragmm}
\Prob{y_{i}=k |\betav } = \displaystyle
  \lambda_{ki}/ (1 + \sum_{l=1}^m  \lambda_{li} ),
 \end{eqnarray}
 where $ \lambda_{1i}, \dots,  \lambda_{mi}$ depend  on unknown parameters $\betav$, \HW{while $\lambda_{0i}=1$}. 
 For the standard MNL regression model (\ref{mraglam}), for instance,
$ \log \lambda_{ki}= \Xbeta_{i} \betav_{k}$ for  $k=1, \dots, m$.

Our starting point is writing the MNL model as a random utility
model (RUM), see \citet{mcf:con}:
\begin{eqnarray}
&& \displaystyle   \ypro_{ki}=  \log \lambda_{ki}  +
\error_{ki}, \quad k=0, \dots,m, \label{obspoida1}\\
&& \displaystyle y_{i }=k \Leftrightarrow   \ypro_{ki} = \max_{l \in \labset}
 \ypro_{li} .\label{osbmunom} 
\end{eqnarray}
  Thus the  observed category is equal to the category with
maximal utility. If the  errors $\error_{0i}, \dots, \error_{mi}$   in (\ref{obspoida1})
  are \iid\ random variables from an extreme value ($\EVfs$) distribution, then the
MNL model (\ref{mragmm}) results as marginal distribution of the categorical variable $y_i$.

Conditional on  
$y_i$, the posterior distribution
   $p(\yprovall_{ki}|\lambda_{ki}, y_i)$  	of the latent utilities $\yprovall_{ki}=(
   \ypro_{0i}, \ldots, \ypro_{mi})$
 is of closed form and easy to sample from, see
 Proposition~\ref{lemmaRUM} which is proven in Appendix A.3.

 \begin{prop}\label{lemmaRUM}
 Given $y_i$, realisations from the  
 distribution 
 $p(\ypro_{0i}, \ldots, \ypro_{mi}|\lambda_{1i}\dots,\lambda_{mi}, y_i)$
 can be represented as:
 \begin{eqnarray} \label{simRUM}
  e ^{- \ypro_{ki}} =
   - \frac{\log U_{i}}{1+ \sum_{l=1}^m \lambda_{li}}  
  -   \frac{\log V_{ki} }{\lambda_{ki}} \indic{y_{i}\neq k} , \quad k=0, \ldots, m,
  \end{eqnarray}
where
$U_{i}$ and  $ V_{0i}, \ldots, V_{mi}$ are $m+1$ iid uniform random numbers.
\end{prop}
Utilizing Proposition~\ref{lemmaRUM} together with a mixture approximation of the extreme value distribution to sample all unknown parameters jointly via two levels of data augmentation \citep{fru-fru:aux} turned out to be inefficient.
 Noting that the choice equation (\ref{osbmunom})  can be rewritten as a choice between any category $k$ and all its alternatives in $\labset_{-k}$, \citet{fru-fru:dat} derive the partial dRUM representation of a
 RUM model and show that it has an explicit form  if the errors  $\error_{0i}, \dots, \error_{mi}$   in (\ref{obspoida1})
  are \iid\ random variables from an extreme value  distribution. 
  
  For a multinomial regression model this yields following  well-known representation
  (see e.g. \citealp{hol-hel:bay}):
\begin{eqnarray} \label{pdlatTR} 
&& \yprodiff_{ki}= \Xbeta_i \betav_k - \xi_{ki} (\betav_{-k})  + \errordiff_{ki}, \quad 
\errordiff_{ki} \sim  \Logistic ,\label{pdchoice}\\
&& \displaystyle y_{i }= 
\left\{ \begin{array}{ll}
k, &   \yprodiff_{ki} > 0,  \\
\neq k, &  \yprodiff_{ki} \leq 0.
\end{array} \right. \label{locTR}
\end{eqnarray}
where the error term $\errordiff_{ki}$  follows a logistic distribution,  $\yprodiff _{ki} =  \ypro_{ki} - \max_{\ell \in \labset_{-k}} \ypro_{\ell i}$ is the utility gap between category $k$ and all its alternatives 
and the offset $\xi_{ki} (\betav_{-k})$  is defined as:
\begin{eqnarray*}
&& \xi_{ki} (\betav_{-k})= \log \left(1+ \sum_{\ell \neq \{k,0\}} \exp ( \Xbeta_i \betav_\ell)\right).
\end{eqnarray*}
While \citet{fru-fru:dat} use a very accurate finite mixture approximation for the logistic distribution for MCMC estimation, in the present paper we derive an ultimate P\'{o}lya-Gamma sampler
based on the partial dRUM representation and proceed  similarly  as in Section~\ref{section:binary}.
We utilize  Proposition~\ref{lemmaRUM} 
to sample the utilities $\ypro_{0i}, \ldots, \ypro_{mi}$ in the RUM model (\ref{obspoida1}) and to define
the utility gap $\yprodiff _{ki} $ between category $k$ and all its alternatives. Given the utility gap $\yprodiff _{ki} $,
we  exploit the P\'{o}lya-Gamma  mixture  representation of the logistic distribution  in (\ref{pdchoice})  with category specific latent variables $\scale_{ki}$ which are sampled from  $\scale_{ki}| \SFS{\betav},
\yprodiff_{ki} 
 \sim 
 \PG{2, |\errordiff_{ki}|}
 $, where 
$\errordiff_{ki}=\yprodiff_{ki}- \Xbeta_i \betav_k 
  + 
  \xi_{ki} (\betav_{-k})
  $.   
 
To handle imbalanced data, we apply location- and
scale-based boosting  as in Section~\ref{section:binary}  with category specific working parameters
$\gamma_k$ and $\delta_k$.
For instance, location-based boosting using $\yproztilde_{ki} = \yprodiff_{ki}  + \gammatilde_k$ where $\Star{\gamma_k} \sim \Normal{0,  G_0}$, yields the following expanded model:
\begin{eqnarray} \label{aggnewpd} 
&& \yproztilde_{ki}=  \gamma_k +\Xbeta_i \betav_k - \xi_{ki} (\betav_{-k})  +  \errordiff_{ki},  \quad   \errordiff_{ki} \sim  \Logistic ,\\ 
&& \displaystyle y_{i }= 
\left\{ \begin{array}{ll}
k, &   \yproztilde_{ki} > \gamma_k,  \\
\neq k, &  \yproztilde_{ki} \leq  \gamma_k.
\end{array} \right. \label{pdlocpd}
\end{eqnarray}
Conditional on the latent variables $\scalev_k=(\scale_{k1}, \dots, \scale_{kN})$ and $\tilde \yprodiffv _k=(\yproztilde_{k1},\dots, \yproztilde_{kN})$, equation (\ref{aggnewpd}) defines
a Gaussian \SFS{posterior} distribution $p(\gamma_k|\SFS{\betav_{-k}}, \scalev_k, \tilde \yprodiffv_k) $, marginally w.r.t.~$\betav_k$. Similarly as in Section~\ref{section:binary}, the choice equation (\ref{pdlocpd}) defines a likelihood function $p(\ym|\gamma_k, \tilde\yprodiffv_k)$ which restricts  $\gamma_k  $ to the interval $[L(\Star{\gamma_k}),U(\Star{\gamma_k}))$, where
$L(\Star{\gamma_k})=  \max _{y_i \neq k} \yprodiff _{ki} +  \Star{\gamma_k} $ and
$U(\Star{\gamma_k})=  \min _{y_i= k} \yprodiff _{ki} +  \Star{\gamma_k} $.
Full details on the UPG sampler for
  multinomial logistic regression models are provided in Appendix A.4.3.

\section{Ultimate P\'{o}lya-{G}amma samplers for binomial data} \label{section:binomial}

 In this section, we consider  models  with binomial outcomes, i.e., models of the form
     \begin{eqnarray}  \label{obsljuhbin}
 y_i \sim \Bino{N_i, \pi_i}, \qquad  \logit \, \pi_i= \log \lambda_i, \qquad i=1,\dots, N,
 \end{eqnarray}
 with $\log \lambda_i=  \Xbeta_{i} \betav$ for a standard binomial regression model.
  As shown in  \citet{joh-etal:mcm},  Bayesian inference for binomial regression models based on the P\'{o}lya-Gamma sampler 
  \citep{pol-etal:bay_inf} is sensitive to imbalanced data.
Similarly, the latent variable representation of binomial models of \citet{fus-etal:eff} is sensitive to imbalanced data, as we will show in Section~\ref{sec:sim_da}.
As for a logit model (which results for $N_i \equiv 1$), applying iMDA would be an option to improve mixing. However,
\citet{fus-etal:eff} provide no explicit choice equation, which is needed for iMDA.  
The goal of this section is to define an UPG 
sampler 
  which combines a new latent variable representation of binomial models, based on P\'{o}lya-{G}amma mixture
 representations of generalized logistic distributions with iMDA  to protect the algorithm against imbalanced data.

\subsection{A new latent variable representation for binomial data}

In Theorem~\ref{theo1}, we introduce a new latent variable representation for binomial outcomes where two latent variable equations, both linear in $\log \lambda_i$, with error terms following generalized  logistic distributions are utilized.  An explicit choice equation is provided
   which relates  latent variables $\yprow_i$ and $\yprov_i$
   to the observed binomial outcome $y_i$.
   We show in Theorem~\ref{lemma2U} that, conditional on  $y_i$,
      the posterior distribution of the latent variables is of closed form and easy to sample from, see  Appendix A.3. for a proof of both theorems.

\begin{thm}{{\bf Latent variable representation of a  binomial model}}\label{theo1}
For $0 < y_i< N_i$,  a binomial logistic model has the following  random utility  representation:
\begin{eqnarray} \label{aggnew3U}
&& \yprow_{i} =  \log \lambda_i  +   \errordiff_{\yprow ,i},  \quad   \errordiff_{\yprow ,i} \sim {\GenLogistic{II}{k}}, \\
&& \yprov_{i}  =  \log \lambda_i  +   \errordiff_{\yprov ,i},  \quad     \errordiff_{\yprov ,i} \sim  {\GenLogistic{I}{N_i-k}}, \nonumber \\
&&  y_i=k  \Leftrightarrow   \yprow_{i}>0,  \, \yprov_{i}  \leq 0,  \nonumber
\end{eqnarray}
where  $\GenLogistic{I}{\nu}$ and $\GenLogistic{II}{\nu}$ are,  respectively,
the generalized logistic distributions of  type I and type II.
For  $y_i=0$, the  model  reduces to
\begin{eqnarray*} 
\yprov_{i}   =  \log \lambda_i  +   \errordiff_{\yprov ,i},  \quad     \errordiff_{\yprov ,i} \sim  {\GenLogistic{I}{N_i}} , \quad  y_i=0 \Leftrightarrow  \yprov_{i}  \leq 0.  \nonumber
\end{eqnarray*}
For  $y_i=N_i$, the  model  reduces to
\begin{eqnarray*}
  \yprow_{i}=  \log \lambda_i  +   \errordiff_{\yprow ,i},  \quad   \errordiff_{\yprow ,i} \sim  {\GenLogistic{II}{N_i}}, \quad  y_i=N_i \Leftrightarrow
  \yprow_{i} >0.  \nonumber
\end{eqnarray*}
\end{thm}

\noindent
For $N_i=1$, the logistic  model results, as both  $\GenLogistic{I}{\nu}$ and $\GenLogistic{II}{\nu}$ reduce to a
 logistic distribution for $\nu=1$. For  $y_i=0$, $z_i = \yprov_{i}$,
whereas  for $y_i=1$, $z_i= \yprow_{i}$, and the choice equation reduces to $y_i=\indic{z_i>0}$.

\begin{thm}{{\bf Sampling the utilities in the binomial RUM}}\label{lemma2U}
  Given \SFS{$y_i$}  
  and holding all model parameters \SFS{in $\lambda_i$} fixed,    the latent variables $\yprow_{i}|  \SFS{\lambda_i, (y_i>0 )} 
  $  and $\yprov_{i}| \SFS{\lambda_i, (y_i <N_i) } 
  $ are conditionally
 independent. The
  distributions of $\yprow_{i}| \SFS{\lambda_i, (y_i>0 )} 
  $  and $\yprov_{i}| \SFS{\lambda_i, (y_i <N_i) } 
  $
are equal in distribution to
\begin{eqnarray}
&&  \yprow_{i} =     \log\left(   ( 1+\lambda_i )  \frac{1}{W_i^{1/y_i}}  -  \lambda_i  \right),   \quad y_i>0, \label{postutW}\\
&& \yprov_{i} =   - \log\left(   \frac{1+\lambda_i}{\lambda_i}   \frac{1}{V_i^{1/(N_i-y_i)}}  -  \frac{1}{\lambda_i}  \right), \quad  y_i<N_i,\label{postutV}
\end{eqnarray}
where $W_i$ and $V_i$ are iid uniform  random numbers.
\end{thm}

\subsection{Ultimate P\'{o}lya-{G}amma samplers for binomial data} \label{section:binom}

The two main building blocks for the 
UPG sampler for binomial data are a Gaussian 
    mixture representation of the involved generalized logistic distributions based
    on the P\'{o}lya-Gamma distribution and the application of iMDA to handle
    imbalanced data.

A random variable  $ \errordiff$  following the  generalized logistic distribution of type I or II can be represented as a normal	mixture, 
 \begin{equation} \ferror(\errordiff) =   c( a,b) \frac{  (e ^{ \errordiff})^a }{(1+ e ^{ \errordiff})^{b}}=
	\frac{ c( a,b) }{ 2 ^{b} } \exp( \kappa \errordiff
 ) \int_0^\infty \exp(-\frac{\omegaH \errordiff^2}{2}) p(\omegaH)d\omegaH, \label{eq:pgH} 
\end{equation}
with  $\kappa=a-b/2$ and the  P\'{o}lya-Gamma distribution $\omegaH  \sim \PG{b,0}$, introduced by
	\cite{pol-etal:bay_inf}  serving as mixing measure,  
 see  Appendix A.2.1 to A.2.3. 
 For $y_i>0$,  the  type II
generalized logistic distribution  $ \errordiff_{\yprow ,i} \sim \GenLogistic{II}{y_i}$ in (\ref{aggnew3U}) has such a representation with:
\begin{eqnarray*} \label{aggnew5wU}
	\kappa_{\yprow ,i}=\frac{1-y_i}{2},   \qquad     \omegaH_{\yprow ,i}  \sim \PG{y_i+1,0},
\end{eqnarray*}
 see (A.11). 
 Similarly, for $y_i< N_i$,  the type I
generalized logistic distribution  $  \errordiff_{\yprov ,i} \sim \GenLogistic{I}{N_i-y_i}$
in (\ref{aggnew3U}) has such a representation with
\begin{eqnarray*} 
  \kappa_{\yprov ,i}=\frac{N_i-y_i-1}{2},   \qquad \omegaH_{\yprov ,i} \sim \PG{N_i-y_i+1,0},
\end{eqnarray*}
 see (A.7). 
%
Note that $ \kappa_{\yprow ,i}=0$ for $y_i=1$ and $\kappa_{\yprov ,i}=0$ for $y_i=N_i-1$. Hence,
for $N_i=1$, the P\'{o}lya-Gamma mixture approximation (\ref{PGHW}) of a logistic model involving $\PG{2,0}$ results.
For $N_i>1$,  $\kappa_{\yprov ,i} > 0$ for $0 \leq y_i \leq N_i-2$ and  $ \kappa_{\yprow ,i}< 0$ for $2 \leq y_i \leq N_i$. 
\HW{This leads to a slightly more challenging sampler than for binary and multinomial models.}

 For each $i=1, \dots,N$,  we introduce the latent variables $\zv_i=(\yprow_{i}, \scale_{\yprow ,i}, \yprov_{i},\scale_{\yprov ,i})$, if $0<y_i<N_i$,
 $\zv_i=( \yprow_{i}, \scale_{\yprow ,i})$, if $y_i=N_i$,  and  $\zv_i=(\yprov_{i},\scale_{\yprov ,i})$, if $y_i=0$.
 Conditional on  
  \SFS{$ \lambda_i$},
 the  latent variables $\yprow_{i}| \lambda_i , (y_i>0)$  and $\yprov_{i}|\lambda_i, (y_i<N_i)$ are sampled  from Theorem~\ref{lemma2U} without conditioning on
$\scale_{\yprow ,i}$ and $\scale_{\yprov ,i}$. Given 
$\yprow_{i}$ and $\yprov_{i}$,
the  parameters $ \omegaH_{\yprow ,i}|\yprow_{i}, (y_i>0), \lambda_i $ and
$ \omegaH_{\yprov ,i}|\yprov_{i}, (y_i<N_i ), \lambda_i$   are
  independent  and  follow (tilted) P\'{o}lya-Gamma distributions:
  \begin{eqnarray}  \label{obsffbinU}
 &&  \omegaH_{\yprow ,i} |\yprow_{i} , y_i ,  \lambda_i  \sim
   \PG{y_i+1,| \yprow_{i}- \log \lambda_i |},      \quad y_i>0, \\
    &&  \omegaH_{\yprov ,i} |\yprov_{i} , y_i,  \lambda_i  \sim
    \PG{N_i-y_i+1,|\yprov_{i}- \log \lambda_i|},  \quad   y_i< N_i. \nonumber
   \end{eqnarray}
   To handle imbalanced data, we apply location- and
scale-based boosting  as in the previous sections, based on the working parameters
$\gamma$ and $\delta$. 
 Location-based boosting, for instance, uses $\Star{\gamma} \sim \Normal{0,  G_0}$ to define 
$\yprowtilde_{i} =  \yprow_{i} + \Star{\gamma}$ and 
$  \yprovtilde_{i} = \yprov_{i}   + \Star{\gamma}$ 
in the following expanded version of  model (\ref{aggnew3U})   
  with an explicit choice equation involving  $\gamma$:
    \begin{eqnarray} \label{boonew1}
&& \yprowtilde_{i} = \gamma + \log  \lambda_i  +  \errordiff_{\yprow ,i}, \quad y_i>0, \\
&& \yprovtilde_{i} = \gamma + \log  \lambda_i  +  \errordiff_{\yprov ,i},
\quad y_i < N_i, \nonumber \\
&& \label{chooebo} y_i= k \Leftrightarrow \left\{ \begin{array}{ll}
            \yprovtilde_{i}   \leq \gamma  <  \yprowtilde_{i}, & 0 < k < N_i ,  \\
            \gamma \geq   \yprovtilde_{i},   & k=0,  \\
            \gamma  <   \yprowtilde_{i},  &  k= N_i . \\
         \end{array} \right.
\end{eqnarray}
Full details on  the UPG sampler for
  binomial data 
  are provided in Appendix A.4.4.

\begin{figure}[t!]
  \centering
\begin{subfigure}{0.25\textwidth}
  \centering
  \includegraphics[width=\linewidth]{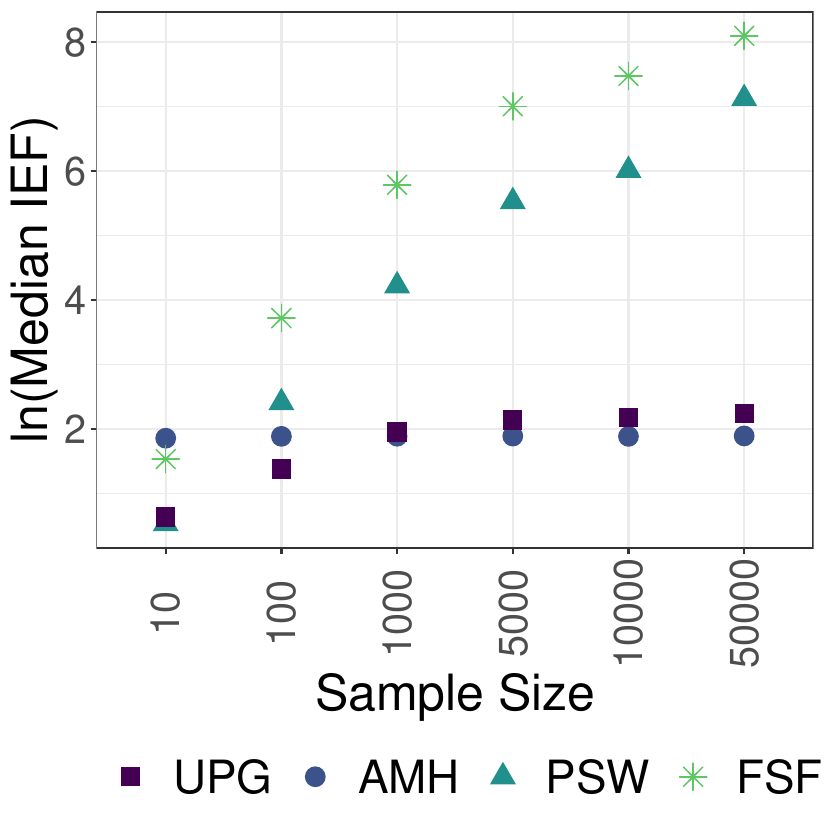}
\end{subfigure}\hspace{3em}
\begin{subfigure}{0.25\textwidth}
  \centering
  \includegraphics[width=\linewidth]{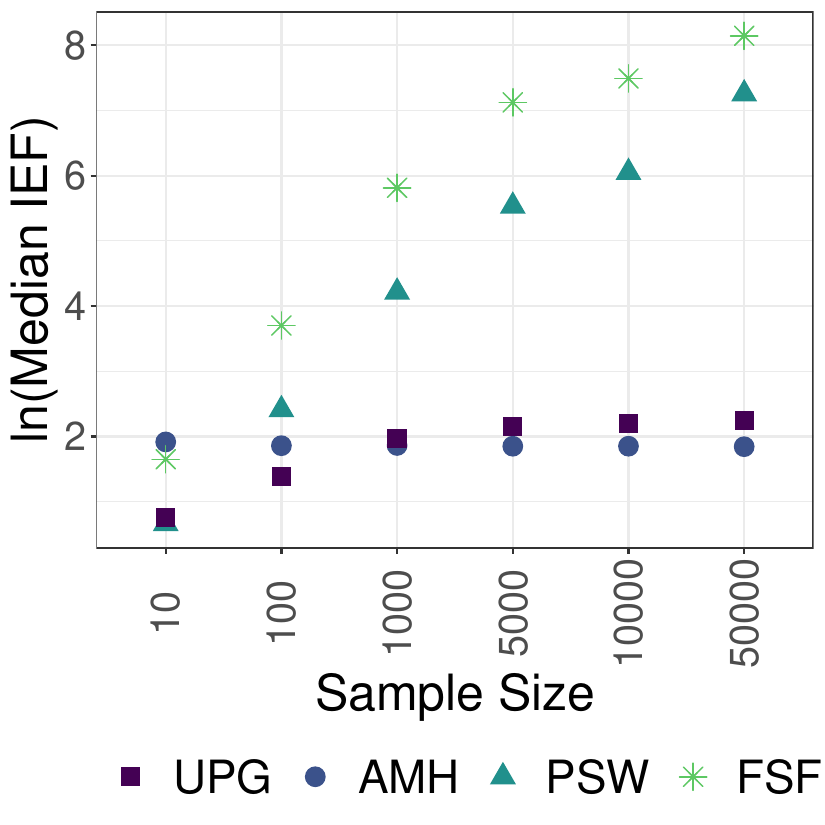}
\end{subfigure}\hspace{3em}
\begin{subfigure}{0.25\textwidth}
  \centering
  \includegraphics[width=\linewidth]{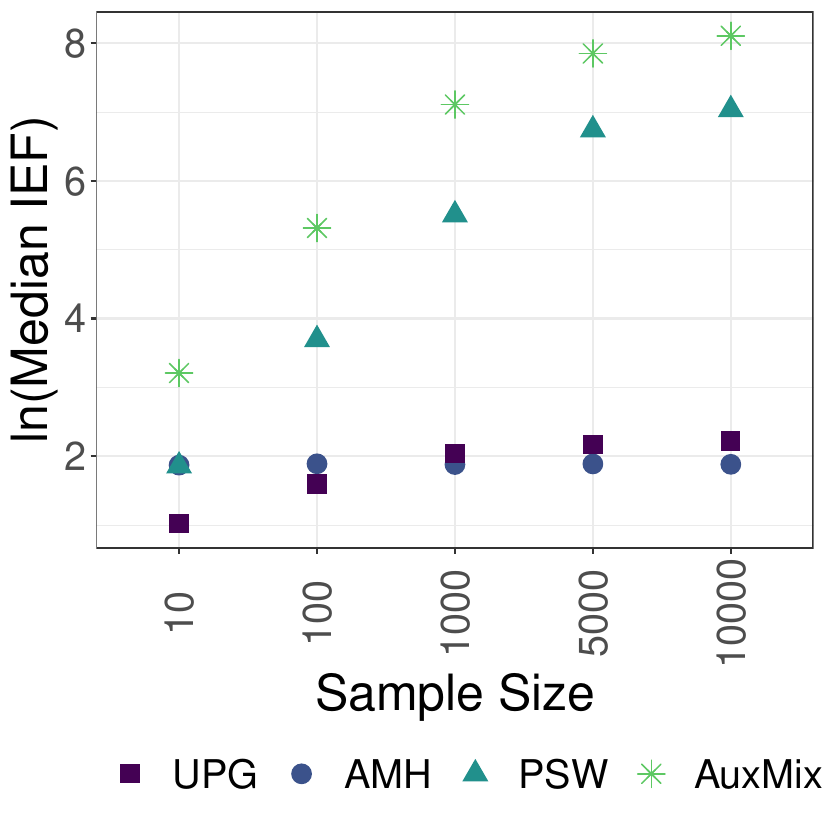}
\end{subfigure}\\
\begin{subfigure}{0.25\textwidth}
  \centering
  \includegraphics[width=\linewidth]{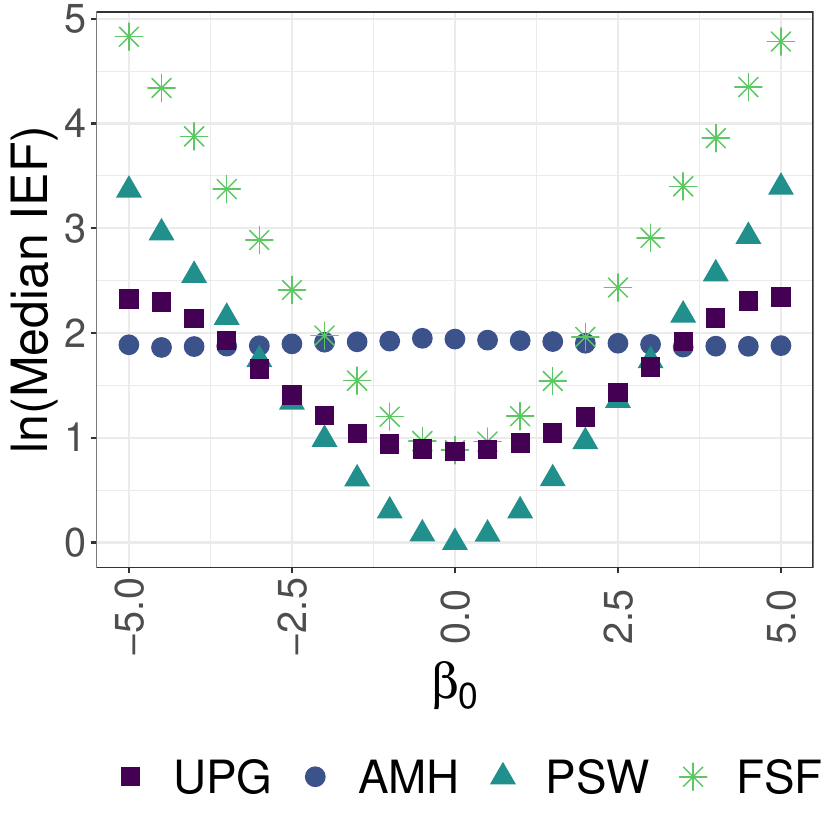}
\end{subfigure}\hspace{3em}
\begin{subfigure}{0.25\textwidth}
  \centering
  \includegraphics[width=\linewidth]{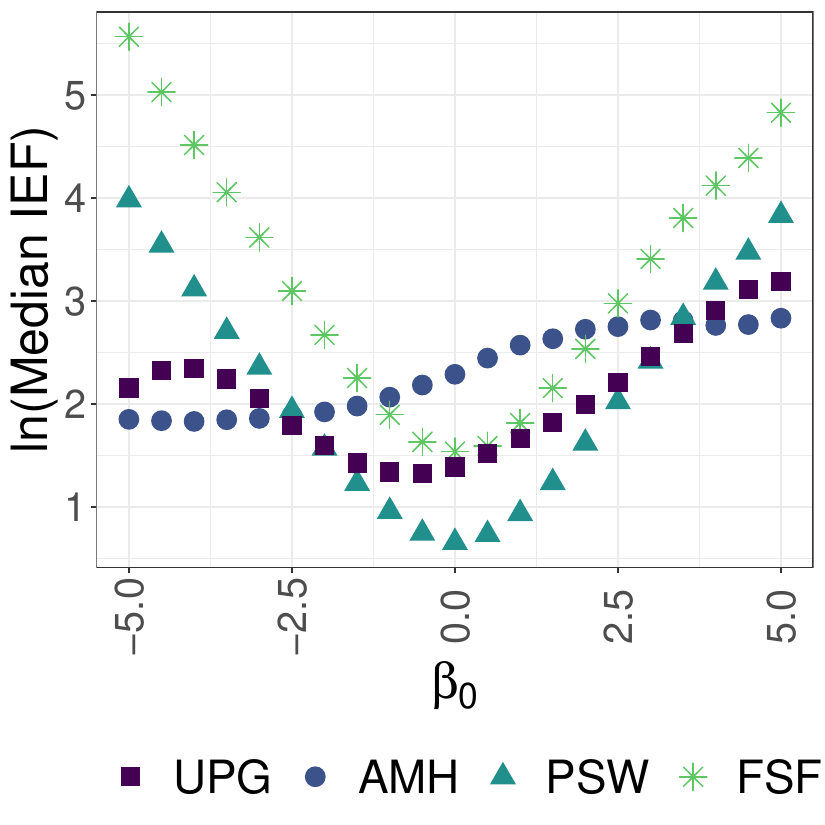}
\end{subfigure}\hspace{3em}
\begin{subfigure}{0.25\textwidth}
  \centering
  \includegraphics[width=\linewidth]{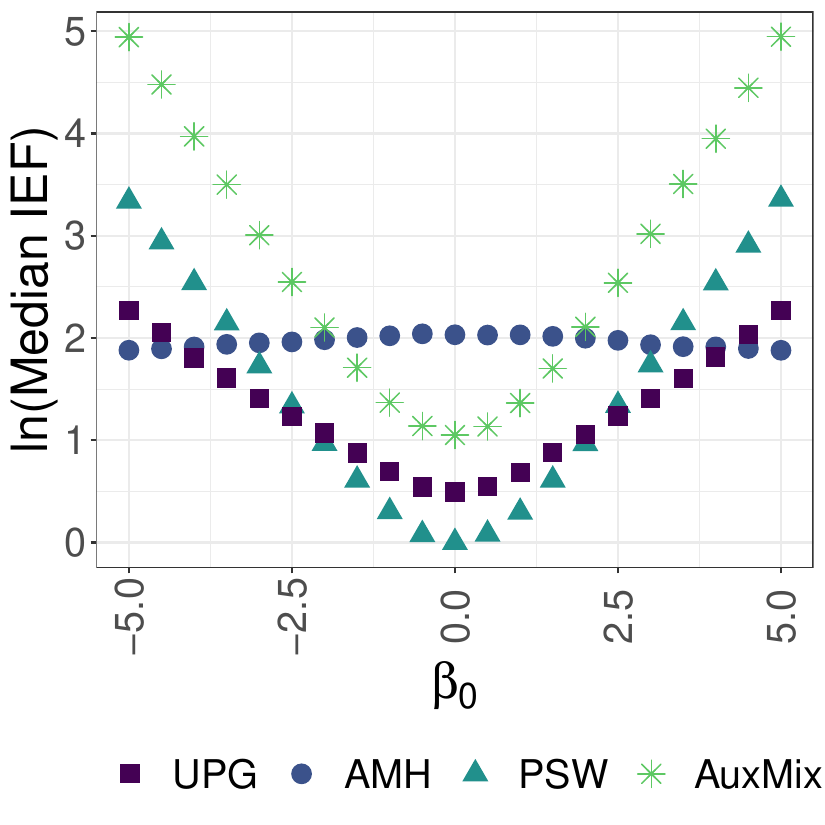}
\end{subfigure}
\caption{Sampling efficiency of intercept $\beta_0$ fitted to data sets with increasing sample size \HW{$N$} and two successes (top) and varying intercepts \HW{$\beta_0$} with $N = 1,000$ (bottom) for binary logistic regression (left), multinomial logistic regression (middle) and binomial logistic regression (right). Y-axis is on the log-scale and results are medians across 100 replications.}
\label{fig:sim_res}
\end{figure}

\section{Comparison with other sampling strategies} \label{sec:sim_da}

This section compares the proposed sampling framework with other 
DA approaches for posterior simulation in binary and categorical 
regression models. Specifically, we conduct a large scale simulation study to establish the efficiency of our approach in imbalanced scenarios relative to other DA approaches. However, from a practical point of view, a number of alternative estimation algorithms that do not rely on DA 
are available for binary and categorical regression modeling. 
These algorithms can be highly efficient, and relying on them is often a reasonable choice. Hence, a thorough discussion of the unique advantages and disadvantages of the DA strategy outlined in this article -- and DA schemes in general -- is warranted, and we provide such a discussion in Appendix A.1.

A set of systematic simulations is carried out to compare the efficiency of our approach to other popular Bayesian sampling schemes that involve DA. The main results are based on simulations with varying levels of imbalancedness, where imbalancedness is either induced by fixing the number of successes at two and increasing the sample size, or fixing the sample size at $N = 1,000$  and varying the intercept term in the data generating process. Each Markov chain was run for 10,000 iterations after an initial burn-in period of 2,000 iterations. To gain robustness with respect to the computed inefficiency factors, each simulation is repeated 100 times and median results across these replications are reported. The computation of the inefficiency factors is based on an estimate of the spectral density of the posterior chain evaluated at zero.\footnote{Estimating the spectral density at zero is accomplished via R package \texttt{coda} \citep{plu-etal:cod} and is based on fitting an autoregressive process to the posterior draws.} In this section, we present results on various logistic regression models, while additional results for probit regression models and tabulated simulation results can be found in Appendix A.6.

For binary logistic regression, we compare the sampling scheme outlined in Section~\ref{sec:mcmcdetails} (UPG), the P\'{o}lya-Gamma sampler of \citet{pol-etal:bay_inf} (PSW) and the auxiliary mixture DA scheme outlined in \citet{fru-fru:dat} (FSF). To assess sampling efficiency for the MNL model, we compare the MNL sampler proposed  in Section~\ref{section:MNL} (UPG) with the sampling scheme of \citet{pol-etal:bay_inf} (PSW) and the partial dRUM sampler of \citet{fru-fru:dat} (FSF) in a setting with three categories. For the simulations with varying sample sizes, the first two categories are observed twice each and the remaining $N - 4$ observations fall into the baseline category. For the varying intercept simulations, the intercept of the first category is varied while the other intercepts are fixed at zero. Finally, to illustrate the efficiency gains in the case of logistic regression analysis of binomial data, we compare the approach outlined in Section~\ref{section:binomial} (UPG) to the sampling scheme of \citet{pol-etal:bay_inf} (PSW) and to the auxiliary mixture sampler introduced in \citet{fus-etal:eff} (AuxMix). For all observations, we assume $N_i = 5$ trials. In all simulations, an adaptive Metropolis-Hastings sampler (AMH) is included as a benchmark as well. Throughout all simulation settings, independent $\mathcal{N}(0,10)$ priors are specified on the regression parameters, and we choose $\gamma \sim \mathcal{N}(0, 100)$ and $\delta \sim \mathcal{IG}(2.5, 1.5)$ as working prior for the iMDA algorithms.

The results of the main simulation exercise are summarized in \autoref{fig:sim_res}. The empirical inefficiency factors confirm that standard DA techniques exhibit extremely inefficient sampling behavior when confronted with imbalanced data, as shown theoretically and empirically in \citet{joh-etal:mcm}. The MDA strategy we propose alleviates this issue and allows for rather efficient estimation also in highly imbalanced data settings. 

\section{Applications to more complex models} \label{section:appext}

\subsection{Application to a binary state space model} \label{section:ssm}

Let  $\left\{y_{t}\right\}$ be a time series of binary observations, observed for  $t=1,\dots, T$, 
taking one of two possible values labelled $\{0,1\}$. 
The    probability that
$y_{t}$ takes the value $1$ depends on covariates $\Xbeta_{t}$, including a constant, 
through  time-varying
parameters $\betat{t}$ as follows: 
 \begin{eqnarray}  \label{ltpoi1ssm}
\Prob{y_{t}=1 |\betat{1},\dots,\betat{T}} = \frac{ \exp( \Xbeta_{t} \betat{t})} {1 + \exp(\Xbeta_{t} \betat{t})}.
\end{eqnarray}
We assume that conditional on knowing $\betat{1},\dots,\betat{T}$, the observations are mutually
independent. A commonly used model for describing the time-variation of $\betat{t}$ reads:
\begin{eqnarray}
&& \betat{t} = \betat{t-1} + \wt{t}, \quad \wt{t} \sim \Normult{\betad}{\bf0,\Qrcm}, \label{misstra}
\end{eqnarray}
with $\betat{0}\sim \Normult{\betad}{\bfz,\Pt{0}{0}}$ and
 $\Qrcm=\Diag{\theta_1, \dots, \theta_d}$,
where $\theta_1, \dots, \theta_d$ are unknown variances. 
MCMC estimation of binary state space models (SSM) is challenging. Single-move sampling of  $\betat{t}$
is potentially very inefficient \citep{she-pit:lik}, while blocked  MH updates require suitable proposal densities in a  high-dimensional space \citep{gam:mar}.  Within the DA framework, a latent utility $\yprodiff_{t}$ of choosing category
1 is introduced for each $y_{t}$:
\begin{eqnarray}
& y_t=1 \Leftrightarrow  \yprodiff_{t} > 0, \qquad \yprodiff_{t} =  \Xbeta_{t} \betat{t} + \varepsilon_{t}. & \label{cGuass} 
 \label{ltpdarcm}
\end{eqnarray}
Given  $\yprodiffv=\{z_t\}$, this SSM  is conditionally Gaussian 
for a probit link, but conditionally non-Gaussian
for a logit link. 
 \citet{fru-fru:aux}
implemented  an auxiliary mixture sampler for a binary
logit SSM. 
Alternatively,  using  the P\'{o}lya-Gamma  mixture representation of the logistic distribution
of $\varepsilon_{t}$ 
yields a
conditionally Gaussian SSM which allows  multi-move sampling of the
  entire state process $\betav=\{\betat{0}, \betat{1}, \dots, \betat{T}\}$ using FFBS \citep{fru:dat,car-koh:ong} in a similar fashion
as for a probit SSM.
To achieve robustness again imbalance,
we extend the iMDA scheme introduced  in Section~\ref{section:binary} to SSMs, see Appendix A.5 for details. 


\begin{figure}[t!]
  \centering
\begin{subfigure}{0.45\textwidth}
  \centering
  \includegraphics[width=\linewidth]{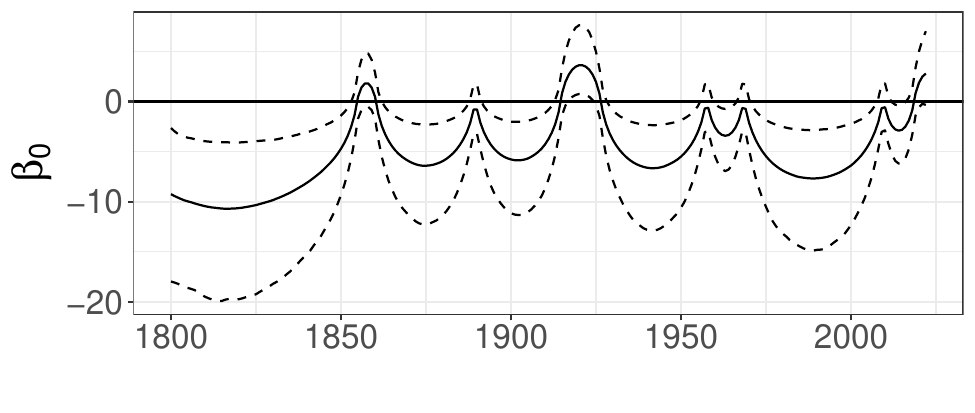}
  \caption{Posterior of Time-Varying Constant.}
\end{subfigure}
\begin{subfigure}{0.45\textwidth}
  \centering
  \includegraphics[width=\linewidth]{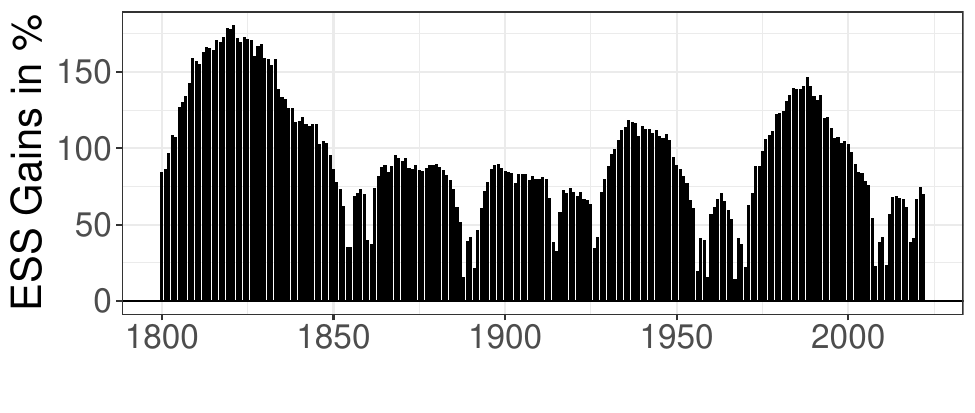}
  \caption{Gains in Effective Sample Size.}
\end{subfigure}
\caption{Panel (a) shows the posterior of a local level model fitted to the global pandemic data
(solid line: posterior mean, dashed lines: 0.05 and 0.95 posterior quantiles). 
Panel (b) shows the percentage gains in effective sample size when  \HW{iMDA} 
is applied, averaged across ten independent chains.}
\label{fig:tvp_logit}
\end{figure}

To illustrate the gains in sampling efficiency for binary SSMs, we apply the UPG framework  to an example data set on severe global pandemics. The data covers $T = 222$ years from 1800 to 2022 and documents disease episodes characterized by a worldwide spread and a death toll of more than 75,000. In addition, we focus on diseases that are characterized by relatively short periods of activity, hence excluding pandemics such as HIV/AIDS. This results in a total of eight pandemic events falling into the sample period, starting with a bubonic plague outbreak between 1855 and 1860 and ending with the global outbreak of COVID-19, starting in 2019.\footnote{The data is sourced from \url{https://en.wikipedia.org/wiki/List_of_epidemics} and the sources therein.} 

For years featuring a global pandemic, $y_t = 1$ and $y_t=0$ otherwise. A pandemic is observed in roughly 1 out of 8 years with high state persistence, rendering the data set relatively imbalanced. We fit a logistic local level model to the data, once with and once without iMDA, using $\theta \sim \mathcal{IG}(5,2)$ and $P_{0|0} = 100$ as prior settings. The Gibbs sampler is iterated 100,000 times after an initial burn-in period of 10,000 iterations. This numerical study is repeated ten times. One of the resulting posterior distributions (based on the boosted sampler) is shown in Panel (a) of Figure~\ref{fig:tvp_logit}. The time-varying intercept evolves smoothly, as is typical for binary state space models. The estimated path is characterized by long periods without severe pandemics, interrupted by short pandemic episodes. In Panel (b), the percentage gains in effective sample size of the sampler with \HW{iMDA}  relative to the plain sampler are plotted for each year. The iMDA scheme described in Appendix A.5
is able to significantly improve sampling efficiency in all years. The most pronounced gains -- up to $200\%$ improvement in effective sample size -- are observed during prolonged `imbalanced' periods where the outcome does not change. Averaging across all periods, the inefficiency factors are roughly halved, from about 96 in the plain sampler to around 45 in the 
\HW{UPG sampling} scheme.




\subsection{Application to logistic mixture-of-experts regression models} \label{sec:moe}

Let $y_i$ ($i = 1, \dots, N$) be a grouped binary outcome with $C_i = j$ denoting that observation $i$ belongs to group $j = 1, \dots, J$. A logistic mixture-of-experts regression model with $H$ ($h = 1, \dots, H$) components takes the form
\begin{equation}
\begin{split}
        p(y_{i}~|~C_i = j, \bm{x}_i, \bm{w}_j) = &{\sum_{h=1}^{H}} \eta_{jh}(\bm{w}_j) Ber(\zeta_{ih}(\bm{x}_i))\\
        \zeta_{ih} \HW{(\bm{x}_i)}= \frac{\text{exp}(\bm{x}_i\bm{\beta}_h)}{1 + \text{exp}(\bm{x}_i\bm{\beta}_h)} \quad \quad &\quad \quad
        \eta_{jh} \HW{(\bm{w}_j)} = \frac{\text{exp}(\bm{w}_j\bm{\psi}_h)}{{\sum_{l=1}^{H}}\text{exp}(\bm{w}_j\bm{\psi}_l)}
\end{split}
\label{moe:mod}
\end{equation}
where $H$ logistic regression `experts' are used to model cluster-specific success probabilities $\zeta_{ih} \HW{(\bm{x}_i)}$ using individual-level covariates $\bm{x}_i$ and a multinomial logistic regression plays the role of a `gating function', modeling the mixture weights $\eta_{jh} \HW{(\bm{w}_j)} $ based on group-level covariates $\bm{w}_j$. This model has good approximation properties \citep{wenxin1999moe} and is popular in model-based clustering and ensemble learning. Furthermore, developing efficient inferential tools is an important research avenue \citep{sharma2019flexible}. A thorough treatment of mixture-of-experts models is given in \citet{gor-fru:mix}.

The model in (\ref{moe:mod}) naturally involves multiple layers of hierarchy, multi-modal posteriors and discrete parameter spaces, potentially rendering inference with general purpose posterior simulation tools difficult.\footnote{See Appendix A.1 
for further discussion.} As a result, DA algorithms are popular tools for the estimation of mixture-of-experts models \citep{gor-fru:mix}. However, imbalanced data and large samples may lead to convergence issues. In model (\ref{moe:mod}), both the success probabilities $\zeta_{ih}$ and the mixture weights $\eta_{jh}$ may be imbalanced.

The methodology proposed in the present article is a potential remedy in such scenarios, as both the logistic regression experts and the gating function can be estimated using DA with additional location-based and scale-based parameter expansion steps. We demonstrate in a numerical exercise in Appendix A.6.3 
that our iMDA scheme indeed leads to sizeable efficiency gains with respect to all involved regression parameters in simulated data. In Appendix A.7,
we further illustrate logistic mixture-of-experts regression models in a large-sample real world application on maternal education and child mortality. Again, effective sample sizes increase as soon as iMDA is introduced.

\section{Concluding Remarks}
\label{section:conclude}

Due to a wide range of applications in many areas of applied science, much attention has been dedicated towards the development of estimation algorithms for generalized linear models. In the past decades, various DA algorithms have been brought forward that have steadily increased accessibility and popularity of Bayesian estimation techniques in the context of regression models for binary and categorical outcomes.
 In this article, we introduce new sampling algorithms based on P\'{o}lya-Gamma mixture representations for estimation of these models. The algorithms are easily implemented, intuitively appealing and allow for a conditionally Gaussian posterior distribution of the regression effects in binary, multinomial and binomial logistic regression frameworks. To counteract potentially inefficient sampling behavior, we develop a novel parameter expansion strategy and apply it to the introduced sampling algorithms as well as to probit frameworks. This results in a comparative level of sampling efficiency, even in scenarios where outcomes are heavily imbalanced, as is demonstrated via extensive simulation studies and real data applications.

A number of future research avenues worth exploring come readily to mind. First, the proposed family of DA and MCMC boosting schemes could be extended to accommodate other types of limited outcomes such as ordered or count data. Second, we approached the problem of efficiency comparisons mostly empirically and left theoretical aspects
largely  unexplored. Extending the theoretical results 
of \cite{choi2013polya} and \cite{joh-etal:mcm}, among others, might be fruitful and assessing convergence rates of the proposed sampling schemes more formally may reveal additional insights. Finally, it is well-known that scale-based parameter expansion leads to faster convergence of expectation-maximization algorithms \citep{liu-etal:par}. It may be worth to investigate whether the proposed location-based expansion leads to additional efficiency gains in this context.

\subsubsection*{Acknowledgements}

The authors would like to thank Darjus Hosszejni, 
two anonymous referees and an associate editor for helpful comments and suggestions.

\bibliographystyle{agsm}
\bibliography{sylvia_kyoto}

\newpage
\appendix

\setcounter{equation}{0}
\setcounter{figure}{0}
\setcounter{table}{0}
\setcounter{page}{0}

\renewcommand{\thesection}{\Alph{section}}
\renewcommand{\thetable}{\Alph{section}.\arabic{table}}
\renewcommand{\thefigure}{\Alph{section}.\arabic{figure}}
\renewcommand{\theequation}{\Alph{section}.\arabic{equation}}

\noindent \textbf{Online supplementary material}\\
\textit{Ultimate {P}\'{o}lya  {G}amma Samplers -- Efficient {MCMC} for possibly imbalanced binary and categorical data}
\ \\
\ \\
Gregor Zens, Sylvia Fr{\"u}hwirth-Schnatter, Helga Wagner
\ \\
\ \\
\HWn{May 2023}
\thispagestyle{empty}
\clearpage

\section{Appendix}

\subsection{Discussion: Advantages and disadvantages compared to approaches without data augmentation}
\label{app:non_da_algos}

In the context of the probit model, \citetsupp{durante2019conjugate} discusses conjugate analysis under unified skew-normal priors. This approach allows to sample from the resulting posterior distribution extremely efficiently and works well in the context of \textit{small $N$, large $P$} settings. However, even for moderately large $N$, the computational burden of the approach makes posterior simulation infeasible. In addition, deviating from the standard probit regression setup and introducing modifications such as time-varying parameters is a non-trivial task in this framework. Finally, this sampling strategy is restricted to the probit link function. In comparison, the approach outlined in this article is much more general, scales better to larger data sets and extensions to more complex setups are often trivial to achieve due to the conditionally Gaussian representation.

A number of recent contributions have established piecewise deterministic Markov processes (\citealpsupp{vanetti2017piecewise}; \citealpsupp{fearnhead2018piecewise}) as a successful tool for posterior simulation. One particularly useful approach arising from this literature is the so-called Zig-Zag sampler \citepsupp{bierkens2019zig}. The advantages of this approach in the context of logistic regression with imbalanced data have been pointed out by \citetsupp{sen2020efficient}. However, methods based on piecewise deterministic Markov processes are rather involved, both from a computational and a mathematical perspective. This makes them relatively inaccessible to applied researchers and renders extensions to customized, complex modeling tasks difficult. 

Compared to that, gradient-based posterior simulation techniques, such as the Metropolis-adjusted Langevin algorithm (MALA) or Hamiltonian Monte Carlo (HMC) are commonly encountered in practice. These methods have become popular due to readily available software implementations such as Stan
\citepsupp{carpenter2017stan}. When applied to simple models with a small to moderate parameter dimension, these approaches are likely to produce posterior samples that are nearly independent of each other, conditional on having access to a well-chosen set of tuning parameters. If tuning parameters are chosen suboptimally, gradient-based methods may fail in scenarios with ill-conditioned likelihoods. A particularly relevant example are highly imbalanced logistic regression problems, see for instance \citetsupp{hird2020fresh}. In comparison, one very convenient property of DA approaches is the absence of tuning parameters. Nonetheless, the issue of searching for good tuning parameters can be facilitated via automatic tuning approaches such as the \textit{No-U-turn sampler} outlined in \citetsupp{hoffman2014no} or via methods that are more robust to tuning parameters, such as the \textit{Barker proposal} \citepsupp{livingstone2022barker}. However, even when good tuning parameters can be automatically obtained, standard gradient-based methods may encounter issues when confronted with complex, hierarchical frameworks involving multi-modal or discrete (i.e., discontinuous) posterior distributions, where specialized solutions have to be employed (\citealpsupp{mangoubi2018does}; \citealpsupp{nishimura2020discontinuous}). 

Compared to that, DA is an easily applicable out-of-the-box tool that often is one of the few available approaches that is easy to implement and achieves convergence, even in more complex scenarios. As a result, DA is still one of the standard tools used by practitioners in many modeling frameworks. Examples include mixture and mixture-of-experts models, where multi-modal posteriors, discrete-valued parameters and imbalanced data are commonly encountered. 
\HW{Expanding Section 6.2, we}
discuss such mixture-of-experts frameworks in more detail in Appendix~\ref{app:moe} and in Appendix~\ref{app:childmort_moe}. Another typical application of DA algorithms are state space models, due to the potentially high-dimensional parameter space. \HW{In Section 6.1, we extended our iMDA approach to logistic state space models and provide further details in  Appendix~\ref{sec:smmUPG}}. 

\subsection{Mixture representations}

\subsubsection{The P\'{o}lya-Gamma  mixture representation} \label{GenLogPG}

For all latent variable representations derived in this paper
 for binary, \SFS{binomial or categorical} data,
the error term  in the latent equations  arises from a distribution $ \errordiff \sim  \Ferror(\errordiff)$, for which the density  $ \ferror(\errordiff)$
 can be represented as a   
  mixture of normals using the P\'{o}lya-Gamma distribution as mixing measure:
   \begin{eqnarray}   \label{mixPG}
\ferror(\errordiff) =   c( a,b) \frac{  (e ^{ \errordiff})^a }{(1+ e ^{ \errordiff})^{b}}=
  \frac{ c( a,b) }{ 2 ^{b} }{ e ^ {\kappa \errordiff}}  \int e ^{- \frac{\omegaH   \errordiff^2}{2}} p (\omegaH ) d  \omegaH,
\end{eqnarray}
where  $\kappa=a-b/2$ and $\omegaH  \sim \PG{b,0}$  follows the P\'{o}lya-Gamma distribution introduced by \citetsupp{pol-etal:bay_inf} with parameter $b$.

 This  new
 representation is very convenient, as the conditional posterior of $\scale| \errordiff $ can be derived from following (tilted) P\'{o}lya-Gamma distribution with the same parameter $b$:
    \begin{eqnarray}  \label{obsHWup}
   \quad  \omegaH |\errordiff \sim \PG{b, |\errordiff|}.
   \end{eqnarray}
\HW{On the other hand,
conditional on $\omegaH$,} the likelihood contribution of $\errordiff$ is proportional to that of a  $\Normal{\kappa /\scale ,1/ \omegaH }$ observation.\footnote{{Based on rewriting
\begin{eqnarray}  \label{obsFSF}
  \kappa \errordiff  - \frac{\omegaH   \errordiff^2}{2} =   \frac{-\omegaH  (  \errordiff- \kappa/ \omegaH)^2}{2} + d,
   \end{eqnarray}
 where $d$ is a constant not depending on   $\errordiff$.}}
   
%
   To  simulate from a (tilted) P\'{o}lya-Gamma $\PG{q , c}$ distribution, the following convolution property is exploited:
    \begin{eqnarray*}
 X_1 \sim  \PG{q_1 , c} , \quad  X_2 \sim  \PG{q _2 , c} \Rightarrow  X_1+ X_2   \sim  \PG{q_1+ q_2 , c} .
   \end{eqnarray*}
  where $X_1$ and $X_2$ are independent. 
  Hence,  to simulate from  $ Y \sim \PG{q, c}$,  use $Y=\sum_{j=1} ^q X_j $ , where $X_j \sim  \PG{1 ,c }$ are $q$ independent draws from the  $\PG{1 ,c }$   distribution.

\subsubsection{The logistic and  the type I generalized  logistic distribution}  \label{GenLogI}

  For  the type I generalized logistic distribution   $\errordiff \sim \GenLogistic{I}{\nu}$ with parameter $\nu>0$,
the density reads
 \begin{eqnarray}   \label{genlogI}
 \ferror  ( \errordiff) =     \frac{ \nu  e ^{-\errordiff} }{  (1+  e ^{-\errordiff} )^{\nu+1}} =  \frac{ \nu  e ^{ \nu \errordiff} }{  (1+  e ^{\errordiff} )^{\nu+1}}.
\end{eqnarray}
$\GenLogistic{I}{\nu}$ reduces to the logistic distribution for $\nu=1$.
The c.d.f. of a type I generalized logistic distribution 
takes  a simple form:
  \begin{eqnarray}   \label{genlogIdist}
 \Ferror  ( \errordiff) =     \frac{1}{  (1+ e ^{-\errordiff})^{\nu}} =     \frac{ e ^{\nu \errordiff} }{  (1+ e ^{\errordiff})^{\nu}} .
\end{eqnarray}
Hence, the quantiles are available in closed form:
 \begin{eqnarray}   \label{genlogIp}
 \errordiff_p =  \Ferror   ^{-1} ( p) =  - \log  \left(  \frac{1}{  p^{1/\nu}} - 1  \right).
\end{eqnarray}
The type I generalized logistic distribution  $ \errordiff \sim \GenLogistic{I}{\nu}$ can
be  represented as a 
mixture of normals  with a P\'{o}lya-Gamma distribution serving as mixing measure,
  where
   \begin{eqnarray}  \label{obstypeI}
\omegaH \sim \PG{\nu+1,0}, \quad   \kappa= \frac{\nu-1}{2},
   \end{eqnarray}
   see Appendix~\ref{GenLogPG}.
For the logistic distribution, $\omegaH \sim \PG{2,0}$ and $\kappa= 0$.

 \subsubsection{The type II generalized  logistic distribution}    \label{GenLogII}

  For  the type II generalized logistic distribution  $ \errordiff \sim \GenLogistic{II}{\nu}$ with parameter $\nu>0$,
the density reads
 \begin{eqnarray}   \label{genlogII}
 \ferror  ( \errordiff) =     \frac{ \nu  e ^{- \nu\errordiff} }{  (1+  e ^{-\errordiff} )^{\nu+1}} =  \frac{ \nu  e ^{ \errordiff} }{  (1+  e ^{\errordiff} )^{\nu+1}}.
\end{eqnarray}
Also $\GenLogistic{II}{\nu}$ reduces to the logistic distribution for $\nu=1$.

The c.d.f. of a type II generalized logistic distribution
takes  a simple form:
  \begin{eqnarray}   \label{genlogIIdist}
 \Ferror  ( \errordiff) =   1-   \frac{1}{  (1+ e ^{\errordiff})^{\nu}} =   1 -  \frac{ e ^{-\nu \errordiff} }{  (1+ e ^{-\errordiff})^{\nu}}.
\end{eqnarray}
Hence, the quantiles are available in closed form:
 \begin{eqnarray}   \label{genlogIIp}
 \errordiff_p =  \Ferror   ^{-1} ( p) =   \log  \left(  \frac{1}{ (1- p)^{1/\nu}} - 1  \right).
\end{eqnarray}
The type II generalized logistic distribution  $ \errordiff \sim \GenLogistic{II}{\nu}$ can
be  represented as  a 
mixture of normals with a P\'{o}lya-Gamma distribution serving as mixing measure,
where
   \begin{eqnarray}  \label{obstypeII}
  \omegaH \sim \PG{\nu+1,0},  \quad   \kappa= \frac{1-\nu}{2},
   \end{eqnarray}
   see Appendix~\ref{GenLogPG}. Again, for the logistic distribution,  $\omegaH \sim \PG{2,0}$ and $\kappa= 0$ results.

\subsection{Proofs} \label{section:proofs}

\subsubsection*{\HW{Proof of Proposition~1}}

The proof of  Proposition~1  
is straightforward. Depending on the observed category $y_i$, the corresponding utility $\ypro_{y_i,i}$ is the maximum among all latent utilities. Equivalently, given that $y_i=k$,
$e ^{- \ypro_{ki}}$ attains the minimum among all random variables $e ^{- \ypro_{0i}},\ldots,
e ^{- \ypro_{mi}}$. Since $e ^{- \ypro_{\ell i}} \sim \Exp{1}$ for $\ell=0, \ldots,m$ are iid standard exponential a priori,
we obtain that a posteriori 
(given that  $y_i=k $) 
the minimum $e ^{- \ypro_{ki}}$ follows an exponential distribution:
 \begin{eqnarray*}
   e ^{- \ypro_{ki}}|y_i=k    \sim     \Exp{\lambda^\star_i},  
\end{eqnarray*}
  where $\lambda^\star_i=1 +\sum_{l=1}^m \lambda_{li}$.
  Furthermore, given the minimum $e ^{- \ypro_{ki}}$ all
  remaining random variables $e ^{- \ypro_{\ell i}}$ with
  $\ell \neq k$ are conditionally independent with following distributions:
   \begin{eqnarray*}
  && e^{ - \ypro_{0i}} = e^{-\ypro_{ki}} +  \Exp{1}, \\
  &&  e^{ - \ypro_{\ell i}} =  e^{-\ypro_{ki}}  + \Exp{ \lambda_{\ell i}}, \quad \forall \ell \in \labset_{-(k,0)}.
  \end{eqnarray*}
For efficient joint sampling of all utilities for all $i=1, \dots, N$, this can be rewritten as  in~(17).

\subsubsection*{Proof of Theorem~2}

A binomial observation $y_i$ can be regarded as the aggregated number of successes among $N_i$ independent binary  outcomes $z_{1i},\dots,z_{N_i,i}$, labelled $\{0,1\}$, and each following the binary logit model $\Prob{z_{ni}=1 |\pi_i} = \pi_i $.
 For each individual binary observation $z_{ni}$,  the logit model can be written as a RUM  \citepsupp{mcf:con}:
\begin{eqnarray*} \label{indivRUMbin}
&& \displaystyle  \ypro_{ni}  = \log \lambda_i  + \error_{ni}, \qquad \error_{ni} \sim \Logistic \label{ut1},\\
&& z_{ni} =\indic{\ypro_{ni}>0}, \nonumber
\end{eqnarray*}
involving a latent variable $\ypro_{ni}$,   where $\error_{ni}$   are \iid\ errors following a logistic distribution.
Among the $N_i$ binary experiment,  $y_i$ outcomes $z_{ni}$ choose the category 1, whereas the remaining $N_i-y_i$ outcomes $z_{ni}$
 choose the category 0.
 The challenge is to aggregate the  latent variables $\ypro_{ni}$ to a few latent variables in such a way that an explicit choice equation is available.
 As it turns out, such an aggregation can be based on the order statistics  $\ypro_{(1),i} < \dots <  \ypro_{(N_i),i} $
  of $\ypro_{1i}, \dots, \ypro_{N_i,i} $.

  Consider first the case that  $ y_i=0$. Such an outcome is observed, iff  $z_{ni}=0$  or, equivalently,  the latent utility is  negative ($\ypro_{ni} \leq 0$) for all $n=1,\dots,N_i$. Hence, a necessary and sufficient condition for   $ y_i=0$ is that the maximum of  all utilities is negative, or equivalently,
  \begin{eqnarray}   \label{choice1}
     y_i=0   \,\, \Leftrightarrow \,\, \ypro_{(N_i),i} \leq 0.
\end{eqnarray}
   Next, consider  the case that  $ y_i=N_i$. Such an outcome is observed, iff  $z_{ni}=1$  or, equivalently,  the latent utility is  positive ($\ypro_{ni}>0$ ) for all $n=1,\dots,N_i$. Hence, a necessary and sufficient condition for   $ y_i=N_i$ is that the minimum of all utilities is  positive, or equivalently,
 \begin{eqnarray} \label{choice2}
     y_i=N_i  \,\, \Leftrightarrow \,\, \ypro_{(1),i}> 0.
\end{eqnarray}
 Also for outcomes $ 0 < y_i <N_i$, the order statistics  $\ypro_{(N_i-k),i}$  and $\ypro_{(N_i-k+1),i}$ provide necessary and sufficient conditions:
   \begin{eqnarray}   \label{choice3}
     y_i=k,\,\,  0< k < N_i  \,\, \Leftrightarrow \,\,  \ypro_{(N_i-k),i} \leq 0,   \ypro_{(N_i-k+1),i}>0.
\end{eqnarray}
 Note that   (\ref{choice1}) --  (\ref{choice3})  are   choice equations involving either a single or two order statistics. Hence, we introduce
 the corresponding order statistics as aggregated latent variables.
  Given \HW{$0 < y_i=k < N_i$, we define
$\yprov_{i}= \ypro_{(N_i-k),i}$ and  
  $\yprow_{i}= \ypro_{(N_i-k+1),i}$. 
   The choice equation then follows from (\ref{choice3}):
   \begin{eqnarray*} 
   y_i=k   \Leftrightarrow     \yprov_{i} \leq 0, \,   \yprow_{i}>0,
\end{eqnarray*}}
with  obvious modifications for  $y_i=0$ and $y_i= N_i$.
 %
 
It remains to prove  that  the latent variables can be  represented as in the aggregated model (23):
 \begin{eqnarray}   \label{chppice1}
&& \yprow_{i}=  \log \lambda_i  +   \errordiff_{\yprow ,i},  \quad   \errordiff_{\yprow ,i} \sim \GenLogistic{II}{y_i},  \quad y_i>0, \\
&& \yprov_{i}  =  \log \lambda_i  +   \errordiff_{\yprov ,i},  \quad     \errordiff_{\yprov ,i} \sim \GenLogistic{I}{N_i-y_i}, \quad y_i< N_i.  \nonumber
\end{eqnarray}
 Note that the order statistics  
 \HW{$\ypro_{(1),i}, \ldots , \ypro_{(N_i),i} $}
 can be represented for $j=1, \dots,N_i$
as $ \ypro_{(j),i} =  \log \lambda_i  +   \errordiff_{(j) ,i}$,
involving  the order statistics    $\errordiff_{(1) ,i}, \dots,  \errordiff_{(N_i) ,i}$ are of $N_i$ iid  realisations $\errordiff_{1,i}, \dots, \errordiff_{N_i,i} $ of a logistic distribution.
Their distribution 
can be derived from the order statistics  $\rvX_{(1) ,i}, \dots, \rvX_{(N_i) ,i}$  of $N_i$ uniform random numbers  $\rvX_{1 i}, \dots, \rvX_{N_i ,i}$ using:
  \begin{eqnarray}   \label{chptr3}
  \errordiff_{(j) ,i} = F^{-1}( \rvX_{(j) ,i} ) = \log \frac{\rvX_{(j) ,i}}{1-\rvX_{(j) ,i}}  \quad \Leftrightarrow \quad  \rvX_{(j) ,i}  = F(  \errordiff_{(j) ,i}  ),
\end{eqnarray}
where $F$ is the cdf of the logistic distribution.

First, for the special cases where $y_i=0$ or $y_i= N_i$,  we use that  $\rvX_{(j) ,i} \sim \Betadis{j,N_i-j+1}$.
Using (\ref{chptr3}), we can derive   the density of  $ \errordiff_{(N_i) ,i}$:
 \begin{eqnarray*}
  p(\errordiff_{(N_i) ,i} )=  N_i F(  \errordiff_{(N_i) ,i}  )^{N_i-1}  f(  \errordiff_{(N_i) ,i}  )=   \frac{ N_i \exp( \errordiff_{(N_i),i})^{N_i} }{(1 + \exp( \errordiff_{(N_i),i}))^{N_i+1}},
\end{eqnarray*}
which is the density of a  $\GenLogistic{I}{N_i}$ distribution, see  (\ref{genlogI}). Hence,  for $y_i=0$,
\begin{eqnarray*}
 \yprov_{i}  =  \log \lambda_i  +   \errordiff_{\yprov ,i},  \quad     \errordiff_{\yprov ,i}= \errordiff_{(N_i) ,i}  \sim \GenLogistic{I}{N_i} .  \nonumber
\end{eqnarray*}
Using (\ref{chptr3}), we can derive  the density of  $ \errordiff_{(1) ,i}$:
 \begin{eqnarray*}
  p(\errordiff_{(1) ,i} )=  N_i  (1-F(  \errordiff_{(1) ,i}  ))^{N_i-1}  f(  \errordiff_{(1) ,i}  )=     \frac{N_i \exp( \errordiff_{(1),i}) }{(1 + \exp( \errordiff_{(1),i}))^{N_i+1}},
\end{eqnarray*}
  which is the density of a  $\GenLogistic{II}{N_i}$ distribution, see  (\ref{genlogII}). Hence,
  for $y_i=N_i$:
\begin{eqnarray*}
\yprow_{i}=  \log \lambda_i  +   \errordiff_{\yprow ,i},  \quad   \errordiff_{\yprow ,i}=  \errordiff_{(1) ,i} \sim \GenLogistic{II}{N_i}.
\end{eqnarray*}
Second, for \HW{any $ 0 < y_i=k < N_i$} we  need the joint distribution of   
$(\errordiff_{(N_i-k),i},   \errordiff_{(N_i-k+1),i}) ^\top $, where  $\errordiff_{(N_i-k+1),i}= \errordiff_{(N_i-k),i} + \Delta \errordiff_{i} $ with  $\Delta \errordiff_{i} >0$.
 Using that  $(\rvX_{(j) ,i}, \rvX_{(j+1) ,i} - \rvX_{(j) ,i}, 1-  \rvX_{(j+1),i} ) \sim \Dir{j,1,N_i-j}$ follow a Dirichlet distribution, see e.g. \citetsupp{rob-cas:mon}, we obtain that
 $(\rvX_{(N_i-k) ,i},    1-  \rvX_{(N_i-k+1) ,i},  \rvX_{(N_i-k+1) ,i} -  \rvX_{(N_i-k) ,i}) \sim \Dir{N_i-k,k,1}$.
To derive  $p(\errordiff_{(N_i-k),i}, \errordiff_{(N_i-k+1),i} )$, we consider the transformations
 \begin{eqnarray}   \label{chptrbiv3}
   \errordiff_{(N_i-k) ,i} =  F^{-1}( \rvX_{(N_i-k) ,i}),  
\quad  \errordiff_{(N_i-k+1) ,i} =  F^{-1}( \rvX_{(N_i-k+1) ,i}) ,
\end{eqnarray}
and their inverse,
 $ \rvX_{(N_i-k) ,i} = F( \errordiff_{(N_i-k) ,i})$ and $\rvX_{(N_i-k+1) ,i} = F( \errordiff_{(N_i-k+1) ,i})$.
  We determine
\begin{eqnarray*}
 \left| \frac{\partial   (\rvX_{(N_i-k) ,i}, \rvX_{(N_i-k+1) ,i}) }{\partial  (\errordiff_{(N_i-k) ,i}) ,\errordiff_{(N_i-k+1) ,i}))} \right| &=&
 \left|    \begin{array}{cc}
  f ( \errordiff_{(N_i-k) ,i})  &  0  \\
 0  &   f( \errordiff_{(N_i-k+1) ,i} )    \\
   \end{array}   \right|  =   f ( \errordiff_{(N_i-k) ,i}) f( \errordiff_{(N_i-k+1) ,i}) ,
   \end{eqnarray*}
where $f$ is the pdf of the logistic distribution. 
\HW{Since} 
 \begin{eqnarray*}
&& p(\errordiff_{(N_i-k),i}, \errordiff_{(N_i-k+1),i} )
=  \\
&& {\frac{ \Gamfun{N_i + 1}}{\Gamfun{k} \Gamfun{N_i-k} }} F ( \errordiff_{(N_i-k) ,i})^{N_i-k-1}
 (1-   F ( \errordiff_{(N_i-k+1) ,i} ))^{k-1}   \cdot    f ( \errordiff_{(N_i-k) ,i})  f( \errordiff_{(N_i-k+1) ,i} ) =   \\
& &
\frac{\exp( \errordiff_{(N_i-k),i})^{N_i-k} }{(1 + \exp( \errordiff_{(N_i-k),i}))^{N_i-k+1}}
 \frac{ \exp( \errordiff_{(N_i-k+1),i})}{(1 + \exp( \errordiff_{(N_i-k+1),i}))^{k+1}} \cdot
 {\frac{ \Gamfun{N_i + 1}}{\Gamfun{k} \Gamfun{N_i-k} }}   
\end{eqnarray*}
%
%
\HW{whenever $\errordiff_{(N_i-k+1),i} > \errordiff_{(N_i-k),i}$, the density $p(\errordiff_{(N_i-k),i}, \errordiff_{(N_i-k+1),i} )$} can be expressed as
\begin{eqnarray*}
p(\errordiff_{(N_i-k),i}, \errordiff_{(N_i-k+1),i} )
& = &  p(\errordiff_{(N_i-k),i})  \, p( \errordiff_{(N_i-k+1),i})   \cdot   C \,  \cdot \indic{ \errordiff_{(N_i-k+1),i} > \errordiff_{(N_i-k),i}},
\end{eqnarray*}
where
 \begin{eqnarray}   \label{margkk1}
p(\errordiff_{(N_i-k),i}) =
 \frac{{(N_i-k)} \exp( \errordiff_{(N_i-k),i})^{N_i-k} }{(1 + \exp( \errordiff_{(N_i-k),i}))^{N_i-k+1}}
\end{eqnarray}
is the density of   a  $\GenLogistic{I}{N_i-k}$ distribution, see  (\ref{genlogI}),
 \begin{eqnarray}   \label{jointpp2}
 p( \errordiff_{(N_i-k+1),i}) =
 \frac{k \exp( \errordiff_{(N_i-k+1),i})}{(1 + \exp( \errordiff_{(N_i-k+1),i}))^{k+1}}
  \end{eqnarray}
 is the density of a  $\GenLogistic{II}{k}$ distribution, see  (\ref{genlogII}), and
 \begin{eqnarray*}
 C={\frac{ \Gamfun{N_i + 1}}{\Gamfun{k+1} \Gamfun{N_i-k+1} }}
  \end{eqnarray*}
  is a  normalising constant.  It is possible to verify that
   \begin{eqnarray*}
 \int_{-\infty} ^{+\infty}  \int  _{\errordiff_{(N_i-k),i} } ^{+\infty}   p(\errordiff_{(N_i-k),i}, \errordiff_{(N_i-k+1),i} ) \, \text{d} \, \errordiff_{(N_i-k+1),i}    \, \text{d} \, \errordiff_{(N_i-k),i} = 1.
  \end{eqnarray*}
  %
%
Defining $ (\errordiff_{\yprov ,i},  \errordiff_{\yprow,i})^\top = (\errordiff_{(N_i-k),i},   \errordiff_{(N_i-k+1),i}) ^\top $,  yields (\ref{chppice1}).

\subsubsection*{Proof of Theorem~3}

Knowing that $y_i=k$, with $0< k < N_i$,  $ \yprow_{i}|\SFS{\lambda_i}, y_i$ and  $\yprov_{i}|\SFS{\lambda_i}, y_i$ are conditionally independent and the following holds:
 \begin{eqnarray*} \label{aggprew3d}
 \yprow_{i} =  \log \lambda_i  +   \errordiff_{\yprow ,i}, \qquad    \yprov_{i}  =  \log \lambda_i  +   \errordiff_{\yprov ,i},
\end{eqnarray*}
where  $ \errordiff_{\yprow ,i}|y_i=k  \sim \GenLogistic{II}{k}$ is  truncated to  $(-\log \lambda_i, +\infty)$, since $ \yprow_{i}>0$
and  $\errordiff_{\yprov ,i}|y_i =k \sim \GenLogistic{I}{N_i-k} $ is truncated to $( - \infty,  -\log \lambda_i]$, since  $ \yprov_{i}\leq 0$. 
For $y_i=0$,  only $\yprov_{i}|\SFS{\lambda_i}, y_i$ is sampled \SFS{using that $\errordiff_{\yprov ,i}|y_i =0 \sim \GenLogistic{I}{N_i} $,  truncated to $( - \infty,  -\log \lambda_i]$.
For  $y_i=N_i$,  only $\yprow_{i}| \SFS{\lambda_i}, y_i$ is sampled using that $\errordiff_{\yprow ,i}|y_i=N_i  \sim \GenLogistic{II}{N_i}$,   truncated to  $(-\log \lambda_i, +\infty)$.}

 Since both $\Ferror $ and   $\Ferror ^{-1}$ are available  in closed form for both types of generalized logistic distributions,
  we obtain:
  \begin{eqnarray*}
  \errordiff_{\yprow ,i}=   \log  \left(  \frac{1}{ (1- p)^{1/y_i}} - 1  \right),   \quad 1- p = W_i ( 1 -  \Ferror ( -\log \lambda_i)) =
 W_i \left(  \frac{ \lambda_i }{ 1+  \lambda_i }\right) ^{y_i},
\end{eqnarray*}
where $W_i$ is a uniform random number,  see (\ref{genlogIIp}).
This proves  equation  (24):
 \begin{eqnarray*}
 \errordiff_{\yprow ,i}=  \log\left(  \frac{1 + \lambda_i }{ \lambda_i }  \frac{1}{W_i^{1/y_i}}  - 1 \right)  \Rightarrow
  \yprow_{i} =  \log \lambda_i  +   \errordiff_{\yprow ,i} =  \log\left(   ( 1+\lambda_i )  \frac{1}{W_i^{1/y_i}}  -  \lambda_i  \right).
\end{eqnarray*}
Furthermore,
  \begin{eqnarray*}
  \errordiff_{\yprov ,i}=  - \log  \left(  \frac{1}{  p^{1/(N_i-y_i)}} - 1  \right),  \quad p = V_i \Ferror (  -\log \lambda_i) = V_i  \frac{1}{(1+ \lambda_i) ^{N_i-y_i}},
\end{eqnarray*}
where $V_i$ is a uniform random number,  see (\ref{genlogIp}).
This proves equation  (25):
 \begin{eqnarray*}
 \errordiff_{\yprov ,i} =   - \log\left(  \frac{1+\lambda_i}{V_i^{1/(N_i-y_i)}}  -  1 \right)
    \Rightarrow
  \yprov_{i} =  \log \lambda_i  +   \errordiff_{\yprov ,i} =   - \log\left(   \frac{1+\lambda_i}{\lambda_i}   \frac{1}{V_i^{1/(N_i-y_i)}}  -  \frac{1}{\lambda_i}  \right).
\end{eqnarray*}
 It is easy to verify that indeed $\yprow_{i}>0$ and  $ \yprov_{i}\leq 0$.

\subsection{Computational details}

\subsubsection{Sampling the utilities in a binary model} \label{section:utibin}

Consider the latent variable representation of a binary model
\begin{equation}\label{eq:binmapp}
  \Prob{y_i=1|\lambda_i}= \Ferror (\log \lambda_i),
  \end{equation}
     involving the latent variables $\yprodiff_{i}$:
 \begin{eqnarray}
y_i=\indic{ \yprodiff_{i}>0}, && \displaystyle   \yprodiff_{i} = \log \lambda_i + \errordiff_{i},  \quad  \errordiff_{i} \sim \ferror(\errordiff_i),\label{eq:binlatapp}
\end{eqnarray}
where  $\ferror(\errordiff)$ is the pdf of the cdf $\Ferror (\errordiff)$.
$\ferror(\errordiff)= \Normalpdf (\errordiff)$ is equal to the standard normal  pdf for a probit model and equal to $\ferror(\errordiff)= e^{\errordiff}/(1+e^{\errordiff})^2$ for a logit model.
 \SFS{Given $\lambda_1, \ldots, \lambda_N$  and $\ym=(y_1, \ldots,y_N)$, 
the latent variables $\yprodiff_{1}, \ldots,
\yprodiff_{N}$ in  the latent variable  representation (\ref{eq:binlatapp})} 
are conditionally independent with conditional posterior 
 \begin{eqnarray*}
&& p(\yprodiff_{i}|y_i,\lambda_i ) \propto  p(y_i|\yprodiff_{i}) \ferror(\yprodiff_{i}-\log \lambda_i).
\end{eqnarray*}
The posterior
 of $\yprodiff_{i}$ is $\ferror(\yprodiff_{i}-\log \lambda_i) $ truncated to $(-\infty,0]$,  if $y_i=0$, and truncated to  $(0, \infty)$, if $y_i=1$,
 hence,   $\yprodiff_{i}= \log \lambda_i + \errordiff_i$, where
$\errordiff_i \sim   \ferror(\errordiff_i ) \indic{\errordiff_i  > - \log \lambda_i} $, if $ y_i=1$, and
$\errordiff_i \sim   \ferror(\errordiff_i ) \indic{\errordiff_i  \leq - \log \lambda_i} $, if $ y_i=0$.
\SFS{Since the quantile function $\Ferror ^{-1}(p)$ is available  in closed form,
it is easy to sample $\yprodiff_i$  from the posterior density $\yprodiff_i| \lambda_i,y_i$.}\footnote{To simulate $\errordiff$ from
a distribution $ \Ferror(\errordiff)$ truncated to $[a,b]$ we
simulate a uniform random number $U$ and define either $\errordiff =  \Ferror ^{-1}  (  \Ferror (a) + U ( \Ferror (b) -  \Ferror (a)))$
or  $\errordiff = \Ferror  ^{-1}  ( \Ferror (b) - U ( \Ferror (b) -  \Ferror (a)))$.}
\HW{Since $\ferror(\errordiff)$ is symmetric around 0,} 
\begin{eqnarray} \label{masamziall}
&& \yprodiff_{i}= \log \lambda_i +\Ferror^{-1} ( y_i +    U_i (1-y_i -\pi_i)),
\end{eqnarray}
where $U_i \sim \Uniform{0,1}$ and $\pi_i= \Prob{y_i=1|\lambda_i}=\Ferror(\log \lambda_i)$,
  where $\Ferror^{-1}(p) = \Normalcdf^{-1}(p)$ for
the probit model and  $\Ferror^{-1}(p)= \log p - \log (1- p)$ for the logit model.


\subsubsection{Proof of (8)}
\label{section:proofbeta}

 From the P\'{o}lya-Gamma mixture representation (5),
 it follows
that
 \begin{eqnarray*}  
 p(\tilde{\yprodiff}_{i}| \omegaH_i, \gamma, \betav ) \propto
 \exp\left\{  - \frac{\omegaH_i}{2} ( \tilde{\yprodiff}_{i} - \gamma - \Xbeta_{i} \betav ) ^2 \right\},
     \end{eqnarray*}
     \HW{see also Appendix~\ref{GenLogPG}.}
     
 Hence, conditional on $\gamma$, the posterior $\betav|\gamma, \tilde{\zv},\scalev  \sim \Normal{\Br_N  (\mr_N  (\tilde{\zv}) - \mbg \gamma),  \Br_N }$ is Gaussian with  moments as in (8). 
Using a well-known result, $p(\tilde{\zv}| \scalev ,\gamma)$
can be expressed as
 \begin{eqnarray}  \label{PGHWrat}
p(\tilde{\zv}| \scalev ,\gamma) = \frac{
\prod_{i=1} ^N p(\tilde{\yprodiff}_{i}| \omegaH_i, \gamma, \betav )
p(\betav)}{
p(\betav|\gamma, \tilde{\zv},\scalev)  }.
\end{eqnarray}
 Evaluating the right hand side of (\ref{PGHWrat}) at $\betav=\bfz$  yields,
 in combination with the Gaussian working prior $p(\gamma)$,
the conditional posterior $p(\gamma|\scalev, \tilde{\zv})$ given in (8).


\subsubsection{Details on the UPG sampler for MNL regression models using the partial dRUM representation}
\label{sec:details_mnl_pdRUM}


In the standard MNL model (\ref{mraglam}), 
we define independent Gaussian priors, $\betav_k   \sim  \Normult{\betad}{\bfz,   \Amult_k}$, for the category specific regression parameters which can be equipped with a hierarchical structure on the prior covariance matrices $\Amult_k$.
Algorithm~\ref{alg:bin}
can be extended in a fairly straightforward manner to MNL models.
 The ultimate P\'{o}lya-Gamma sampler for \SFS{categorical} data is summarized in Algorithm~\ref{alg:MNLpdRUM}.
 
To implement category specific boosting, we proceed in the following way.
The  priors for the working parameters  $\gamma_k$ and  $\delta_k$   
 are chosen similarly as for the binary model, namely
 \begin{eqnarray} \label{priormnlTR}
  \gamma_k \sim \Normal{0,  G_0},\qquad  \delta_k \sim \Gammainv{d_0,D_0}.
\end{eqnarray}
In location-based boosting,  we sample $ \gammatilde_k \sim \Normal{0,  G_0}$, and define 
$\yproztilde_{ki} = \yprodiff_{ki}  + \gammatilde_k$.
This leads to the expanded model 
\begin{eqnarray} \label{pdlateqTR} 
&& \yproztilde_{ki}=  \gamma_k + \Xbeta_i \betav_k - \xi_{ki} (\betav_{-k})  +   \errordiff_{ki}, \\ 
&& \displaystyle y_{i }= 
\left\{ \begin{array}{ll}
k, &   \yproztilde_{ki} > \gamma_k,  \\
\neq k, &  \yproztilde_{ki} \leq  \gamma_k,
\end{array} \right. \label{pdlocTR}
\end{eqnarray}
where the error term $\errordiff_{ki}$  follows a logistic distribution.
Based on \SFS{this expanded model},
we derive the posterior 
$\gamma_k| \SFS{\betav_{-k}}, \scalev_k, \tilde \zv_k$ marginalized w.r.t.~the regression parameter $\betav_k$, sample a new location parameter $\gamma_k \new$ 
from this posterior and define the shifted utility gap 
$\yprodiff_{ki} ^{L}= 
\yproztilde_{ki} - \gamma_k
\new$,
or equivalently, 
$\yprodiff_{ki}^{L}= 
\yprodiff_{ki} + \left( \gammatilde_k - \gamma_k\new \right)$.

Given the outcomes $\ym=(y_1, \dots, y_N)$, the choice equation (\ref{pdlocTR}) implies the constraint $ L( \gammatilde_k )  \leq   \gamma_k <  U(  \gammatilde_k )$,
conditional on  $\tilde{\zv}_k$, 
where 
 \begin{eqnarray} \label{pdchoiMNLBTR}
&& \displaystyle L(\gammatilde_k)=   \max _{i: y_i \neq k} \yproztilde_{ki}=
\max _{i : y_i \neq k} \yprodiff_{ki} +  \gammatilde_k,
   \\
&&  \displaystyle U( \gammatilde_k) =   \min _{i: y_i=k} \yproztilde_{ki} = \min _{i:y_i=k} \yprodiff_{ki} +   \gammatilde_k.    \nonumber
\end{eqnarray}
If   $\Count{ y_i=k}=0$, then $U(  \gammatilde_k ) =+\infty$
 and the constraint reads $\gamma_k> L(\gammatilde_k)$; if $\Count{ y_i \neq k}=0$, then
 $L( \gammatilde_k )=-\infty$ and  the constraint reads $\gamma_k< U( \gammatilde_k)$.
%
 It can be shown that  the posterior $\gamma_k|\SFS{\betav_{-k}},  \scalev_k, \SFS{\tilde{\zv}_k, \ym } $ takes the following form,
  \begin{eqnarray}  \label{pdGspMNLtrTR}
\gamma_k| \SFS{\betav_{-k}}, \scalev_k, \tilde{\zv}_k,\ym  \sim \Normal{g_k ,G_k} \indic{ L( \gammatilde_k )  \leq   \gamma_k <  U(  \gammatilde_k ) } , 
\end{eqnarray}
where $L(\gammatilde_k)$ and  $U( \gammatilde_k)$ are the boundaries defined in (\ref{pdchoiMNLBTR}) and $g_k$ and $G_k$  are defined in (\ref{pdGSpMNLTR}).
To derive (\ref{pdGspMNLtrTR}), the  likelihood  function $p(\ym|\gamma_k, \SFS{\tilde{\zv}_k})$ of the 
 location shift parameter 
  $\gamma_k$
 is combined with the
 conditional distribution $\gamma_k|\SFS{\betav_{-k}},  \scalev_k,  \SFS{\tilde{\zv}_k} \sim \Normal{g_k,G_k } $, marginalized w.r.t.~$\betav_k$. \SFS{The moments of this distribution are} given by:
 \begin{eqnarray}
& \displaystyle  G_k = \left(G_0^{-1} + 
\sum_{i=1}^N \scale_{ki}
  -  \trans{\mbgk}  \Br_k \mbgk \right)^{-1}, \quad 
 & \label{pdGSpMNLTR}\\
 & g_k  = G_k \left( 
 \mgk - \trans{\mbgk} \Br_k \mr_k (\tilde{\zv}_k) \right), & \nonumber \\
& \displaystyle \Br_k = (\Amult_k^{-1} +  \sum_{i=1}^N \scale_{ki} \trans{\Xbeta_{i}} \Xbeta_{i}  )^{-1},& \nonumber\\
& \displaystyle  \mr_k (\tilde{\zv}_k)   = \sum_{i=1}^N \trans{\Xbeta_{i}} \scale_{ki}(\yproztilde_{ki} + 
\xi_{ki} (\betav_{-k}) ) ,
&  \nonumber \\
& \displaystyle   \quad  
\mbgk= \sum_{i=1}^N  
\scale_{ki} \trans{\Xbeta_{i}}, \quad 
\mgk= \sum_{i=1}^N 
 \scale_{ki} (\yproztilde_{ki} +
\xi_{ki} (\betav_{-k})
) . & \nonumber
\end{eqnarray}
(\ref{pdGSpMNLTR}) is derived from the latent equation 
(\ref{pdlateqTR})
  under the Gaussian working prior $\gamma_k \sim \Normal{0,  G_0}$ similarly as for the logit model.
 Conditional on $\gamma_k$, the posterior $p(\betav_k|\SFS{\betav_{-k}}, \gamma_k, \tilde{\zv}_k,\scalev_k)$ is Gaussian,
 $$\betav_k|\SFS{\betav_{-k},} \gamma_k, \tilde{\zv}_k,\scalev_k  \sim \Normal{\Br_k  (\mr_k  (\tilde{\zv}_k) - \mbgk \gamma_k),  \Br_k }$$  with  moments as in (\ref{pdGSpMNLTR}).
$p(\tilde{\zv}_k|\SFS{\betav_{-k}} ,\scalev _k,\gamma_k)$
can be expressed as
 \begin{eqnarray}  \label{PGHWMNLTR}
p(\tilde{\zv}_k| \SFS{\betav_{-k}} ,\scalev _k,\gamma_k) = \frac{
\prod_{i=1} ^N p(\tilde{\yprodiff}_{ki}| \omegaH_{ki}, \gamma_k, \SFS{\betav_{-k}}, \betav_k  )
p(\betav_k)}{
p(\betav_k| \SFS{\betav_{-k}}, \gamma_k, \tilde{\zv}_k,\scalev_k)  }.
\end{eqnarray}
 Evaluating the right hand side of (\ref{PGHWMNLTR}) at $\betav_k=\bfz$  yields,
 in combination with the Gaussian  prior $p(\gamma_k)$,
the conditional   
\SFS{distribution $p(\gamma_k|\SFS{\betav_{-k},} \scalev_k, \tilde{\zv}_k )\sim \Normal{g_k,G_k } $  with moments 
given in
(\ref {pdGSpMNLTR})}.
  Taking the  \SFS{choice equation (\ref{pdlocTR})} into consideration, \SFS{the posterior} 
  $p(\gamma_k| \SFS{\betav_{-k},}\scalev_k, \tilde{\zv}_k,\ym ) \propto
  p(\ym|\gamma _k,\tilde{\zv}_k) p(\gamma_k| \SFS{\betav_{-k},} \scalev_k, \tilde{\zv}_k)
  $ \SFS{given the outcomes $\ym$}
  is  a truncated version of the Gaussian  
  \SFS{distribution $p(\gamma_k|\SFS{\betav_{-k}}, \scalev_k,  \tilde{\zv}_k )\sim \Normal{g_k,G_k }$}
  and yields the posterior $\SFS{p(\gamma_k|\betav_{-k}, \scalev_k, \tilde{\zv}_k,\ym)}$  given
in  (\ref{pdGspMNLtrTR}).
An updated working parameter $\gamma _k \new$ is sampled from (\ref{pdGspMNLtrTR}) and the proposed location-based move is corrected by defining the shifted utility gap
$ \yprodiff_{ki} \shift = \tilde{\yprodiff}_{ki} - \gamma_k \new = \yprodiff_{ki} + \Star{\gamma_k}-\gamma_k \new$.

This location-based move is followed by a scale-based move
using a scale parameter $\delta_k$ following 
the inverse Gamma prior  defined in (\ref{priormnlTR}). 
More specifically, 
we move to
$$\yprodiff_{ki}^{LS} =
\sqrt{\frac{\deltatilde_k}{\delta_k}} \yprodiff_{ki}^L,   $$
for all  $i =1, \ldots,N$.  
 To  perform the move  from $z_{ki}^L$ to 
${z}_{ki}^{LS}$, 
first $\Star{\delta_k}  $ is sampled from the prior $p(\delta_k)$,  which is the $\Gammainv{d_0,D_0}$-distribution,  and 
the  scale move   is then corrected 
	by sampling  $\delta_k$ from the posterior $p(\delta_k|\SFS{\betav_{-k}, \scalev_k,  \tilde{\zv}_k} )$ where
	$\tilde{\zv}_k =
	(\tilde{\yprodiff}_{k1}, \dots, \tilde{\yprodiff}_{kN})$
	and 
$\tilde{\yprodiff}_{ki}=\sqrt{\deltatilde_k } \yprodiff_{ki}^L$
for all $i =1, \ldots, N$.
This leads to the expanded model 
\begin{eqnarray} \nonumber 
&& \yproztilde_{ki}= 
 \sqrt{\delta_k} \big( \Xbeta_i \betav_k - \xi_{ki}  (\betav_{-k})
 \big)  +   \sqrt{\delta_k}  \, \errordiff_{ki}, \\ 
 && \displaystyle y_{i }= 
 \left\{ \begin{array}{ll}
k, &   \yproztilde_{ki} > 0,  \\
\neq k, &  \yproztilde_{ki} \leq  0,
\end{array} \right. 
 \label{pdscaTrtr}
 \end{eqnarray}
 where the error term  $\errordiff_{ki}$ follows a logistic distribution. 
Given the outcomes $\ym=(y_1, \dots, y_N)$, the choice equation in (\ref{pdscaTrtr}) does not impose any  constraint  on $ \delta_k $.
%
 The   likelihood of $\tilde  z_{ki}$  conditional on the 
scale  parameter 
  $\delta_k$ is given for all  $i=1, \ldots,N$ as:
  \begin{align*}  
	p(\tilde z_{ki}| \omegaH_{ki}, \delta_k, \betav_k, \SFS{\betav_{-k}} ) &  \propto \\
	&\propto
	\frac{1}{\sqrt{\delta_k}}\exp\left\{ - \frac{\omegaH_{ki}}{2}
	 \left( \sqrt{\frac{\tilde{\delta}_k}{\delta_k}}z^L_{ki} - \Xbeta_{i}  \betav_k + \xi_{ki} (\betav_{-k}) 
	 \right)^2 
	   \right\}.
\end{align*}
Hence, conditional on $\delta_k$, $\Star{\delta_k}$ and all latent variables  the vector of regression coefficients
$\betav_k| \SFS{\betav_{-k}}, \delta_k, \Star{\delta_k}, \zv^L_k , \scalev_k \sim
\Normal{ \br_k  ,  \Br_k} $ follows a Gaussian distribution with parameters     %
$ \br_k = \Br_k  \mr_k(\zv_k^L)$ where  $\Br_k$   is given in   (\ref{pdGSpMNLTR}) 
and $\mr_k(\zv_k^L)$ is given as:
\begin{align} \nonumber
	\mr_k(\zv_k^L)  
		&=
\mr_{bk} + \sqrt{\frac{\tilde \delta_k}{\delta_k}}
	\mr_{ak} , \\
	\mr_{ak} &= \sum_{i=1}^N \scale_{ki} z_{ki}^L \trans\Xbeta_{i}, \label{pdmabkTR}\\ 
	\nonumber
\mr_{bk} &=  
\sum_{i=1}^N \scale_{ki}\, 
\xi_{ki} (\betav_{-k})  
\, \trans\Xbeta_{i}.
\end{align}
We sample $\delta_k$ conditional on
$\zv^L_k$, \SFS{$\Star{\delta_k}$ and $\betav_{-k}$}, but marginally w.r.t.~$\betav_k$,
from the posterior $p(\delta_k| \SFS{\betav_{-k}, \Star{\delta_k}, \zv^L_k, \scalev_k})$. To determine the likelihood $p(\zv_k^L | \SFS{\betav_{-k},\scalev_k, \Star{\delta_k}}, \delta_k)$ marginally w.r.t.~$\betav_k$, we 
evaluate  the right hand side of following ratio at $\betav_k=\bfz$ 
$$p({\zv}^L_k | \SFS{\betav_{-k}, \scalev_k, \Star{\delta_k}},  \delta_k) \propto \frac{p(\betav_k)
	\prod_{i=1}^N 
	p(z_{ki}^L| \SFS{\betav_{-k}, \Star{\delta_k}, \omegaH_{ki},\delta_k, \betav_k} ) p(\omegaH_{ki})}{p(\betav_k|
 \SFS{\betav_{-k}, \Star{\delta_k},\delta_k,\zv^L_k, \omegav_k})}
=\frac{A(\delta_k)}{B(\delta_k)}.  $$
We have
\begin{align*}
\log A(\delta_k) = & -0.5 \left(  \frac{\Star{\delta_k}}{\delta_k} \sum_{i=1}^N
\omegaH_{ki} (z^L_{ki})^2+ 2\sqrt{\frac{\tilde \delta_k}{\delta_k}}
\sum_{i=1}^N
\omegaH_{ki}z^L_{ki} \, 
\xi_{ki} (\betav_{-k}) 
\right)\\
	& -  N/2 
\log(\delta_k)+ c_A,
\end{align*}
and 
\begin{align*}
\log B(\delta_k) & =   -0.5\, \mr_k(\zv_k^L) ^\top \Br_k	\mr_k(\zv_k^L) =\\
	&=
	-0.5 \Big( 
	( \mr_{bk} + \sqrt{\frac{\tilde \delta_k}{\delta_k}}
	\mr_{ak}) ^\top  \, \Br_k \, \big(	\mr_{bk} + \sqrt{\frac{\tilde \delta_k}{\delta_k}}
	\mr_{ak} \big)\Big) + c_B=\\
&=
	-0.5 \frac{\tilde \delta_k}{\delta_k} \trans{\mathbf{m}}_{ak} \Br_k \mathbf{m}_{ak}-\sqrt{\frac{\tilde \delta_k}{\delta_k}}\trans{\mathbf{m}}_{ak} \Br_k \mathbf{m}_{bk} +c_C,
	\end{align*}
where $\mr_{ak}$ and $\mr_{bk}$ have been defined in (\ref{pdmabkTR}) and $c_A$,
 $c_B$ and $c_C$ are terms independent of $\delta_k$. 
 \SFS{Due to the presence of the off-set  $\xi_{ki} (\betav_{-k}) $, 
the likelihood \SFS{$p({\zv}^L_k | \SFS{\betav_{-k}, \scalev_k, \Star{\delta_k}},  \delta_k)$}  does not admit a conjugate analysis
as opposed to a binary model. This would be possible iff  $m=1$ and the MNL  reduces to a binary logit model in which case all $\xi_{ki} (\betav_{-k}$)s are zero. }
 Combined with the prior $\delta_k \sim \Gammainv{d_0,D_0}$,
the posterior for $\delta_k|
\SFS{\betav_{-k},\tilde \delta_k,\zv^L_k, \omegav_k}$ is given as
\begin{equation} 
p(\delta_k|\SFS{\betav_{-k},\tilde \delta_k,\zv^L_k, \omegav_k})\propto 
\left( \frac{1}{\delta_k} \right) ^{d_k+1} \exp \left(- \frac{D_k}{\delta_k}\right)
\exp \left( \frac{B_k}{\sqrt{\delta_k}} \right),\label{pddeltapost}
\end{equation}
 with parameters 
\begin{align*}
d_k &=d_0+  \frac{N}{2},\\
D_k  &= D_0 + \frac{\Star{\delta_k}}{2} \left( \sum_{i=1}^N
\scale_{ki}(z^L_{ki})^2
	- \mr_{ak}^\top\Br_k \mr_{ak} \right),\\
B_k & =  \sqrt{\Star{\delta_k}}
\left( -\sum_{i=1}^N \scale_{ki} z^L_{ki}
\, 
\xi_{ki} (\betav_{-k}) 
+  \trans{\mr}_{ak} \Br_k \mr_{bk} \right).
\end{align*}
 Details  how to sample 
 \SFS{$\delta_k \new$} from such a distribution are provided in Appendix~\ref{sect:booBinU}. 

Finally, after the  scale boosting step, we
sample the regression parameter $\betav_k$
from following Gaussian distribution:
$ \betav_k |\SFS{\betav_{-k},\zv_k ^{L},\delta_k \new, \tilde \delta_k},\scalev_k
   \sim \Normult{\betad}{ \Br_k \mr_k^S,\Br_k},$ where
   $\Br_k$
  is given by (\ref{pdGSpMNLTR})
  and 
  \begin{align} \label{pdmabkTRmr}
  \mr_k^S= \mr_{bk} + \sqrt{\frac{\tilde \delta_k}{\delta \new_k}}
	\mr_{ak},
	\end{align}  
  where  $\mr_{bk}$ and $\mr_{ak}$ 
  are given by  (\ref{pdmabkTR}). This is  equivalent to sampling $\betav_k$ from the 
  partial dRUM  (\ref{pdlatTR})  based on the boosted utility gap 
  $$z^{LS}_{ki}=\sqrt{\frac{\tilde \delta_k}{\delta \new_k}} z^{L}_{ki}= \sqrt{\frac{\tilde \delta_k}{\delta \new_k}} 
   \Big(\yprodiff_{ki} + \gammatilde_k - \gamma_k\new \Big).$$
   The ultimate P\'{o}lya-Gamma sampler for \SFS{categorical} data based on the partial dRUM representation is summarized in Algorithm~\ref{alg:MNLpdRUM}.
   
      \begin{algorithm}[t]
  \caption{The ultimate P\'{o}lya-Gamma sampler for \SFS{categorical} data based on the partial dRUM model.}  \label{alg:MNLpdRUM}
  \footnotesize
  Choose starting values for  
  $\betav=(\betav_1, \dots, \betav_m)$.
  For each MCMC sweep, loop over the categories $k=1, \dots,m$ and perform the following steps: \vspace*{-2mm}
    \begin{itemize} \itemsep -1mm
  \item[(Z)]   For each $i=1, \dots, N$,  sample  the  latent variables
   $\yprovall_{i}=(\ypro_{0i}, \ypro_{1i},  \ldots, \ypro_{mi} ) \sim  p(\yprovall_{i}|\betav, y_i)$,  
   using
 $m+1$ independent uniform random numbers  $U_{i}$ and $V_{1i}, \ldots, V_{mi} $, i.e. for all
 $\ell=0, \ldots, m$, sample 
\begin{eqnarray*} 
\ypro_{\ell i} = - \log \left( - \frac{\log (U_{i})}{1+ \sum_{k=1}^m \lambda_{ki}} - \frac{\log
(V_{\ell i})}{\lambda_{\ell i}} \indic{y_{i}\neq \ell} \right),
\end{eqnarray*}
where $\log \lambda_{ki }= \Xbeta_i \betav_k $ and $\lambda_{0i}=1$. 
Define
$\yprodiff _{ki} =  \ypro_{ki} - \max_{k'\neq k} \ypro_{k',i}$ and define 
 $\xi_{ki} (\betav_{-k})= \log \left(1+ \sum_{\ell \neq \{k,0\}} \exp ( \Xbeta_i \betav_\ell)\right)$. 
Given 
$\errordiff_{ki}=\yprodiff_{ki}- \Xbeta_i \betav_k 
  + 
  \xi_{ki} (\betav_{-k})
  $, sample  
 $\scale_{ki}| \SFS{\betav} , 
 \yprodiff_{ki} 
 \sim 
 \PG{2, |\errordiff_{ki}|}
 $.

   \item[(B-L)] Location-based parameter expansion:
   sample $\Star{\gamma_k}  \sim \Normal{0,G_0} $ and propose $\tilde \yprodiff_{ki} = \yprodiff_{ki} + \Star{\gamma_k}$
 for $i=1,\dots,N$, while all other latent variables remain unchanged.
 Sample  $\gamma_k \new $ from  the truncated Gaussian-posterior 
 $p(\gamma_k|\SFS{\betav_{-k}}, \scalev_k, \tilde{\zv}_k,\ym )$  given by (\ref{pdGspMNLtrTR}) and
 define the shifted utility gap $\yprodiff_{ki} \shift  =  \tilde \yprodiff_{ki}  - \gamma_k \new $,
 for $i=1, \dots, N$,
 and $\zv_k \shift=(\yprodiff_{k1} \shift, \dots, \yprodiff_{kN} \shift)$.

 \item[(B-S)] Scale-based parameter expansion:
sample  $\Star{\delta_k} \sim \Gammainv{d_0,D_0}$ and
sample  $\delta_k \new $ from the posterior $p(\delta_k|\SFS{\betav_{-k},\tilde \delta_k,\zv^L_k, \omegav_k})$ given in 
(\ref{pddeltapost}) using the resampling technique described in Appendix~\ref{sect:booBinU}.

  \item [(P)] Sample
   $ \betav_k | \SFS{\betav_{-k}}, \zv_k \shift,\delta_k \new ,\Star{\delta_k}, \scalev_k
   \sim \Normult{\betad}{\Br_k \mr_k ^S,\Br_k}$ where $\Br_k$ is given by (\ref{pdGSpMNLTR})
 and  $\mr_k^S =\mr_{bk} + \sqrt{\tilde \delta_k /\delta \new_k}	\mr_{ak}$, where  
   $\mr_{bk}$ and $\mr_{ak}$ are given by  (\ref{pdmabkTR}).
\end{itemize}
\end{algorithm}


\subsubsection{Details on \HW{the UPG sampler for binomial logistic regression models}}
\label{sec:details_binom}

In the standard binomial regression model, the prior $\betav  \sim  \Normult{\betad}{\bfz,   \Amult_0}$ is assumed,
 where  $\Amult_0$ can be equipped with a hierarchical structure.
The working  priors are the same as in a logit model, namely 
$\gamma \sim \Normal{0,  G_0}$ and $\delta \sim \Gammainv{d_0,D_0}$.
 The ultimate P\'{o}lya-Gamma sampler for binomial regression models is summarized in Algorithm~\ref{alg:binomial}.

  \begin{algorithm}[t]
  \caption{The ultimate P\'{o}lya-Gamma sampler for binomial  data.}  \label{alg:binomial}
  \footnotesize
  Choose starting values for $\lambdav=(\lambda_1, \dots \lambda_N)$   and repeat the
  following steps: \vspace*{-2mm}
\begin{itemize} \itemsep -1mm
  \item[(Z)]   For each $i=1, \dots, N$, sample $\yprow _{i} |\lambda_i,( y_i>0) $  and  $\yprov _{i} |\lambda_i,
  ( y_i<N_i )$ 
   using Theorem~\ref{lemma2U} 
   and  sample $ \scale_{\yprow ,i}| \yprow _{i}, \lambda_i,( y_i>0 ) $ and
  $ \scale_{\yprov ,i}| \yprov _{i},  \lambda_i, (y_i<N_i) $ using  (\ref{obsffbinU}).
  
    \item[(B-L)] Location-based parameter expansion:
   sample $\Star{\gamma}  \sim \Normal{0,G_0} $ and propose $\yprowtilde_{i} =  \yprow_{i} + \Star{\gamma}$ and
$  \yprovtilde_{i} = \yprov_{i}   + \Star{\gamma}$, for $i=1,\dots,N$.
 Sample  $\gamma \new $ from $ \gamma|  \tilde{\zv},\ym$, conditional on
$ \tilde{\zv} = (\tilde{\zv} _1, \dots, \tilde{\zv} _N)$, where  $\tilde{\zv} _i=( \yprowtilde_{i}, \omegaH_{\yprow ,i}, \yprovtilde_{i} , \omegaH_{\yprov ,i})$
 and
 define shifted utilities $ \yprow_{i} \shift = \yprowtilde_{i} - \gamma \new $
 and $ \yprov_{i} \shift = \yprovtilde_{i} - \gamma \new $.
 For a standard binomial regression model,   $p(\gamma|\tilde{\zv},\ym )$ is a truncated Gaussian posterior, given in  (\ref{GVBintr}).

 \item[(B-S)] Scale-based parameter expansion:
sample  $\Star{\delta} \sim \Gammainv{d_0,D_0}$ and
sample  $\delta \new $ from $ \delta | \Star{\delta}, \zv \shift$. Define rescaled utilities  $\yprow_{i} \newS = \sqrt{\Star{\delta}/\delta \new} \yprow \shift_{i}$ and
  $\yprov_{i} \newS = \sqrt{\Star{\delta}/\delta \new} \yprov \shift_{i} $.
 For a standard binomial regression model,
  $ p(\delta | \Star{\delta} ,
     \zv \shift) \propto  \left( \frac{1}{\delta} \right) ^{d_I+1} e ^{- D_I/\delta}  e ^{ B_I /\sqrt{\delta}}  $, where  $ d_I= d_0  + d_L$ and  $ D_I= D_0+  D_L$, with $d_L, D_L$ and $B_I$ given by  (\ref{GVBin}).
     Use resampling as described in Appendix~\ref{sect:booBinU} to sample $\delta \new $.
  
   \item [(P)] Sample the unknown parameter in  $\lambda_{i}$ conditional on
   $ \zv \newS$.   For a standard binomial regression model, \SFS{this is equivalent to sampling from} $\betav|\delta \new, \Star{\delta}, \zv \shift 
 \sim \Normal{\Br_N \mr_N ,  \Br_N }$ where $\Br_N$ is defined in   (\ref{GVVVBin})  and
 $\mr_N  =  \sqrt{\Star{\delta}} /\sqrt{\delta \new}  \mb_a  -  \mb_b$, with 
  $\mb_a$
 and $\mb_b$  being defined in  (\ref{GVBin}).
\end{itemize}
\end{algorithm}

\HW{Using DA, it follows from the P\'{o}lya-Gamma mixture representation (26) 
and Appendix~\ref{GenLogPG} } %
that  for all $i$ with $y_i>0$,
 \begin{eqnarray*}  
 p(\yprowtilde_{i}| \omegaH_{\yprow ,i}, \gamma, \betav ) \propto
 \exp\left\{  - \frac{\omegaH_{\yprow ,i}}{2} ( \yprowtilde_{i}
 - \frac{ \kappa_{\yprow ,i}}{  \scale_{\yprow,i}} - \gamma - \Xbeta_{i} \betav ) ^2 \right\},
     \end{eqnarray*}
     while for all $i$ with $y_i< N_i$,
\begin{eqnarray*}  
 p(\yprovtilde_{i}| \omegaH_{\yprov ,i}, \gamma, \betav ) \propto
 \exp\left\{  - \frac{\omegaH_{\yprov ,i}}{2} ( \yprovtilde_{i}
 - \frac{ \kappa_{\yprov ,i}}{  \scale_{\yprov,i}} - \gamma - \Xbeta_{i} \betav ) ^2 \right\}.
     \end{eqnarray*}
   Conditional on $\gamma$ and the latent variables
  $ \tilde{\zv} = (\tilde{\zv} _1, \dots, \tilde{\zv} _N)$, where $\tilde{\zv} _i=( \yprowtilde_{i}, \omegaH_{\yprow ,i}, \yprovtilde_{i} , \omegaH_{\yprov ,i})$, the posterior $\betav|\gamma, \tilde{\zv} \sim \Normal{\Br_N  (\mr_N  (\tilde{\zv}) - \mbg \gamma),  \Br_N }$ is Gaussian with  moments
given in (\ref{GVVVBin}).
 Evaluating the right hand side of following ratio at $\betav=\bfz$, 
 \begin{eqnarray*}  
p(\tilde{\zv} | \gamma) \propto \frac{p(\betav)
 \prod_{i:y_i>0} p(\yprowtilde_{i}| \omegaH_{\yprow ,i}, \gamma, \betav ) \HW{p(\omegaH_{\yprow ,i})}
 \prod_{i:y_i< N_i} p(\yprovtilde_{i}| \omegaH_{\yprov ,i}, \gamma, \betav ) \HW{p(\omegaH_{\yprov ,i})}
}{
p(\betav|\gamma, \tilde{\zv})  },
\end{eqnarray*}
%
  yields, in combination with the Gaussian prior $p(\gamma)$,
the conditionally  Gaussian distribution
$ \gamma|\tilde{\zv} \sim \Normal{g_N,G_N}$, marginalized w.r.t. $\betav$, 
 where
\begin{eqnarray}
& 
G_N= (G_0^{-1} + \sum_{i=1}^N M_i  
-  \trans{\mbg}  \Br_N \mbg )^{-1}, \quad
g_N = G_N ( \mg - \trans{\mbg} \Br_N \mr_N (\tilde{\zv}) ),
 & \label{GVVVBin}\\
&\Br_N = (\Amult_0^{-1} +  \sum_{i=1}^N M_i \trans{\Xbeta_{i}} \Xbeta_{i}  )^{-1}, \quad \mr_N (\tilde{\zv})   = \sum_{i=1}^N m_i (\tilde{\zv}_i) \trans{\Xbeta_{i}} , &\nonumber
\\
 & M_i  = \indic{y_i>0}  \scale_{\yprow ,i} + \indic{y_i<N_i} \scale_{\yprov ,i},
\quad  \mbg =  \sum_{i=1}^N M_i \trans{\Xbeta_{i}}, 
    \quad \mg= \sum_{i=1}^N m_i (\tilde{\zv}_i), & \nonumber\\
 & m_i (\tilde{\zv}_i) =
 \indic{y_i>0} \left(  \yprowtilde_{i}  \scale_{\yprow ,i} -  \kappa_{\yprow ,i}\right)  + \indic{y_i<N_i} \left( \yprovtilde_{i}  \scale_{\yprov ,i}  -  \kappa_{\yprov ,i}\right). & \nonumber
\end{eqnarray}
Given the observed choices $\ym=(y_1, \dots,y_N)$, the choice equation (29) 
implies the constraint $L( \Star{\gamma})  \leq   \gamma  <  U( \Star{\gamma})$ for $\gamma$, where 
conditional on $\tilde{\zv}$:
 \begin{eqnarray} \label{choiBinU}
& \displaystyle
 L(\Star{\gamma})= \max_{i=1, \dots,N}  \yprovtilde_{i}=
  \max_{i=1, \dots,N}  \yprov_{i} + \Star{\gamma}, &\\ 
  & \displaystyle  U(\Star{\gamma}) =
 \min_{i=1, \dots,N} \yprowtilde_{i} = \min_{i=1, \dots,N} \yprow_{i} + \Star{\gamma}. & \nonumber
\end{eqnarray}
Hence,
  $p(\gamma|\tilde{\zv},\ym ) \propto
  p(\ym|\gamma,\tilde{\zv}) p(\gamma|\tilde{\zv})
  $ is a truncated version of the Gaussian posterior (\ref{GVVVBin}):
  \begin{eqnarray}  \label{GVBintr}
\gamma| \tilde{\zv},\ym  \sim \Normal{g_N,G_N} \indic{
L(\Star{\gamma})  \leq \gamma <  U(\Star{\gamma}) }.
\end{eqnarray}
An updated working parameter $\gamma \new$ is sampled from (\ref{GVBintr}) and the proposed location-based move is corrected by defining the shifted utilities
$ \yprow_{i} \shift = \yprowtilde_{i} - \gamma \new =  \yprow_{i} +  \Star{\gamma}-\gamma \new$
 and $ \yprov_{i} \shift = \yprovtilde_{i} - \gamma \new = \yprov_{i} + \Star{\gamma}-\gamma \new$ \SFS{and, correspondingly, 
 $ \zv ^L = (\zv ^L_1, \ldots, \zv ^L _N)$, where $\zv ^L _i=( \yprow_{i} \shift, \omegaH_{\yprow ,i}, \yprov_{i} \shift , \omegaH_{\yprov ,i})$}.

This location-based move is followed by a scale-based expansion.
 $\Star{\delta} \sim \Gammainv{d_0,D_0} $ is sampled from $p(\delta)$
 to propose, for each $i=1,\dots,N$, the scale moves 
$\yprowtilde_{i} = \sqrt{\Star{\delta}} \yprow_{i}  \shift $ and 
  $ \yprovtilde_{i} = \sqrt{\Star{\delta}}  \yprov_{i}  \shift $
 in the expanded model
 \begin{eqnarray} \label{booBnew2}
&& \yprowtilde_{i} =  \sqrt{\delta}  \Xbeta_{i} \betav + \sqrt{\delta} \errordiff_{\yprow ,i}, \quad y_i>0, \\
&& \yprovtilde_{i} =  \sqrt{\delta}  \Xbeta_{i} \betav + \sqrt{\delta} \errordiff_{\yprov ,i},
\quad y_i < N_i, \nonumber 
\end{eqnarray}
where the choice equation is independent of  $\delta$. 
 Using similar arguments as above, it follows from the P\'{o}lya-Gamma mixture representation  (26) 
 that  for all $i$ with $y_i>0$,
 \begin{eqnarray*}  
 p(\yprowtilde_{i}| \omegaH_{\yprow ,i}, \delta, \betav ) \propto
 \frac{1}{\sqrt \delta} \exp\left\{  - \frac{\omegaH_{\yprow ,i}}{2} \left(  \frac{\yprowtilde_{i}}{\sqrt\delta}
 - \frac{ \kappa_{\yprow ,i}}{  \scale_{\yprow,i}} - \Xbeta_{i} \betav\right) ^2 \right\},
     \end{eqnarray*}
     while for all $i$ with $y_i< N_i$,
      \begin{eqnarray*}  
 p(\yprovtilde_{i}| \omegaH_{\yprov ,i}, \delta, \betav ) \propto
 \frac{1}{\sqrt \delta} \exp\left\{  - \frac{\omegaH_{\yprov ,i}}{2} \left( \frac{\yprovtilde_{i}}{\sqrt\delta}
 - \frac{ \kappa_{\yprov ,i}}{  \scale_{\yprov,i}}  - \Xbeta_{i} \betav \right) ^2 \right\}.
     \end{eqnarray*}
Completing the squares yields that 
  $\betav|\delta, \Star{\delta}, \zv \shift   \sim
 \Normal{\Br_N \mr_N  ,  \Br_N } $ is Gaussian
 with $\Br_N$  as in (\ref{GVVVBin}) and 
 $ \mr_N  =  \sqrt{\Star{\delta}/\delta}  \mb_a  -  \mb_b $, where $\mb_a$
 and $\mb_b$ are defined in (\ref{GVBin}).

Evaluating the right hand side of following ratio at $\betav=\bfz$ yields a closed form expression for the likelihood $p(\tilde{\zv} | \delta)$: 
 \begin{eqnarray} 
 p(\tilde{\zv} | \delta) &\propto& \frac{p(\betav)
 \prod_{i:y_i>0} p(\yprowtilde_{i}| \omegaH_{\yprow ,i}, \delta, \betav ) \HW{p(\omegaH_{\yprow ,i})}
 \prod_{i:y_i< N_i} p(\yprovtilde_{i}| \omegaH_{\yprov ,i}, \delta, \betav ) \HW{p(\omegaH_{\yprov ,i})}
}{p(\betav|\delta, \tilde{\zv}) } \nonumber \\
&\propto& \left( \frac{1}{\delta} \right) ^{d_L} \exp \left(- \frac{D_L}{\delta}\right)
\exp \left( \frac{B_I }{\sqrt{\delta}} \right), \label{eq:pgBIN} \end{eqnarray}
where
\begin{eqnarray} 
&& d_L=\frac{1}{2}  \sum_{i=1} ^N \left( \indic{y_i>0}  + \indic{y_i<N_i} \right), \label{GVBin} \\ 
&& D_L= \frac{\Star{\delta}}{2} \left( \sum_{i=1} ^N ( \indic{y_i>0}  (\yprow_{i }\shift)^2 \scale_{\yprow ,i} + \indic{y_i<N_i} (\yprov \shift_{i})^2 \scale_{\yprov ,i} )
 - \mb_a^\top  \Br_N  \mb_a \right), \nonumber\\
 && B_I= \sqrt{\Star{\delta}} \left(\sum_{i=1} ^N  \left(  \indic{y_i>0}  \yprow \shift_{i} \kappa_{\yprow ,i} +
\indic{y_i<N_i} \yprov_{i}\shift \kappa_{\yprov ,i}  \right) -  \mb_a^\top  \Br_N  \mb_b \right), \nonumber\\
&& \mb_a   = \sum_{i=1}^N  \left( \indic{y_i>0}  \yprow \shift_{i} \scale_{\yprow ,i}  + \indic{y_i<N_i}  \yprov \shift_{i} \scale_{\yprov ,i} \right)  \trans{\Xbeta_{i}} , \nonumber\\
&& \mb_b = \sum_{i=1}^N  \left( \indic{y_i>0} \kappa_{\yprow ,i}   + \indic{y_i<N_i}  \kappa_{\yprov ,i} \right) \trans{\Xbeta_{i}}, \nonumber
 \end{eqnarray}
 and where $\Br_N$ is the same as in (\ref{GVVVBin}).
 However, due to the presence of the (fixed) location
 parameters  $\kappa_{\yprow ,i} $ and $\kappa_{\yprov ,i}$, the likelihood $p(\tilde{\zv} | \delta)$  does not  take a conjugate form
\HW{as opposed to a binary logistic model. Combined with the inverse gamma prior $\delta \sim \Gammainv{d_0,D_0}$,
the posterior $p(\delta|\Star{\delta}, \zv \shift)$
belongs for a binomial regression model to the same generalized distribution family as the posterior $p(\delta_k|\Star{\delta}_k, 
\SFS{\betav_{-k}, \zv_k \shift, \scalev_k})$ arising for a  MNL model,}
with density 
$$ p( \delta| \Star{\delta}, \zv \shift ) \propto  \left( \frac{1}{\delta} \right) ^{d_I+1} e ^{- D_I/\delta}  e ^{ B_I /\sqrt{\delta}} ,$$
where $ d_I= d_0  + d_L$ and  $ D_I= D_0+  D_L$.
$\delta \new $ is sampled from $ \delta | \Star{\delta}, \zv \shift$ to define rescaled utilities  $\yprow_{i}
\newS = \sqrt{\Star{\delta}/\delta \new} \yprow \shift_{i}$ and
  $\yprov_{i} \newS = \sqrt{\Star{\delta}/\delta \new} \yprov \shift_{i} $.
  The  posterior $p(\delta|\Star{\delta}, \zv \shift)$
reduces
to the
inverse gamma distribution $\Gammainv{d_I,D_I}$, iff  $N_i=1$ for all $i=1,\dots, N$ and the binomial model reduces to a logit model
 in which case all $\kappa_{\yprow ,i} $s and $\kappa_{\yprov ,i}$s
 are zero
 and   $B_I=0$.
It can be shown that in any case $D_I>0$, provided that $D_0>0$.  
Details  how to sample from such a distribution are provided in Appendix~\ref{sect:booBinU}.

 This concludes the boosting step and $\betav|\zv \newS$ is sampled conditional on
 $\zv \newS$, or equivalently from
 the Gaussian posterior  
 $\betav|\delta \new, \Star{\delta}, \zv \shift
 \sim \Normal{\Br_N \mr_N ,  \Br_N }$ where $\Br_N$ is defined in   (\ref{GVVVBin})  and
 $\mr_N  =  \sqrt{\Star{\delta}} /\sqrt{\delta \new}  \mb_a  -  \mb_b$, with
  $\mb_a$
 and $\mb_b$  being defined in  (\ref{GVBin}).

\subsubsection{\HW{Sampling the scale parameter in scale boosting for binomial and MNL models}} \label{sect:booBinU}

In general, the marginal density for the scale parameter arising in scale boosting for  
binomial and MNL models
does not belong to a well-known distribution family unless these models reduce to a logistic model.


\begin{figure}[!t]
  \centering
\begin{subfigure}{0.45\textwidth}
  \centering
  \includegraphics[width=\linewidth]{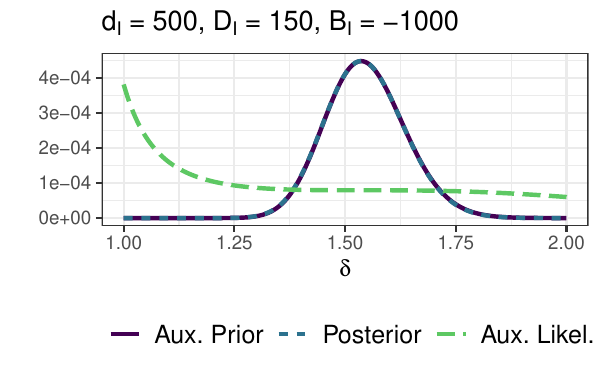}
\end{subfigure}
\begin{subfigure}{0.45\textwidth}
  \centering
  \includegraphics[width=\linewidth]{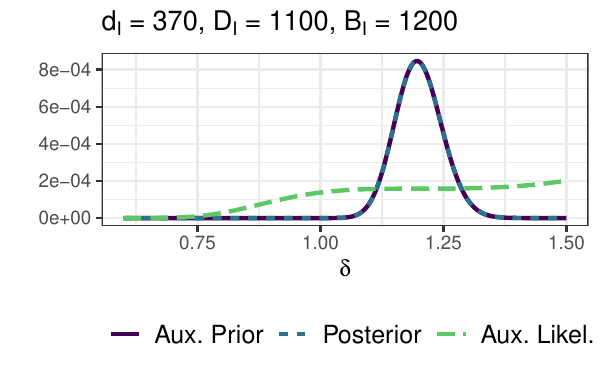}
\end{subfigure}
\caption{Approximating  $p( \delta |\cdot )$ by the \SFS{inverse Gamma} density  $\delta \sim \Gammainv{d_I^\star ,D_I^\star}$ for a case where $B_I<0$ (left) and $B_I>0$ (right). The posterior densities are covered by the auxiliary prior densities.}\label{figbinnew}
\end{figure}


The  following  resampling technique is used to sample in Algorithm~\ref{alg:binomial} from $$ p( \delta|  \SFS{\Star{\delta}, \zv \shift}) \propto  \left( \frac{1}{\delta} \right) ^{d_I+1} e ^{- D_I/\delta}  e ^{ B_I /\sqrt{\delta}} $$  with   obvious modifications to 
sample in  Algorithm~\ref{alg:MNLpdRUM} from  the posterior
\begin{equation*} 
p(\delta_k|\Star{\delta}_k,
\SFS{\betav_{-k}, \zv_k \shift, \scalev_k})\propto 
\left( \frac{1}{\delta_k} \right) ^{d_k+1} \exp \left(- \frac{D_k}{\delta_k}\right)
\exp \left( \frac{B_k}{\sqrt{\delta_k}} \right).
\end{equation*}
 Choose an  `auxiliary  prior'  $\pi(\delta)$ for resampling such that the mode  and the curvature of $\pi(\delta)$
     coincide with the  mode $\delta_M$ and the curvature $I_p$  of the posterior
      $p ( \delta | \SFS{\Star{\delta}, \zv \shift} )$ which are  given by:
   \begin{eqnarray*}
\delta_M= \frac{16 D_I^2}{\left[B_I+\sqrt{B_I^2+16 D_I(d_I+1)}\right]^2}, 
\qquad I_p=  -   \frac{\sqrt{B_I^2+16 D_I(d_I+1)}}{4 \cdot (\delta_M) ^{\frac{5}{2}}}.
\end{eqnarray*}
 Resampling works as follows. $L$ draws  $\delta \im{l} \sim \pi(\delta)$, $l=1,\dots,L$,   from
 the  auxiliary prior  are resampled using  weights proportional to the `auxiliary  likelihood'  $\ell(\delta)= p( \delta| \SFS{\Star{\delta}, \zv \shift} )/\pi(\delta)$, given by:
     \begin{eqnarray*}
  \log \ell(\delta) &\propto& \log p ( \delta |  \tilde \zv, \ym) - \log \pi(\delta) \\
& \propto&
 -(d_I+1) \log \delta - \frac{D_I}{\delta}  + \frac{B_I}{\sqrt{\delta}}  - \log \pi(\delta).
     \end{eqnarray*}
The desired draw  from $p( \delta |\cdot ) $ is given by
 $\delta \im{l ^\star}  $, where $ l ^\star   \sim \Mulnom{1; w_1, \dots, w_L}$ and
the weights $w_l \propto \ell(\delta \im{l})$   are normalized to 1.

 The auxiliary likelihood $\ell(\delta)$ is expected to be rather flat over the support of  $\pi(\delta)$, see Figure~\ref{figbinnew} for illustration. Hence, $w_1, \dots, w_L$ is expected to be close to a uniform distribution and $L$ can be pretty small ($L=5$ or $L=10$ should be enough).

The factorization of the posterior suggests two distribution families as auxiliary prior $\pi(\delta)$. First,
 the inverse Gamma prior $\delta \sim \Gammainv{d_I^\star ,D_I^\star}$ with
   mode $\delta_{IG}$ and   curvature $I_{IG}$ of the pdf given by:
   \begin{eqnarray*}
&& \displaystyle \delta_{IG} = \frac{D_I ^\star}{d_I ^\star +1},  \qquad  I_{IG} = 
-   \frac{(d_I ^\star +1)^3}{(D_I ^\star)^2}.
\end{eqnarray*}
Matching the mode, i.e. $\delta_{IG}=\delta_{M} $,  and the curvature, i.e. $I_{IG}=I_p$,  to the posterior, i.e.,
 \begin{eqnarray*}
&& I_{IG} =  -  \frac{(d_I ^\star +1)^3}{(D_I ^\star)^2} = -  \frac{d_I ^\star +1}{(\delta_{IG})^2}  =  - \frac{d_I ^\star +1}{(\delta_{M})^2} =  I_p , \\
&& \delta_{IG}= \frac{D_I ^\star}{d_I ^\star +1} = \delta_{M} ,
\end{eqnarray*}
yields following optimal choice for the parameters $(d_I^\star ,D_I^\star)$:
 \begin{eqnarray*}
d_I ^\star  = - I_p \cdot(\delta_{M})^2 -1 ,  \quad  D_I ^\star = \delta_{M} (d_I ^\star +1).
\end{eqnarray*}
The log likelihood ratio reads:
 \begin{eqnarray*}
  \log \ell(\delta)  \propto
 -(d_I- d_I ^\star) \log \delta - \frac{D_I - D_I ^\star}{\delta}  + \frac{B_I}{\sqrt{\delta}}.
     \end{eqnarray*}
Second, provided that $B_I^\star>0$, the translated inverse Gamma prior $\sqrt{\delta} \sim \Gammainv{2 b_I^\star, B_I^\star}$ \footnote{As follows from the law of transformation of densities, a random variable $\delta$, where the transformed variable $\sqrt{\delta} \sim \Gammainv{2a,b}$, has density
 \begin{eqnarray} \label{dSQdelta}
 p_{SQ}(\delta; 2a,b)= \frac{b^{2a}}{2 \Gamfun{2a}} \left( \frac{1}{\delta}\right)^{a+1}e ^{-\frac{b}{\sqrt{\delta}}} .
 \end{eqnarray}
To sample from such a density, we sample $ X \sim \Gammainv{2a,b}$  and take the square, i.e.  $\delta = X^2$.}
   with mode $\delta_{SQ}$ and the curvature $I_{SQ}$ of the pdf given by:
   \begin{eqnarray*}
&& \displaystyle \delta_{SQ} =
\frac{(B_I^\star )^2}{4 (b_I ^\star+1)^2} ,  \qquad  I_{SQ} = 
-   \frac{8 (b_I ^\star +1)^5}{(B_I ^\star)^4}.
\end{eqnarray*}
Matching the mode, i.e. $\delta_{SQ}=\delta_{M} $,  and the curvature, i.e. $I_{SQ}=I_p$,  to the posterior, i.e.,
 \begin{eqnarray*}
&& I_{SQ} =  -   \frac{8 (b_I ^\star +1)^5}{(B_I ^\star)^4} = -  \frac{b_I ^\star +1}{2 (\delta_{SQ})^2}  =
-  \frac{b_I ^\star +1}{2 (\delta_{M})^2} =  I_p , \\
&& \delta_{SQ}= \frac{(B_I^\star )^2}{4 (b_I ^\star+1)^2}  = \delta_{M} ,
\end{eqnarray*}
yields following optimal choice for the parameters $(b_I^\star ,b_I^\star)$:
 \begin{eqnarray*}
b_I ^\star  = - 2 I_p \cdot (\delta_{M})^2 -1 ,  \quad  B_I ^\star =  2 (b_I ^\star +1) \sqrt{\delta_{M}}.
\end{eqnarray*}
The log likelihood ratio reads:
 \begin{eqnarray*}
  \log \ell(\delta)  \propto
 -(d_I- b_I ^\star) \log \delta - \frac{D_I}{\delta}  + \frac{B_I  + B_I ^\star}{\sqrt{\delta}}.
     \end{eqnarray*}





\subsection{The UPG sampler for binary state space models} \label{sec:smmUPG}

First,
  location-based boosting based on a $\Normal{0,G_0}$ working prior is applied.
$\Star{\gamma}  \sim \Normal{0,G_0} $ is sampled from this  prior
 to propose, for each $t=1,\dots,T$, a location move $\tilde{\yprodiff}_{t} = \yprodiff_{t}+ \Star{\gamma}$  in an expanded state space model, where the observation equation is affected in the following way:
 \begin{eqnarray} 
&  y_t=1 \Leftrightarrow  \tilde \yprodiff_{t} > \gamma, \quad
 \tilde \yprodiff_{t} =  \gamma + \Xbeta_{t} \betat{t}  +  \varepsilon_{t},  \label{BBssm}
\end{eqnarray}
while the transition equation 
(31) 
and the initial distribution $\betat{0}\sim \Normult{\betad}{\bfz,\Pt{0}{0}}$ remain the same.
Conditional on the latent variables
  $\scalev=( \scale_1,\dots, \scale_T)$, where $\omegaH_t \sim \PG{2,0}$, it follows
 from the P\'{o}lya-Gamma mixture representation (5) 
that the observation density $p(\tilde{\yprodiff}_{t}| \omegaH_t, \gamma, \betav_t ) $
takes the form
 \begin{eqnarray*}  
 p(\tilde{\yprodiff}_{t}| \omegaH_t, \gamma, \betav_t ) \propto
 \exp\left\{  - \frac{\omegaH_t}{2} ( \tilde{\yprodiff}_{t}- \Xbeta_{t} \betav_t   - \gamma ) ^2 \right\}.
     \end{eqnarray*}
Hence, in combination with the  transition density 
(31) 
and the initial distribution $\betat{0}\sim \Normult{\betad}{\bfz,\Pt{0}{0}}$, a conditionally Gaussian state space model is obtained and the Kalman filter can be applied  to determine the moments of the filtering density conditional on $\gamma$, given  $\tilde{\zv}^t =(\tilde{\yprodiff}_{1}, \dots, \tilde{\yprodiff}_{t})$. 
These moments can be expressed as:  \begin{eqnarray} \label{KFboo}
\betat{t}  | \tilde{\zv} ^{t}, \scalev, \gamma, \theta_1, \dots, \theta_d \sim 
 \Normult{\betad}{\xthat{t}{t} + \filtg_t \gamma,\Pt{t}{t}},
\end{eqnarray}
where 
$\filtg_t$ will be defined below in (\ref{recg}) and
$\xthat{t}{t}$ and $\Pt{t}{t}$ are the moments of the filtering density
$\betat{t}  | \gamma=0, \tilde{\zv} ^{t}, \scalev, \theta_1, \dots, \theta_d \sim 
 \Normult{\betad}{\xthat{t}{t},\Pt{t}{t}}$ for the specific state space model where $\gamma=0$:
\begin{eqnarray}
&&\betat{t} = \betat{t-1} + \wt{t}, \quad \wt{t} \sim \Normult{\betad}{\bf0,\Qrcm}, 
\quad \tilde \yprodiff_{t} =  \Xbeta_{t} \betat{t}  +  \varepsilon_{t}.   \label{movesssm}
\end{eqnarray}
Starting with $\xthat{0}{0}=\bfz$, these moments are  given for $t=1, \dots, T$ by the Kalman filter:
  \begin{eqnarray}
&& 
  \Pt{t}{t-1}  = \Pt{t-1}{t-1}   + \Diag{\theta_1, \dots, \theta_\betad},  \label{S2wooKFmain} \\
&& \tilde{\yprodiff}_{t} |\tilde{\zv}^{t-1},\scalev,  \theta_1, \dots, \theta_d  \sim \Normal{\yprohat{t}{t-1}, \Stc{t}{t-1}}, 
\nonumber \\ && \yprohat{t}{t-1} = \Xbeta_{t}\xthat{t-1}{t-1}  ,  \quad \Stc{t}{t-1}  = \Xbeta_{t} \Pt{t }{t-1}  \trans{\Xbeta_{t}} + 1/\omega_t , \nonumber \\
 && \xthat{t}{t} =\xthat{t-1}{t-1} + \Kt{t} (\tilde{\yprodiff}_{t} - \yprohat{t}{t-1} ), 
 \nonumber \\ &&\Pt{t}{t} = (\identm - \Kt{t}\Xbeta_{t} ) \Pt{t}{t-1}, \nonumber \quad \Kt{t} = { \Pt{t}{t-1}} \trans{\Xbeta_{t}} \Stc{t}{t-1}^{-1}. \nonumber 
\end{eqnarray} 
In addition, the weight  $\filtg_t$ in (\ref{KFboo}) satisfies the following recursion for $t=1, \dots, T$ with  $\filtg_0=0$:
 \begin{eqnarray} \label{recg}
 \filtg_t=  (\identm - \Kt{t}\Xbeta_{t} )  \filtg_{t-1} - \Kt{t}.
 \end{eqnarray}
The representation (\ref{KFboo}) is easy to prove. Based on the Kalman filter for a model with arbitrary $\gamma$, we obtain for $t=1$: 
\begin{eqnarray*}
&& \betat{1}| \tilde{\zv}^{1}, \scalev, \gamma, \theta_1, \dots, \theta_d  \sim  \Normult{\betad}{\xthat{1}{1} (\gamma), \Pt{1}{1}}, \\
&& \xthat{1}{1} (\gamma) = \xthat{0}{0} + \Kt{1} ( \tilde{\yprodiff}_{t}- \gamma - \Xbeta_{1}\xthat{0}{0}) = 
\Kt{1}  \tilde{\yprodiff}_{t} - \Kt{1} \gamma= \xthat{1}{1}  + \filtg_1 \gamma,
\end{eqnarray*}
where $\xthat{1}{1} $ and $\Pt{1}{1}$ are the same as in (\ref{S2wooKFmain}) and 
$\filtg_1=- \Kt{1}$.  
Assuming that (\ref{KFboo}) holds up to $t-1$ and based on the Kalman filter for a model with arbitrary $\gamma$, we obtain at time-point $t$:
\begin{eqnarray*}
&& \betat{t}| \tilde{\zv}^{t}, \scalev, \gamma, \theta_1, \dots, \theta_d  \sim  \Normult{\betad}{\xthat{t}{t} (\gamma), \Pt{t}{t}}, \nonumber \\
 \xthat{t}{t} (\gamma) & = &  \xthat{t-1}{t-1} (\gamma) + \Kt{t} ( \tilde{\yprodiff}_{t}- \gamma - \Xbeta_{t}\xthat{t-1}{t-1} (\gamma)) =   (\identm - \Kt{t}\Xbeta_{t} ) \xthat{t-1}{t-1} (\gamma) 
+ \Kt{t} \tilde{\yprodiff}_{t}- \Kt{t} \gamma=\\
& = &   (\identm - \Kt{t}\Xbeta_{t} ) \xthat{t-1}{t-1} + \Kt{t} \tilde{\yprodiff}_{t} +
 (\identm - \Kt{t}\Xbeta_{t} ) \filtg_{t-1} \gamma  - \Kt{t} \gamma=  \xthat{t}{t} + \filtg_t \gamma,
\end{eqnarray*}
where  $\xthat{t}{t} $ and $\Pt{t}{t}$ are the same as in (\ref{S2wooKFmain}) and  
$\filtg_t$ satisfies recursion (\ref{recg}).

To derive the likelihood $p(\tilde \zv| \scalev,\gamma,\theta_1, \dots, \theta_d)$, we exploit the well-known representation of the likelihood as a product of the one-step-ahead predictive densities resulting from Kalman filtering:
\begin{eqnarray*} 
  p(\tilde \zv| \scalev,\gamma,\theta_1, \dots, \theta_d) & = & \prod_{t=1}^T  p(\tilde{\yprodiff}_{t} | \tilde \zv^{t-1 }, \scalev, \gamma,\theta_1, \dots, \theta_d)  \propto
  \exp \left\{- \frac{1}{2} \sum_{t=1}^T \frac{(\tilde{\yprodiff}_t -\yprohat{t}{t-1}(\gamma))^2 }{\Stc{t}{t-1}} \right\}.
\end{eqnarray*}    
Since the mean of the one-step-ahead predictive  distribution is given by $$\yprohat{t}{t-1}(\gamma) = \Xbeta_{t}\xthat{t-1}{t-1} (\gamma) = 
\yprohat{t}{t-1} + \Xbeta_{t}\filtg_{t-1} \gamma,
$$ while the variance is the same as for $\gamma=0$, 
we obtain:
\begin{eqnarray*} 
  p(\tilde \zv| \scalev,\gamma,\theta_1, \dots, \theta_d) \propto
  \exp \left\{- \frac{1}{2} \sum_{t=1}^T \frac{(\tilde{\yprodiff}_t -\yprohat{t}{t-1}- \Xbeta_{t}\filtg_{t-1} \gamma)^2 }{\Stc{t}{t-1}} \right\},   
\end{eqnarray*}   
where $\yprohat{t}{t-1}$ and $\Stc{t}{t-1}$ are defined in (\ref{S2wooKFmain}). 
Combining this likelihood with the Gaussian working prior $p(\gamma)$,
yields the conditional  Gaussian  posterior  $ \gamma|\scalev, \tilde{\zv}, \theta_1, \dots, \theta_d  \sim \Normal{G_T \mg,G_T}$ where:
\begin{eqnarray}
\displaystyle G_T= \left(G_0^{-1} + \sum_{t=1}^T \frac{(\Xbeta_{t}\filtg_{t-1})^2}{\Stc{t}{t-1}} \right)^{-1}, \quad
 \mg= \sum_{t=1}^T \frac{(\tilde{\yprodiff}_t -\yprohat{t}{t-1}) \Xbeta_{t}\filtg_{t-1}}{\Stc{t}{t-1}}.
  \nonumber
\end{eqnarray}
  Since the choice equation in (\ref{BBssm}) depends on $\gamma$,
  $p(\gamma|\scalev, \tilde{\zv},\theta_1, \dots, \theta_d)$ has to be combined with the likelihood $p(\ym|\gamma,\tilde{\zv} )$
  of the observed outcomes $\ym=(y_1,\dots,y_T)$  as before to define the posterior $p(\gamma|\scalev, \tilde{\zv}, \theta_1, \dots, \theta_d,\ym )$:
  \begin{eqnarray}  \label{GVVVssm}
\gamma|\scalev, \tilde{\zv}, \theta_1, \dots, \theta_d , \ym  \sim \Normal{G_T \mg,G_T} \indic{ L(\Star{\gamma})  \leq \gamma  <  U(\Star{\gamma}) },
\end{eqnarray}
where  $L(\Star{\gamma})=\max_{t:y_t=0} \tilde{\yprodiff}_{t} = \max_{t:y_t=0} \yprodiff_{t} + \Star{\gamma} $ and
 $U(\Star{\gamma})=\min_{t:y_t=1} \tilde{\yprodiff}_{t}= \min_{t:y_t=1} \yprodiff_{t} + \Star{\gamma}$.  
 %
An updated working parameter $\gamma \new$ is sampled from (\ref{GVVVssm}) and the proposed location-based move is corrected by defining the shifted utilities $ \yprodiff_{t} \shift = \tilde{\yprodiff}_{t} - \gamma \new = \yprodiff_{t} + \Star{\gamma}-\gamma \new$.

This location-based move is followed by a scale-based expansion, using
an inverse gamma distribution, $\Gammainv{d_0,D_0}$, as  working prior $p(\delta)$.
 $\Star{\delta} $ 
 is sampled from $p(\delta)$
 to propose, for each $t=1,\dots,T$, a scale move $\tilde{\yprodiff}_{t} = \sqrt{\Star{\delta}} \yprodiff_{t} \shift $ in the expanded state space model
 \begin{eqnarray} \nonumber
&  y_t=1 \Leftrightarrow  \tilde \yprodiff_{t} > 0, \quad
 \tilde \yprodiff_{t} =  \sqrt{\delta} \Xbeta_{t} \betat{t}  + \sqrt{\delta} \varepsilon_{t},  \quad  \label{cGuasstild}
\end{eqnarray}
while the transition equation (31) 
and the initial distribution $\betat{0}\sim \Normult{\betad}{\bfz,\Pt{0}{0}}$ remain the same.
From the P\'{o}lya-Gamma mixture representation of the error terms $\varepsilon_t$, it follows that the observation equation takes the form
 \begin{eqnarray*}  
 p(\tilde{\yprodiff}_{t}| \omegaH_t, \delta, \betav_t ) \propto
 \frac{1}{\sqrt \delta}\exp\left\{  - \frac{\omegaH_t}{2} \left(  \frac{\tilde{\yprodiff}_{t}}{\sqrt \delta} - \Xbeta_{t} \betat{t}\right) ^2 \right\}.
    \end{eqnarray*}
Again, in combination with the  transition density 
(31) 
and the initial distribution $\betat{0}\sim \Normult{\betad}{\bfz,\Pt{0}{0}}$, a conditionally Gaussian state space model is obtained and the Kalman filter can be applied  to determine the moments of the filtering density conditional on $\delta$, given  $\tilde{\zv}^t$.
These moments can be expressed as:  
\begin{eqnarray} \label{KFboodel}
\betat{t}  | \tilde{\zv} ^{t}, \scalev, \delta, \theta_1, \dots, \theta_d \sim 
 \Normult{\betad}{\frac{1}{\sqrt \delta}\xthat{t}{t},\Pt{t}{t}},
\end{eqnarray}
where $\xthat{t}{t}$ and $\Pt{t}{t}$ are the moments of the filtering density
$\betat{t}  | \delta=1, \tilde{\zv} ^{t}, \scalev, \theta_1, \dots, \theta_d \sim 
 \Normult{\betad}{\xthat{t}{t},\Pt{t}{t}}$ for the specific state space model where $\delta=1$. This model takes the same form as in (\ref{movesssm}), however with a different outcome variable  $\tilde \yprodiff_{t}$ than before.
Its  moments are  given by the Kalman filter outlined in 
(\ref{S2wooKFmain}), where all (co)variances, i.e.
$ \Pt{t}{t-1}$, $\Pt{t}{t}$, and $\Stc{t}{t-1}$, and the Kalman gain  $\Kt{t}$ are the same as for the location boost,
whereas $\xthat{t}{t}$ and $\yprohat{t}{t-1}$ depend on
$\tilde \yprodiff_{t}$ and have to be recomputed.

The representation (\ref{KFboodel}) is easy to prove. Based on the Kalman filter for a model with arbitrary $\delta$, we obtain for $t=1$: 
\begin{eqnarray*}
&& \betat{1}| \tilde{\zv}^{1}, \scalev, \delta, \theta_1, \dots, \theta_d  \sim  \Normult{\betad}{\xthat{1}{1} (\delta), \Pt{1}{1}}, \\
&& \xthat{1}{1} (\delta) = \xthat{0}{0} + \Kt{1} ( \tilde{\yprodiff}_{1}/\sqrt{\delta} - \Xbeta_{1}\xthat{0}{0}) =  \Kt{1}  \tilde{\yprodiff}_{1} /\sqrt{\delta} = \frac{1}{\sqrt{\delta}} \xthat{1}{1},
\end{eqnarray*}
where $\xthat{1}{1} $ and $\Pt{1}{1}$ are the moments for $\delta=1$.   
Assuming that (\ref{KFboodel}) holds up to $t-1$ and based on the Kalman filter for a model with arbitrary $\delta$, we obtain for time $t$:
\begin{eqnarray*}
&& \betat{t}| \tilde{\zv}^{t}, \scalev, \delta, \theta_1, \dots, \theta_d  \sim  \Normult{\betad}{\xthat{t}{t} (\delta), \Pt{t}{t}}, \nonumber \\
 \xthat{t}{t} (\delta) & = &  \xthat{t-1}{t-1} (\delta) + \Kt{t} ( \tilde{\yprodiff}_{t}/\sqrt{\delta}  -  \Xbeta_{t}\xthat{t-1}{t-1} (\delta)) \\
 &=&  \frac{1}{\sqrt{\delta}} 
 \left( \xthat{t-1}{t-1}+ \Kt{t} ( \tilde{\yprodiff}_{t}  -  \Xbeta_{t}\xthat{t-1}{t-1} ) \right)
 =  \frac{1}{\sqrt{\delta}} \xthat{t}{t}.
\end{eqnarray*}
%
 To derive the likelihood $p(\tilde \zv| \scalev,\delta,\theta_1, \dots, \theta_d)$, we use once more
 the product of the one-step-ahead predictive densities resulting from Kalman filtering:
\begin{eqnarray*} 
  p(\tilde \zv| \scalev,\delta,\theta_1, \dots, \theta_d) & = & \prod_{t=1}^T  p(\tilde{\yprodiff}_{t} | \tilde \zv^{t-1 }, \scalev,\delta,\theta_1, \dots, \theta_d).
\end{eqnarray*}    
Since the one-step-ahead predictive  distribution is given by:
\begin{eqnarray*}
\tilde{\yprodiff}_{t} | \tilde \zv^{t-1 }, \scalev,\delta,\theta_1, \dots, \theta_d \sim\Normal{\yprohat{t}{t-1}(\delta),\Stc{t}{t-1} (\delta)},
\end{eqnarray*}
where
\begin{eqnarray*}
\yprohat{t}{t-1}(\delta) = \sqrt{\delta} \Xbeta_{t}\xthat{t-1}{t-1} (\delta) = \Xbeta_{t}\xthat{t-1}{t-1}=  \yprohat{t}{t-1},
\qquad \Stc{t}{t-1} (\delta)= \delta \Stc{t}{t-1}, \end{eqnarray*}
we obtain:
\begin{eqnarray*} 
  p(\tilde \zv| \scalev,\delta,\theta_1, \dots, \theta_d) \propto
   \left(\frac{1}{\delta} \right) ^{T/2}
  \exp \left\{- \frac{1}{2 \delta} \sum_{t=1}^T \frac{(\tilde{\yprodiff}_t -\yprohat{t}{t-1})^2}{\Stc{t}{t-1}} \right\}.   
\end{eqnarray*}   
Combining this likelihood 
 with the inverse gamma working prior, the posterior $\delta|  \tilde \zv ,\scalev, \theta_1, \dots, \theta_d \sim \Gammainv{d_T,D_T}$ with the following moments results:
 \begin{eqnarray} \label{scdelH}
&& \delta| \tilde \zv ,\scalev, \theta_1, \dots, \theta_d \sim \Gammainv{d_T,D_T}, \\
 && d_T= d_0 + \frac{T}{2}, \quad D_T= D_0 +  \nonumber
  \frac{1}{2} \sum_{t=1}^T \frac{(\tilde \yprodiff_{t}  -\yprohat{t}{t-1})^2 }{\Stc{t}{t-1}}. \end{eqnarray}
  An updated working parameter $\delta \new$ is sampled from
(\ref{scdelH}) and the proposed scale-based move is corrected by defining the rescaled utilities
 $ \yprodiff_{t} \newS = \tilde \yprodiff_{t} /\sqrt{\delta \new} = \sqrt{\Star{\delta}/\delta \new}\yprodiff_{t} \shift$ for all $t$.
This concludes the boosting step. 
 
The  state process $\betat{0},\dots,\betat{T}$ is then  sampled conditional on 
 $\zv \newS$  from the smoothing density
 $p(\betat{0},\dots,\betat{T}| \zv \newS, \scalev, \theta_1, \dots, \theta_d )$ of the boosted state space model
 \begin{eqnarray}
&&\betat{t} = \betat{t-1} + \wt{t}, \quad \wt{t} \sim \Normult{\betad}{\bf0,\Qrcm}, 
\quad  \yprodiff_{t} \newS =  \Xbeta_{t} \betat{t}  +  \varepsilon_{t}   \label{booboossm}
\end{eqnarray}
using FFBS \citepsupp{fru:dat}. 
 It is easy to verify that the moments of the filtering density in the boosted state space model (\ref{booboossm}) are identical to 
 \begin{eqnarray} \label{KFboosmod}
\betat{t}  | \SFS{\zv ^{LS,t}}, \scalev, \theta_1, \dots, \theta_d \sim 
 \Normult{\betad}{\frac{1}{\sqrt{\delta \new}}\xthat{t}{t},\Pt{t}{t}},
\end{eqnarray}
where $\SFS{\zv ^{LS,t}=(z_1^{LS}, \ldots, z_t^{LS})}$.
Indeed,  
$ \xthat{1}{1} (\delta \new) =  \Kt{1}  \tilde{\yprodiff}_{1} /\sqrt{\delta \new} = \Kt{1}  \yprodiff \newS_{1}$ is equal to the mean of the filtering density at $t=1$. 
Assuming this identity holds up to $t-1$, we obtain that
\begin{eqnarray*}
 \xthat{t}{t} (\delta \new)  &=& \xthat{t-1}{t-1} (\delta \new) + \Kt{t} ( \tilde{\yprodiff}_{t}/\sqrt{\delta \new}  -  \Xbeta_{t}\xthat{t-1}{t-1} (\delta \new)) \\
 &=& 
 \xthat{t-1}{t-1} (\delta \new) + \Kt{t} ( \yprodiff \newS_{t}  -  \Xbeta_{t}\xthat{t-1}{t-1} (\delta \new)) 
\end{eqnarray*}
is equal to the mean of the filtering density at time $t$.

 Hence, the moments $\xthat{t}{t}$ 
 and $\Pt{t}{t}$ of the Kalman filter (\ref{KFboodel}) 
 underlying the  scale-based parameter expansion can be
 recycled in the backward sampling step
of FFBS. Starting with a draw $\betat{T}$ from
the filter density 
\begin{eqnarray} \label{KFboosmo}
\betat{T}  | \SFS{\zv ^{LS,T}} , \scalev,  \theta_1, \dots, \theta_d \sim 
 \Normult{\betad}{\frac{1}{\sqrt{\delta \new}}\xthat{T}{T},\Pt{T}{T}},
\end{eqnarray}
the state $\betat{t}$ is sampled backwards in time for $t=T-1, \dots, 0$ 
using\footnote{The conditional density $\xt{t}|\xt{t+1},\dots, \xt{T}, \zv \newS , \scalev, \theta_1, \dots, \theta_d$ is proportional to
\begin{eqnarray*}
&& p(\xt{t}|\xt{t+1},\dots, \xt{T}, \zv \newS , \scalev, \theta_1, \dots, \theta_d) \propto
p(\xt{t}|\SFS{\zv ^{LS,t}}, \scalev, \theta_1, \dots, \theta_d)  p(\xt{t+1} |\xt{t}, \theta_1, \dots, \theta_d) \propto \\
&&  \exp \left\{ -\frac{1}{2} 
\trans{\left( \xt{t} - \frac{1}{\sqrt{\delta \new}}\xthat{t}{t}\right)} \Pt{t}{t} ^{-1}  
\left( \xt{t} - \frac{1}{\sqrt{\delta \new}}\xthat{t}{t}\right)  -\frac{1}{2}  
\trans{ (\xt{t+1} -\xt{t})}  \Diag{\theta_1, \dots, \theta_\betad}^{-1} (\xt{t+1} -\xt{t})  \right\} .
\end{eqnarray*}
Completing squares yields:
\begin{eqnarray*}
&& \Pt{t}{T} = \left(  \Pt{t}{t}^{-1} + \Diag{\theta_1, \dots, \theta_\betad}^{-1}  \right)^{-1}, \\
&&\xthat{t}{T}(\xt{t+1} ) 
= \Pt{t}{T} \left(\Pt{t}{t}^{-1} \frac{1}{\sqrt{\delta \new}}\xthat{t}{t}  + 
\Diag{\theta_1, \dots, \theta_\betad}^{-1} \xt{t+1}  \right).
\end{eqnarray*}
These moments can be expressed as in (\ref{posmmc}).}
\begin{eqnarray}
&& \xt{t}|\xt{t+1},\dots, \xt{T}, \zv \newS , \scalev, \theta_1, \dots, \theta_d \sim
\Normult{\betad}{\xthat{t}{T}(\xt{t+1} ),\Pt{t}{T}},\label{posmmc}\\
&& \xthat{t}{T}(\xt{t+1} )= \frac{1}{\sqrt{\delta \new}}(\identm- \Bt{t+1}) \xthat{t}{t} + \Bt{t+1} \xt{t+1} , \nonumber
\\&&\Pt{t}{T}= (\identm- \Bt{t+1})
\Pt{t}{t}, \qquad \Bt{t+1} = \Pt{t}{t}  \left( \Pt{t}{t}
+   \Diag{\theta_1, \dots, \theta_\betad} \right)^{-1}. \nonumber
\end{eqnarray}
Finally, the unknown variances $\theta_1, \dots, \theta_d$ are updated conditional on the state process $\betat{0},\dots,\betat{T}$ and $\zv \newS$.
This requires the choice of a  prior
$p(\theta_j)$ and  the inverse gamma prior $ \theta_j \sim  \Gammainv{c_0, C_0}$  is used for illustration. However, this prior is easily substituted by 
variance selection priors, such as the triple gamma prior \citepsupp{cad-etal:tri}.

The ultimate P\'{o}lya-Gamma sampler for binary SSMs is summarized in  Algorithm~\ref{alg:SSM}.

\begin{algorithm}
  \caption{The ultimate P\'{o}lya-Gamma sampler for binary state space models.}  \label{alg:SSM}
  \footnotesize
  Choose starting values for  $(\theta_1, \dots, \theta_d)$, and  $\lambda_1, \dots, \lambda_T$,
  where $ \lambda_t=\Xbeta_{t} \betat{t}$ and repeat the following steps:
\begin{itemize} 
  \item[(Z)] For each $t=1, \dots, T$,  sample $\yprodiff_t =
  \log \lambda_t +\Ferror^{-1} ( y_t +    U_t (1-y_t -\pi_t))$ in the SSM (34), 
where $U_t \sim \Uniform{0,1}$ and $\pi_t= \Ferror(\log \lambda_t)$.
  Sample
  $ \omegaH_t | \yprodiff_{t}, \lambda_t ,y_t \sim \PG{2, |\yprodiff_{t}-\log \lambda_t |}$
 for the logit SSM.

   \item[(B-L)] Location-based parameter expansion: sample $\Star{\gamma}  \sim \Normal{0,G_0} $ and propose $\tilde{\yprodiff}_{t} = \yprodiff_{t}+ \Star{\gamma}$ for $t=1, \dots,T$.  Sample  $\gamma \new $ from $ \gamma| \scalev, \tilde{\zv},\theta_1, \dots, \theta_d ,\ym$ given in (\ref{GVVVssm})
       and  define 
       the shifted utilities $ \yprodiff_{t} \shift = \tilde{\yprodiff}_{t} - \gamma \new = \yprodiff_{t} + \Star{\gamma}-\gamma \new$.
       
        \item[(B-S)] Scale-based parameter expansion:
sample  $\Star{\delta} \sim \Gammainv{d_0,D_0}$ and
propose $\tilde{\yprodiff}_{t} = \sqrt{\Star{\delta}} \yprodiff_{t} \shift $ for all $t$. Sample  $\delta \new $ from the inverse Gamma density $ \delta|  \tilde \zv ,\scalev, \theta_1, \dots, \theta_d$
given in (\ref{scdelH}). Define rescaled utilities
 $ \yprodiff_{t} \newS = \tilde{\yprodiff}_{t}/\sqrt{\delta \new} = \sqrt{\Star{\delta}/\delta \new}\yprodiff_{t} \shift$ for all $t$.

    \item[(F)] Sample the  state process $\betat{0},\dots,\betat{T} | \zv \newS, 
    \SFS{\scalev}, \theta_1, \dots, \theta_d  $ using backward-sampling, based on the Kalman filter from step (B-S), see
    \SFS{(\ref{KFboosmo}) and}    (\ref{posmmc}). 
    
  \item [(P)] Sample   $ \theta_j |  \{\beta_{jt}\}
    \sim  \Gammainv{c_0 +  T/2 ,  C_j}$, for $j=1, \dots,d$, where
    \begin{eqnarray*} 
    C_j = C_0 + \frac{1}{2} \sum_{t=1}^T (\beta_{jt}- \beta_{j,t-1})^2  . 
   \end{eqnarray*}
\end{itemize}
\end{algorithm}

\subsection{Additional Simulation Results} 
\label{app:sim}

This section provides some additional simulation results for probit, logit, multinomial and binomial regression models. In addition, a numerical study on mixture-of-experts models is presented. The results for the probit model are discussed in detail in Appendix~\ref{probres}. Selected tabulated simulation results for all models are given in Appendix~\ref{sec:add}. The mixture-of-experts results are given in Appendix~\ref{app:moe}.

\begin{figure}[t!]
\centering
\begin{subfigure}{0.3\textwidth}
  \centering
  \includegraphics[width=\linewidth]{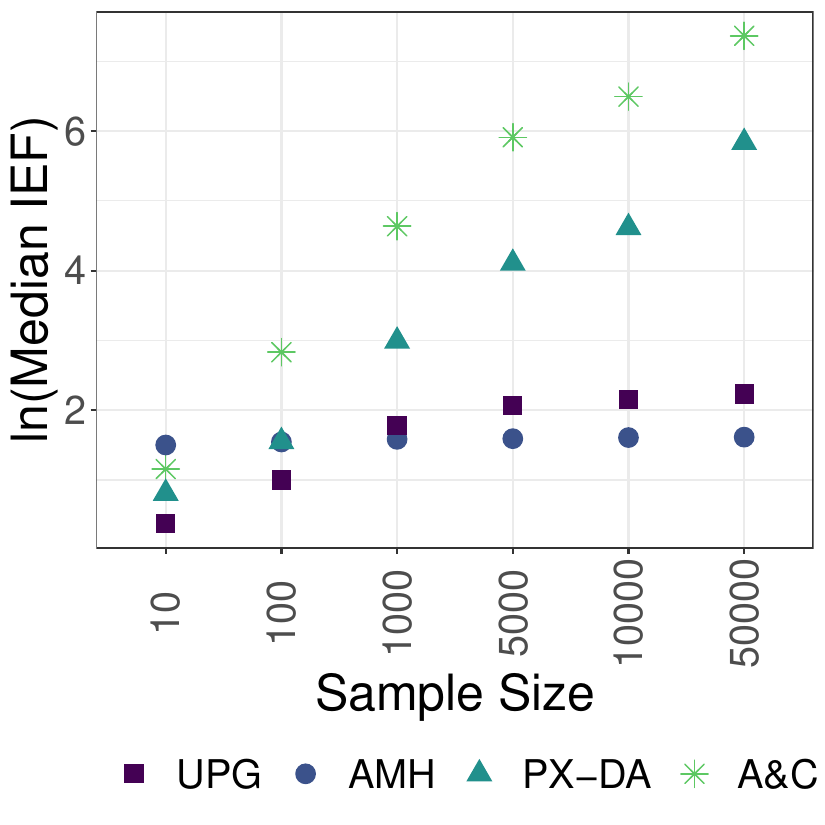}
  \caption{Varying sample sizes with two successes.}
\end{subfigure}\hspace{4em}
\begin{subfigure}{0.3\textwidth}
  \centering
  \includegraphics[width=\linewidth]{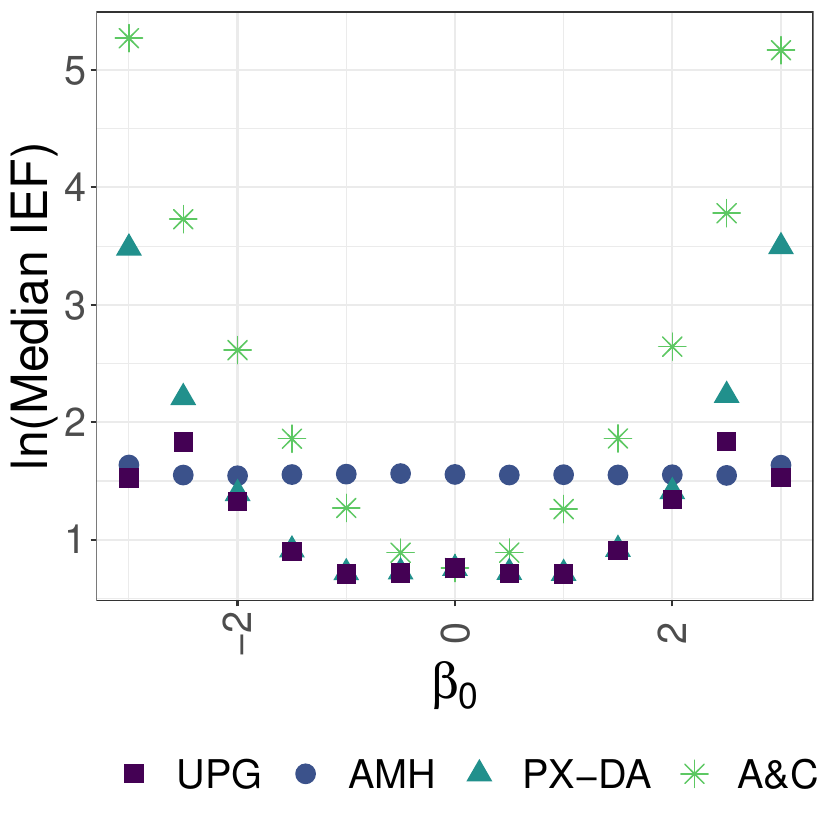}
  \caption{Varying Intercepts with $N = 1,000$.}
\end{subfigure}
\caption{Log median inefficiency factors across 100 simulation runs for binary probit models.}
\label{fig:probit_n}
\end{figure}

\subsubsection{Probit Results}  
\label{probres}

For binary probit models, we compare the 
\HW{iMDA} approach outlined in this article \HW{(UPG)} to an adaptive Metropolis-Hastings sampler (AMH), the MDA scheme from \citetsupp{liu-wu:par} (PX-DA), as well as the original probit Gibbs sampler outlined in \citetsupp{alb-chi:bay_ana} (A\&C). The results for are summarized visually in \autoref{fig:probit_n} and numerical results are provided in \HW{\autoref{tab:probit_n}}. It is obvious that the commonly used standard DA approach from A\&C has severe efficiency problems when increasing the level of imbalancedness of the data. In these cases, increasing inefficiency can also be observed for the PX-DA algorithm. On the other hand, constantly high levels of efficiency are characterizing both UPG and the AMH algorithm.


\subsubsection{Tabulated Simulation Results}
\label{sec:add}

This subsection provides some additional tables summarizing results from various simulation studies. \autoref{tab:probit_n} to \autoref{tab:binomial_n} give detailed results on sampling efficiency for probit as well as binary, multinomial and binomial logistic regression models in scenarios with changing sample size $N$. For the probit case, we choose the PX-DA algorithm of \citet{liu-wu:par} as benchmark, while the algorithms of \citet{pol-etal:bay_inf} serve as baseline for the logistic regression algorithms. The results of all other competing models are reported relative to these benchmarks for ease of interpretation.

\newcommand{\ADDval}{Values of, respectively, ESS / $\text{ESS}^{\text{B}}>1$ and IE / IE$^{\text{B}} <1$ indicate superiority compared to the benchmark}

\begin{table}[h]
\centering
\begin{adjustbox}{max width=\textwidth}
\begin{threeparttable}
\caption{Varying Sample Sizes for Binary Probit Models}
\label{tab:probit_n}
\setlength\tabcolsep{10pt}
\renewcommand\arraystretch{1.3}
\begin{tabular}{llrrrrrr}
\toprule
Binary Probit &  & N = 10 & N = 100 & N = 1,000 & N = 5,000 & N = 10,000 & N = 50,000\\
\midrule
PX-DA & ESS$^{\text{B}}$ & 4451.285 & 2120.910 & 502.611 & 164.747 & 98.851 & 29.174\\
 & IE$^{\text{B}}$ & 2.247 & 4.715 & 19.896 & 60.699 & 101.162 & 342.791\\
\addlinespace
UPG & ESS / ESS$^{\text{B}}$ & 1.546 & 1.735 & 3.341 & 7.638 & 11.693 & 36.785\\
 & IE / IE$^{\text{B}}$ & 0.647 & 0.576 & 0.299 & 0.131 & 0.086 & 0.027\\
\addlinespace
AMH & ESS / ESS$^{\text{B}}$ & 0.501 & 1.006 & 4.088 & 12.355 & 20.265 & 68.240\\
 & IE / IE$^{\text{B}}$ & 1.998 & 0.994 & 0.245 & 0.081 & 0.049 & 0.015\\
\addlinespace
A\&C & ESS / ESS$^{\text{B}}$ & 0.707 & 0.278 & 0.193 & 0.164 & 0.153 & 0.217\\
 & IE / IE$^{\text{B}}$ & 1.414 & 3.599 & 5.185 & 6.083 & 6.534 & 4.608\\
\bottomrule
\end{tabular}
\begin{tablenotes}
\item \small \textit{Note:} Values are medians across 100 simulation runs. Every simulated dependent variable has two successes. 
\HW{IE} 
= inefficiency factors, ESS = effective sample size. \HW{$\text{IE}^{\text{B}}$}  and $\text{ESS}^{\text{B}}$ correspond to the simulation results of the benchmark sampler PX-DA of \citetsupp{liu-wu:par}. Results for all other samplers are reported relative to this benchmark. \HW{\ADDval.}
\end{tablenotes}
\end{threeparttable}
\end{adjustbox}
\end{table}  

\begin{table}[h]
\centering
\begin{adjustbox}{max width=\textwidth}
\begin{threeparttable}
\caption{Varying Sample Sizes for Binary Logit Models}
\label{tab:logit_n}
\setlength\tabcolsep{10pt}
\renewcommand\arraystretch{1.3}
\begin{tabular}{llrrrrrr}
\toprule
Binary Logit &  & N = 10 & N = 100 & N = 1,000 & N = 5,000 & N = 10,000 & N = 50,000\\
\midrule
PSW & ESS$^{\text{B}}$ & 5851.204 & 895.380 & 146.321 & 39.673 & 24.457 & 8.027\\
 & IE$^{\text{B}}$ & 1.709 & 11.168 & 68.343 & 252.062 & 408.884 & 1245.794\\
\addlinespace
UPG & ESS / ESS$^{\text{B}}$ & 0.901 & 2.802 & 9.649 & 29.521 & 46.139 & 132.403\\
 & IE / IE$^{\text{B}}$ & 1.110 & 0.357 & 0.104 & 0.034 & 0.022 & 0.008\\
\addlinespace
AMH & ESS / ESS$^{\text{B}}$ & 0.266 & 1.687 & 10.304 & 37.911 & 61.811 & 187.009\\
 & IE / IE$^{\text{B}}$ & 3.763 & 0.593 & 0.097 & 0.026 & 0.016 & 0.005\\
\addlinespace
FSF & ESS / ESS$^{\text{B}}$ & 0.367 & 0.270 & 0.209 & 0.228 & 0.231 & 0.378\\
 & IE / IE$^{\text{B}}$ & 2.723 & 3.707 & 4.782 & 4.378 & 4.322 & 2.643\\
\bottomrule
\end{tabular}
\begin{tablenotes}
\item \small \textit{Note:} Values are medians across 100 simulation runs. Every simulated dependent variable has two successes. \HW{IE}  = inefficiency factors, ESS = effective sample size. \HW{$\text{IE}^{\text{B}}$}  and $\text{ESS}^{\text{B}}$ correspond to the simulation results of the benchmark sampler PSW. Results for all other samplers are reported relative to this benchmark. \HW{\ADDval.}
\end{tablenotes}
\end{threeparttable}
\end{adjustbox}
\end{table}

\begin{table}[h]
\centering
\begin{adjustbox}{max width=\textwidth}
\begin{threeparttable}
\caption{Varying Sample Sizes for Multinomial Logit Models}
\label{tab:MNL_n}
\setlength\tabcolsep{10pt}
\renewcommand\arraystretch{1.3}
\begin{tabular}{llrrrrrr}
\toprule
Multinomial Logit &  & N = 10 & N = 100 & N = 1,000 & N = 5,000 & N = 10,000 & N = 50,000\\
\midrule
PSW & ESS$^{\text{B}}$ & 5148.712 & 898.435 & 148.772 & 40.204 & 24.609 & 7.654\\
 & IE$^{\text{B}}$ & 1.943 & 11.154 & 67.690 & 252.732 & 424.122 & 1414.031\\
\addlinespace
UPG & ESS / ESS$^{\text{B}}$ & 0.905 & 2.779 & 9.422 & 28.961 & 45.219 & 138.636\\
 & IE / IE$^{\text{B}}$ & 1.105 & 0.359 & 0.106 & 0.034 & 0.021 & 0.007\\
\addlinespace
AMH & ESS / ESS$^{\text{B}}$ & 0.286 & 1.734 & 10.454 & 39.047 & 63.901 & 206.778\\
 & IE / IE$^{\text{B}}$ & 3.503 & 0.576 & 0.095 & 0.025 & 0.015 & 0.004\\
\addlinespace
FSF & ESS / ESS$^{\text{B}}$ & 0.372 & 0.277 & 0.211 & 0.218 & 0.248 & 0.417\\
 & IE / IE$^{\text{B}}$ & 2.686 & 3.638 & 4.947 & 4.912 & 4.224 & 2.435\\
\bottomrule
\end{tabular}
\begin{tablenotes}
\item \small \textit{Note:} Values are medians across 100 simulation runs. Every simulated dependent variable has two successes. \HW{IE}  = inefficiency factors, ESS = effective sample size. \HW{$\text{IE}^{\text{B}}$}  and $\text{ESS}^{\text{B}}$ correspond to the simulation results of the benchmark sampler PSW. Results for all other samplers are reported relative to this benchmark. \HW{\ADDval.}
\end{tablenotes}
\end{threeparttable}
\end{adjustbox}
\end{table}

\begin{table}[h]
\centering
\begin{adjustbox}{max width=0.8\textwidth}
\begin{threeparttable}
\caption{Varying Sample Sizes for Binomial Logit Models}
\label{tab:binomial_n}
\setlength\tabcolsep{10pt}
\renewcommand\arraystretch{1.3}
\begin{tabular}{llrrrrr}
\toprule
Binomial Logit &  & N = 10 & N = 100 & N = 1,000 & N = 5,000 & N = 10,000\\
\midrule
PSW & ESS$^{\text{B}}$ & 1556.512 & 248.392 & 40.558 & 11.797 & 8.808\\
 & IE$^{\text{B}}$ & 6.425 & 40.259 & 246.568 & 847.677 & 1135.303\\
\addlinespace
UPG & ESS / ESS$^{\text{B}}$ & 2.314 & 8.137 & 32.121 & 96.350 & 123.117\\
 & IE / IE$^{\text{B}}$ & 0.432 & 0.123 & 0.031 & 0.010 & 0.008\\
\addlinespace
AMH & ESS / ESS$^{\text{B}}$ & 0.988 & 6.076 & 37.598 & 128.136 & 172.595\\
 & IE / IE$^{\text{B}}$ & 1.012 & 0.165 & 0.027 & 0.008 & 0.006\\
\addlinespace
AuxMix & ESS / ESS$^{\text{B}}$ & 0.260 & 0.198 & 0.202 & 0.330 & 0.342\\
 & IE / IE$^{\text{B}}$ & 3.853 & 5.054 & 4.961 & 3.031 & 2.923\\
\bottomrule
\end{tabular}
\begin{tablenotes}
\item \small \textit{Note:} Values are medians across 100 simulation runs. Every simulated dependent variable has two successes. \HW{IE}  = inefficiency factors, ESS = effective sample size. \HW{$\text{IE}^{\text{B}}$} and $\text{ESS}^{\text{B}}$ correspond to the simulation results of the benchmark sampler PSW. Results for all other samplers are reported relative to this benchmark. \HW{\ADDval.}
\end{tablenotes}
\end{threeparttable}
\end{adjustbox}
\end{table}

\clearpage

\subsubsection{Mixture-of-Experts Simulations}
\label{app:moe}

\begin{figure}
    \centering
    \includegraphics[width=0.8\textwidth]{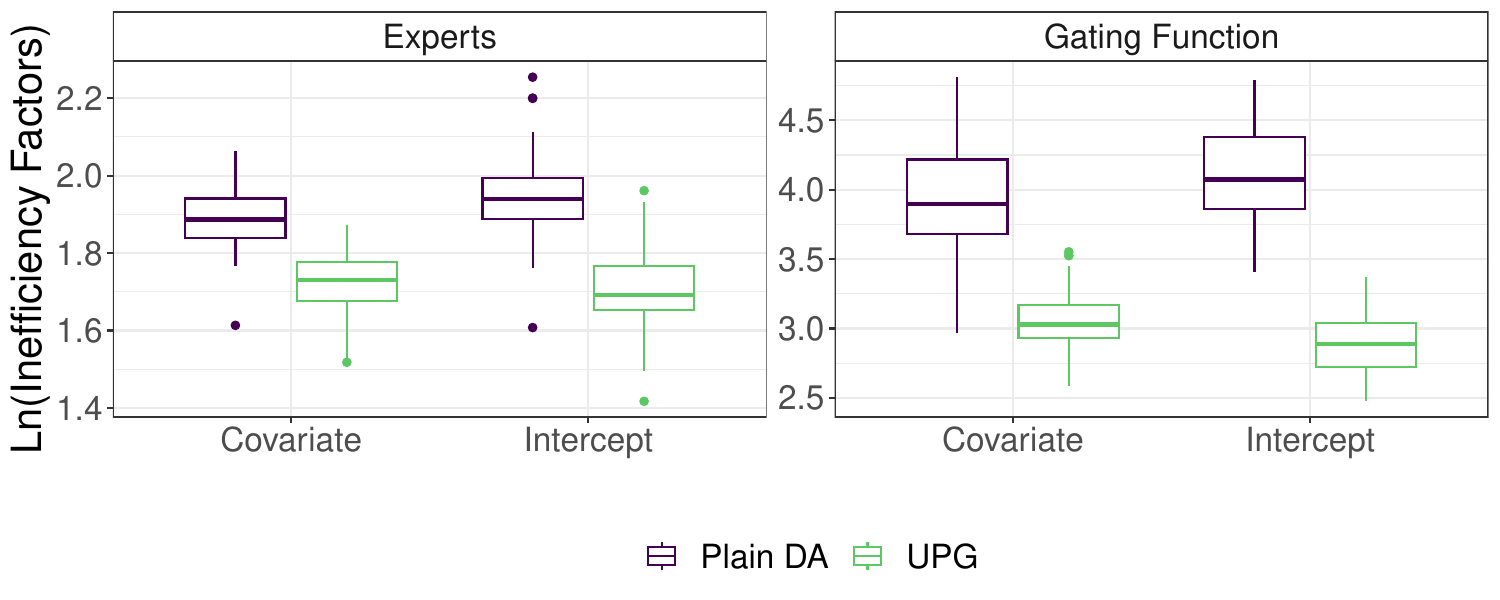}
    \caption{Log inefficiency factors for simulated mixture-of-experts data. \textit{UPG} is a data augmentation scheme with location-based and scale-based parameter expansion steps.}
    \label{fig:moe_sim}
\end{figure}

To evaluate whether the proposed MDA scheme can be expected to increase efficiency in complex settings such as mixture-of-experts models, we conduct a numerical experiment as follows. As in the application presented in Section 6.2, 
we consider logistic regression experts and a multinomial logistic gating function. We simulate $H = 3$ mixture components and $N = 1,000$ units that are observed $N_i = 20$ times each. Both for the gating function and the logistic regression experts, an intercept term and a $\mathcal{N}(0,1)$ covariate are used to simulate the respective outcomes. The regression coefficients of the gating function are $\bm{\psi}_1 = (-4, 2)'$, $\bm{\psi}_2 = (-4, -2)'$ and $\bm{\psi}_3 = (0, 0)'$. The regression parameters in the logistic regression experts are set to $\bm{\beta}_1 = (-4, 2.5)'$, $\bm{\beta}_2 = (4, 2.5)'$ and $\bm{\beta}_3 = (0, 0)'$. This corresponds to a setup where both the classification and the success probabilities within the logistic regression components are relatively imbalanced. Specifically, the class membership probabilities that the gating function produces are on average between 6\% and 7\% for the first two categories and 87\% for the baseline category. Within the logistic experts, the first category has an average success probability of around 10\%, the second one of around 90\% and the baseline category is balanced with 50\% success probability.

We simulate 50 replicate data sets using these parameters and estimate the model once with and once without \HW{iMDA}, collecting inefficiency factors in each simulation run. Throughout the simulation runs, we assume independent $\mathcal{N}(0,4)$ priors for all elements of $\bm{\psi}_1$, $\bm{\psi}_2$, $\bm{\beta}_1$, $\bm{\beta}_2$ and $\bm{\beta}_3$. Figure~\ref{fig:moe_sim} summarizes the results of these 50 runs and shows that the proposed MDA scheme offers significant performance gains when compared to a plain DA sampler. These gains are expected to become larger as the data becomes more imbalanced.

Finally, it is worth to note that mixture-of-experts models with logistic regression experts potentially suffer from inefficiencies stemming from two sources. First, the degree of dependency of the parameters during sampling is potentially high due to class membership being dependent on the parameters in the gating function and vice versa. Second, there may be inefficiencies due to imbalanced outcomes, either in the gating function or the logistic experts or both. The iMDA scheme introduced in the present article is concerned with the latter type of inefficiency.

\subsection{Mixture-of-experts illustration using child mortality data}
\label{app:childmort_moe}

The goal of this illustration is to examine heterogeneity and non-linearities in the relationship of child mortality and maternal education in developing countries. For this, we construct a large data set on $N = 99,641$ births in eight countries in sub-Saharan Africa using household survey data from the \textit{Demographic and Health Survey (DHS)} program. For each birth, the survey data indicates whether the child died before its fifth birthday. To measure maternal education, a categorical indicator on whether a child's mother has no formal education or achieved a primary, secondary or tertiary level of education is extracted. The data itself has been collected at $J = 5,558$ distinct geographical locations. At an average location, around 18 children are observed, and a location can be thought of as, e.g., a small village or a neighborhood within a city. The data set contains information on malaria incidence and urban/rural status for all $J$ survey clusters. All data is publicly available from IPUMS-DHS (\citealpsupp{ipumsDHS}).

We denote the mortality status of child $i = 1, \dots, N$ as $y_{i}$ and use $C_i = j$ to indicate that child $i$ has been observed at location $j$. To examine potential non-linearities in the relationship of child mortality and maternal education, we assume that $y_{i}$ can be modeled using a $H$-component mixture ($h = 1, \dots, H$) of logistic regression models
\begin{equation} \label{A55}
    y_{i}~|~C_i = j \sim {\sum_{h=1}^{H}} \eta_{jh}(\bm{w}_j) 
    Ber(\zeta_{ih}  \HW{(\bm{x}_i)}).
\end{equation}
%
\SFS{In model (\ref{A55}), 
 where $\sum_h \eta_{jh}  \HW{(\bm{w}_j)} = 1$, 
$\eta_{jh} \HW{(\bm{w}_j)}$} is the probability that location $j$ is a member of mixture component $h$. The \SFS{corresponding categorical}  class membership indicator $S_j$ takes on values $h = 1, \dots, H$ and is modeled as the outcome of a multinomial logistic regression model such that
\begin{equation}
\label{eq:moe_mnl}
    \Prob{S_j = h~|~\bm{w}_j} =  \eta_{jh} \HW{(\bm{w}_j)} = \frac{\text{exp}(\bm{w}_j\bm{\psi}_h)}{ {\sum_{l=1}^{H}}\text{exp}( \bm{w}_j \bm{\psi}_l)},
\end{equation}
where the regression parameters $\bm{\psi}_h$ of one category are set to zero for identification purposes. The covariate vector $\bm{w}_j$ contains an intercept, the malaria incidence at the survey location in the closest available year to the survey year, as well as a binary variable indicating whether a survey cluster is located in an urban area. In addition, to let country-specific factors such as the effectiveness of public health care systems influence class membership, binary vectors indicating whether the cluster is located in Ethiopia, Kenya, Malawi, Mozambique, Rwanda, Tanzania or Zimbabwe are included, with the DR Congo serving as baseline. The country-specific intercepts also control for the fact that the surveys in the countries have been conducted in different years.

\SFS{For  each  component $h$ in model (\ref{A55})}, $\zeta_{ih} \HW{(\bm{x}_i)} $ is the mortality rate, i.e., the probability of a child $i$ dying before the age of five, conditional on survey location $j$ being a member of mixture component $h$ \SFS{(i.e. $S_j = h$)}. 
$\zeta_{ih} \HW{(\bm{x}_i)} $ is modeled as a function of maternal education using the logistic link
\begin{equation}
\label{eq:moe_log}
        \zeta_{ih} \HW{(\bm{x}_i)} = \Prob{y_{i} = 1~|~C_i = j, S_j = h, \HW{\bm{x}_i} } = \frac{\text{exp}(\beta_{0h} + \sum_{s = 1}^3 \beta_{sh}\text{EDUC}^s_{i})}{1 + \text{exp}(\beta_{0h} + \sum_{s = 1}^3 \beta_{sh}\text{EDUC}^s_{i})},
\end{equation}
where $\HW{\bm{x}_i=(\text{EDUC}^1_{i}, \text{EDUC}^2_{i}, \text{EDUC}^3_{i})}$ 
are binary indicators for primary, secondary and tertiary education level of the mother \HW{of child $i$} and \textit{no formal education} serves as reference category. 

Both the component-specific logistic regressions and the multinomial logistic regression that serves as class membership prior are estimated using the methodology introduced in this paper, with $\mathcal{N}(0,4)$ priors on all regression parameters. We examine models of order $H = 1, \dots, 7$ and use the full mixture likelihood to determine the appropriate number of components.\footnote{For a formal discussion on model selection in mixture models, refer to \citetsupp{cel-etal:mod}.} The BIC selects $H = 2$ as the most suitable number of components, and we discuss results for this case in more detail below. Posterior estimates are based on 25,000 posterior draws after a burn-in period of 10,000 iterations using a random permutation sampler to deal with label switching. Identification of the mixture parameters is achieved via k-means clustering of the posterior draws in the point process representation, see for instance \citetsupp{mal-etal:mod} for more details. In terms of sampling efficiency, the effective sample sizes of the parameters of the logistic experts and the gating function coefficients in the boosted sampler are on average larger than in the plain DA approach. As the results presented here are based on a single MCMC chain, we refer to Appendix~\ref{app:moe} for more robust evidence on efficiency gains in mixture-of-experts settings.

\begin{figure}[!t]
  \centering
\begin{subfigure}{0.45\textwidth}
  \centering
  \includegraphics[width=\linewidth]{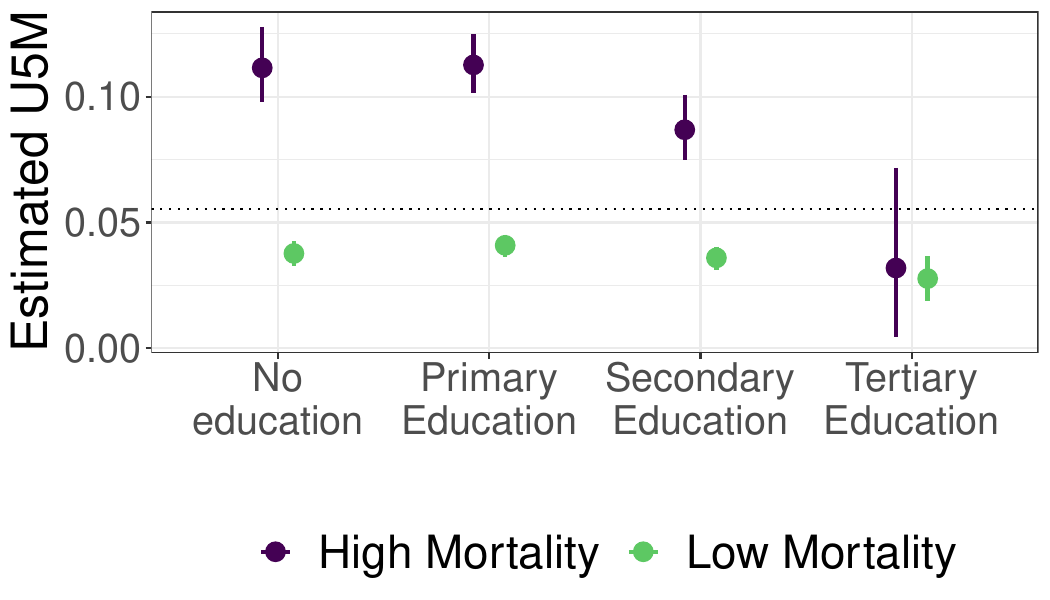}
  \caption{Estimated \HWn{under five }mortality rates  from logistic experts.}
\end{subfigure}
\begin{subfigure}{0.4\textwidth}
  \centering
  \includegraphics[width=\linewidth]{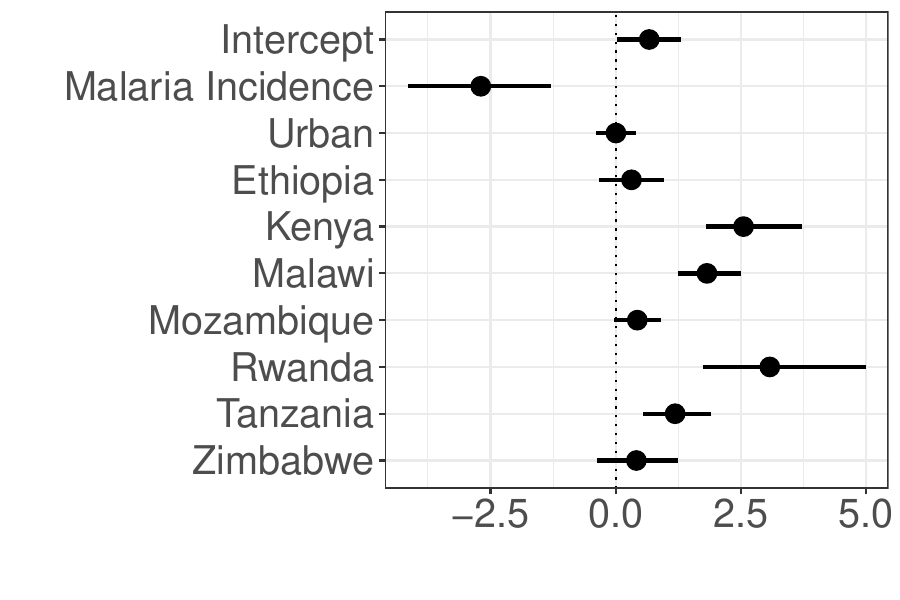}
  \caption{Estimated coefficients gating function.}
\end{subfigure}
\caption{(left) Component-specific child mortality estimates by maternal education groups. (right) Estimated gating function coefficients. Baseline category is the low mortality component. Uncertainty bounds correspond to 95\% credible intervals, and the dotted line in the left panel is the sample average mortality rate.}
\label{fig:mortality}
\end{figure}

Panel (a) in \autoref{fig:mortality} shows the estimated relationship of maternal education and \SFS{under five} mortality rates in the two mixture components. The 'low mortality' component exhibits relatively low mortality rates that do not systematically decrease with maternal education. The 'high mortality' component is characterized by significantly higher mortality rates and a pronounced, negative relationship of maternal education and child mortality. Panel (b) in \autoref{fig:mortality} shows posterior summaries of the gating function coefficients $\bm{\psi}_h$ that determine the mixture weights $\eta_{jh}  \SFS{(\bm{w}_j)} $. The baseline category corresponds to the high mortality component. The most striking observation is that a high malaria incidence at the survey cluster location strongly decreases the odds of being a member of the low mortality component. This is in line with the fact that malaria is particularly dangerous for children under five, who for instance accounted for 67\% of malaria deaths worldwide in 2019 (\citealpsupp{who2020}). A potential channel behind the implied significantly negative relationship of maternal education and mortality in the high mortality component is that maternal education is a critical factor in preventing malaria infections in young children (\citealpsupp{njau2014investigating}).



%

\bibliographystylesupp{agsm}
\bibliographysupp{sylvia_kyoto.bib}


\end{document}